\documentclass[journal]{new-aiaa}
\usepackage[utf8]{inputenc}
\usepackage{textcomp}
\usepackage{graphicx}
\usepackage{amsmath}
\usepackage[version=4]{mhchem}
\usepackage{siunitx}
\usepackage{longtable,tabularx}
\usepackage{multirow}
\usepackage{textalpha}
\usepackage[table,xcdraw]{xcolor}
\usepackage{lineno}
\usepackage{floatrow}
\usepackage[labelformat=brace,labelfont=normalsize]{subfig}
\usepackage{caption}
\usepackage{utfsym}
\definecolor{myred}{RGB}{206, 90, 103}
\definecolor{mygreen}{RGB}{0, 126, 0}
\definecolor{myblue}{RGB}{0, 102, 204}
\definecolor{darkgreen}{RGB}{156, 10, 100}
\usepackage[final,commandnameprefix=ifneeded,defaultcolor=myblue]{changes}
\newcommand{\Tau}{\mathcal{T}} 
\title{Uncertainty-aware Surrogate Models for Airfoil Flow Simulations with Denoising Diffusion Probabilistic Models \\ \textcolor{darkgreen}{\small \\ Addendum Sept. 20024: This version contains updates regarding accuracy-per-performance and flow matching. \\ The added sections are highlighted with subscript "$\star$." The original version can be found at \href{https://arc.aiaa.org/doi/10.2514/1.J063440}{here}.}}

\author{Qiang Liu \footnote{
Ph.D. Candidate, School of Computation, Information and Technology, qiang7.liu@tum.de} and Nils Thuerey\footnote{Professor, School of Computation, Information and Technology,  nils.thuerey@tum.de}}
\affil{Technical University of Munich, Garching, Germany, D-85748}

\begin{document}
\maketitle

\begin{abstract}
Leveraging neural networks as surrogate models for turbulence simulation is a topic of growing interest. At the same time, embodying the inherent uncertainty of simulations in the predictions of surrogate models remains very challenging. The present study makes a first attempt to use denoising diffusion probabilistic models (DDPMs) to train an uncertainty-aware surrogate model for turbulence simulations. Due to its prevalence, the simulation of flows around airfoils with various shapes, Reynolds numbers, and angles of attack is chosen as the learning objective. Our results show that DDPMs can successfully capture the whole  distribution of solutions and, as a consequence, accurately estimate the uncertainty of the simulations. The performance of DDPMs is also compared with varying baselines in the form of Bayesian neural networks and heteroscedastic models. Experiments demonstrate that DDPMs outperform the other methods regarding a variety of accuracy metrics. Besides, it offers the advantage of providing access to the complete distributions of uncertainties rather than providing a set of parameters. As such, it can yield realistic and detailed samples from the distribution of solutions. \textcolor{darkgreen}{Addendum$^\star$:} We also evaluate an emerging generative modeling variant, flow matching, in comparison to regular diffusion models. The results demonstrate that flow matching addresses the problem of slow sampling speed typically associated with diffusion models. As such, it offers a promising new paradigm for uncertainty quantification with generative models.
\end{abstract}

\section*{Nomenclature}

{\renewcommand\arraystretch{1.0}
\noindent\begin{longtable*}{@{}l @{\quad=\quad} l@{}}
$A$  & wing area, $\mathrm{m}^2$ \\
$C_d$  & drag coefficient \\
$\mathbf{d}$  & training dataset \\
$\mathbb{E}$ & expectation of a distribution \\
$\mathbf{F_d}$  & drag force, $\mathrm{N}$ \\
$h$ & cell size of the data field, $\mathrm{m}$ \\
$\mathbf{I}$ & unit tensor \\
$KL$ & Kullback–Leibler divergence \\
$l$ & chord length, $\mathrm{m}$ \\
$L$ & number of basic blocks in the U-Net \\
$\mathcal{L}_{\mathrm{NN}}$ & training loss function of a network \\
$M$ & number of simulation cases in the training dataset \\
$\mathbf{n}$ & unit normal vector of an airfoil shape \\
$N$ & number of snapshots samples in training dataset \\
$\widehat{N}$ & number of snapshots samples in test dataset \\
$\mathcal{N}$ & Gaussian distribution \\
$p,q$ & probability density function of a distribution \\
$P$ & probability distribution \\
$\mathcal{P}$  & physical system of airfoil flow\\
$\mathrm{p}$ & pressure, $\mathrm{pa}$ \\
$Re$  & Reynolds number \\
$s$  & resolution of a field data\\
$\mathcal{S}$  & simulator of airfoil flow\\
$t$  & index of Markov chain steps in DDPM\\
$T$  & number of Markov chain steps in DDPM\\
$\mathbf{u}$ & velocity vector, m/s \\
$\mathbf{u_f}$ & freestream velocity vector, m/s \\
$\mathrm{u}_f$ & velocity component in freestream direction, m/s \\
$\mathrm{u}_x$ & velocity component in chord direction, m/s \\
$\mathrm{u}_y$ & velocity component perpendicular to chord direction, m/s \\
$U$ & uniform distribution \\
$\mathbf{x}$  & physical parameters of airfoil flow, $\mathbf{x}=[\Omega,\alpha,Re]$\\
$\mathbf{y}$  & flow field, $\mathbf{y}=[\mathrm{p^*},\mathbf{u}^*]$\\
$x$  & coordinates in chord direction\\
$y$  & coordinates perpendicular to chord direction\\
$\alpha$  &  angle of attack, $^\circ$ \\
$\beta^t$ &  hyperparameter of DDPM controlling the noise schedule \\
$\beta_1,\beta_2$ &  hyperparameters of AdamW optimizer  \\
$\gamma^t$ &  hyperparameter of DDPM, $\gamma^t=1-\beta^t$  \\
$\bar{\gamma}^t$ & hyperparameter of DDPM, $\bar{\gamma}^t = \prod_{i=1}^t \gamma^i$  \\
$\boldsymbol{\epsilon}$ &  Gaussian noise  \\
$\theta$ & parameters of a neural network\\
$\boldsymbol{\mu}_{v}$ & expected value of variable $v$ where $v \in \{\mathbf{y},\mathrm{p^*},\mathrm{u_x^*},\mathrm{u_y^*}\}$\\
$\nu$ & viscosity of air, $\mathrm{m^2/s}$ \\
$\xi$ & parameters of the surrogate model \\
$\rho$ & density of air, $\mathrm{kg/m^3}$ \\
$\boldsymbol{\sigma}_{\mathrm{v}}$ & standard deviation of variable $v$ where $v \in \{\mathbf{y},\mathrm{p^*},\mathrm{u_x^*},\mathrm{u_y^*,|\mathbf{u^*}|}\}$  \\
$\tau$ & iteration numbers of a simulation\\
$\Tau$ & random variable corresponding to $\tau$\\
$\phi$ & parameters of a variational distribution\\
$\psi$ & numerical parameters of a simulation\\
$\Omega$ & shape of an airfoil\\
\multicolumn{2}{@{}l}{Subscripts}\\
$a$ & average value of a field \\
$g$ & ground truth data \\
$1,2,\cdots i \cdots N$ & $i$th sample in a distribution \\
$k$ & $k$th data point of a field \\
$\theta$ & predictions of a network parameterized by $\theta$ \\
$\xi$ & predictions of the surrogate model parameterized by $\xi$ \\
$\phi$ & predictions of the variational distribution parameterized by $\phi$\\
\multicolumn{2}{@{}l}{Superscript}\\
$1,2,\cdots t \cdots$  T& $t$th intermediate state in a DDPM Markov chain\\
$*$ & dimensionless variable 
\end{longtable*}}

\section{Introduction}
From fuel combustion in car engines~\cite{ICTaylor1995,ReviewLumley2001} to supersonic flow around aircraft airfoils~\cite{Nieuwland1973,Drela1987}, turbulence is ubiquitous in modern engineering. 
Despite the rapidly advancing power of modern computers, 
simplified turbulence models, like Reynolds-averaged Navier–Stokes simulations (RANS)~\cite{Alfonsi_RANS2009} and Large eddy simulations (LES)~\cite{Georgiadis_LES2010}, are still prevalent for turbulence simulation in the engineering community~\cite{Argyropoulos_turbluence_2015}. Simplified turbulent models bring uncertainties to the simulation results by introducing hypotheses and parameters to be determined~\cite{Duraisamy_review_2019}. 
Estimating and mitigating these uncertainties is essential for turbulence simulation, and numerous approaches such as perturbations methods~\cite{Iaccarino2017,Mishra2019}, random matrix approaches~\cite{wang_2016,xiao2017}, and polynomial chaos techniques~\cite{Najm2009,Roberts2011} have been proposed.

Meanwhile, deep learning techniques have permeated the ﬁeld of fluid dynamics research in the last few years~\cite{Brunton2020,Vinuesa2022,lino2023current}. On the one hand, many successful applications, e.g., for turbulence closure modeling~\cite{Brendan2015,Durbin2018} and detecting regions of high uncertainty~\cite{Ling2015}, have shown the potential of deep learning methods in modeling the uncertainty of turbulence simulations. On the other hand, deep learning methods have demonstrated promising capabilities as surrogate models for turbulent phenomena~\cite{thuerey2020,chen2021,Sabater2022,CHEN2023}. Considering the inherent uncertainty of the underlying simulations, the prediction of surrogate models ideally encompasses a \textit{probabilistic distribution} containing all possible solutions rather than a single-point estimation for the simulation result.  
Bayesian inference~\cite{John1992} offers an efficient tool for probabilistic predictions through inferring the surrogate models with parameters sampled from a probabilistic distribution conditional on the observed data, i.e., the posterior distribution~\cite{Moloud2021}. 
Directly employing a neural network as a surrogate model within Bayesian inference gives rise to Bayesian Neural Networks (BNNs) \cite{Denker1990, MacKay1992, neal1996, Wang2020}. BNNs perform posterior sampling based on a prior distribution of the network parameters. An analogous alternative is Stochastic Weight Averaging (SWA) \cite{2018averaging}, which also employs a neural network as the surrogate model but samples parameters during the training iterations. While these methods and their variants have found application in fluid simulations \cite{Masaki2022, tang_data-driven_2023, qiu_transient_2023, geneva_quantifying_2019, SUN2020161}, subtle distinctions endure in the uncertainty they capture compared to the inherent uncertainty in target simulations.

Researchers often consider two kinds of uncertainty, aleatoric and epistemic~\cite{Duraisamy_review_2019,Elisabeth1996,Armen2009,hullermeier2021}. Aleatoric uncertainty, or data uncertainty, captures the inherent uncertainty in the data, e.g., observation and measurement noise. In contrast, epistemic or model uncertainty denotes the model's confidence in its output. Generally, we can  reduce aleatoric uncertainty only by providing more precise data, while a better model can decrease the epistemic uncertainty. Most of the uncertainty in a turbulence simulation is epistemic in nature since an improved parametrization of a simulation could enhance the reliability of simulation results. However, the epistemic uncertainty of a turbulence simulation will turn into aleatoric uncertainty of the training dataset $\mathbf{d}$ once the simulation data is used to train a surrogate model. Although fully disentangling aleatoric and epistemic uncertainty is hard in Bayesian deep learning, it is important to note that BNNs construct probabilistic distributions on the network parameters, aiming to capture the epistemic uncertainty inherent in the neural network rather than the aleatoric uncertainty of the simulation-generated dataset~\cite{hullermeier2021}.
It is thus challenging to use the uncertainty of the prediction of a BNN to directly represent the inherent uncertainty of the simulation, as will be shown in the present study. 
In parallel, methods for estimating aleatoric uncertainty such as mixture density networks~\cite{bishop1994mixture} and their simplified variant, heteroscedastic models~\cite{Nix1994,Kendall2017}, are a widely used solution for directly predicting data uncertainty. In the fluid mechanics community, They have been demonstrated to be able to quantify the uncertainty in many problems, including reduced-order modeling and spatial data recovery~\cite{Maulik2020}.

Meanwhile, generative deep learning methods have employed Generative Adversarial Networks (GANs)~\cite{GAN2014} or variational autoencoders (VAEs)~\cite{VAE2014} to sample from a latent space to generate outputs. Linking latent space sampling to posterior sampling provides a new potential solution for assessing the prediction uncertainty through generative methods~\cite{Moloud2021}. However, these approaches were shown to have problems generating details and covering whole  distributions of solutions~\cite{Kaddoury2019,Creswell2018}. Recently, denoising diffusion probabilistic models (DDPMs)~\cite{Sohl2015,Song2019,Ho2020}, a state-of-the-art family of generative models, have been shown to outperform previous generative approaches in synthesizing highly impressive results in a variety of adjacent contexts~\cite{Dhariwal2021,Rombach_stablediffusion2022}. Despite the vibrant developments in many other research areas like 
material design~\cite{xie2022crystal,luo2022antigenspecific} and medical image reconstruction~\cite{CHUNG2022102479,Peng2022}, only very few studies in fluid dynamics have employed DDPM. Exceptions are works that investigate the performance of DDPM for super-resolution tasks~\cite{SHU2023111972} and inverse problem solving~\cite{holzschuh2023score}, while the capabilities of DDPMs as a surrogate model in fluid dynamics have not been investigated. 

In the present research, we leverage DDPMs to train an uncertainty-aware surrogate model for inferring the 
solutions of RANS-based airfoil flow simulations. Simulations of airfoil flow with RANS turbulence models are a fundamental problem and a widely studied use case of turbulence research~\cite{thuerey2020,hui2020,Sun2021,Yang2022,duru2022,CHEN2023}. As such, they provide a very good basis for assessing the capabilities of DDPM. The uncertainty considered in the present study is represented by a \textit{distribution} of solutions that encapsulates the inherent unpredictability associated with the RANS model when addressing flow
separations and other flow instabilities. We compare the performance of DDPMs with varying baselines like BNNs and heteroscedastic models. The capabilities of DDPMs and other baseline methods are measured in terms of their ability to accurately reconstruct the target distribution of solutions. Additionally, our study distinguishes itself from common applications such as image and speech generation by providing a clear \textit{ground truth} for the distribution to be learned. This means its uncertainty can be quantified, and the accuracy of the learned distribution of solutions can be estimated in a non-trivial setting. To ensure reproducibility, the source code and datasets of the present study are published at \url{https://github.com/tum-pbs/Diffusion-based-Flow-Prediction}.


The remainder of the paper is organized as follows: the definition of the problem and the data generation process are described in the next section; an introduction of the methods used in the present study, including DDPMs, BNNs, and heteroscedastic uncertainty estimation method, is given in Sec.~\ref{sec:method}; a single-parameter and a multi-parameter experiment are performed in Sec.~\ref{sec:experiment} to evaluate the performance of different methods; finally, the conclusions are summarized in the last section.

\section{Problem Statement\label{sec:problem_statement}}

\subsection{Learning Target}
For all commonly used formulations, the steady-state flow around an airfoil $\mathbf{y}=[p^*,\mathbf{u}^*]$ is uniquely determined by a set of physical parameters $\mathbf{x}=[\Omega,\alpha,Re]$ which parametrize a physical model in the form of a PDE $\mathcal{P}$, 
i.e., $\mathbf{y}=\mathcal{P}(\mathbf{x})$.
In our case, $\mathcal{P}$ represents the time-averaged Navier-Stokes equations with the corresponding boundary conditions,
while $\mathrm{p}^*$ and $\mathbf{u}^*$  denote the dimensionless pressure field and velocity field, respectively. 
The physical parameters $\mathbf{x}$ consist of the airfoil shape $\Omega$, the angle of attack $\alpha$, and the Reynolds number $Re$. The Reynolds number 
is defined as $Re=|\mathbf{u_f}|l/\nu$ where $\mathbf{u_f}$ is the freestream velocity, $l$ is the chord length, and $\nu$ is the viscosity of air.

The present study considers discrete, numerically approximated solutions of a turbulent RANS simulation $\mathcal{S}$ for the physics system $\mathcal{P}$. Besides the physical parameters $\mathbf{x}$, additional numerical parameters $\psi$ are introduced in $\mathcal{S}$ to determine the flow field, i.e., $\mathbf{y} = \mathcal{S}(\mathbf{x},\mathbf{\psi})$. Examples of $\psi$ include the choice of discretization, numerical schemes, and turbulence model parameters. These numerical parameters contain inherent uncertainty since they are, in practice, determined via knowledge obtained from experiments, resource constraints, and human experience. As their specifications vary, and some parameters can even prevent a unique choice, they represent a probabilistic distribution
$\Psi\sim P(\Psi)$. Thus, a solution from the numerical simulation can be seen as drawing a sample $\psi$ from $P(\Psi)$ and computing $\mathbf{y} = \mathcal{S}(\mathbf{x},\mathbf{\psi})$. The simulated flow field corresponding to given physical parameters $\mathbf{x}$ is then represented by a distribution as 

\begin{equation}
p(\mathbf{y}|\mathbf{x})=\int p(\mathbf{y}|\mathbf{x},\psi)p(\psi) d\psi 
\end{equation}
Since this full distribution is usually highly complex, it is often simplified: 
after drawing $N$ samples from the distribution as $\left\{ \mathbf{y}_1, \mathbf{y}_2, \cdots \mathbf{y}_N \right\}=\left\{ \mathcal{S}(\mathbf{x},\mathbf{\psi}_1), \mathcal{S}(\mathbf{x},\mathbf{\psi}_2), \cdots, \mathcal{S}(\mathbf{x},\mathbf{\psi}_N)\right\}$,  the expectation $\boldsymbol{\mu}_{\mathbf{y}}$ and standard deviation $\boldsymbol{\sigma}_{\mathbf{y}}$ can easily be computed to characterize the distribution, and the $\boldsymbol{\sigma}_{\mathbf{y}}$ is often used to quantify the uncertainty. While these first two moments of the distribution of solutions are important quantities, they can hide important aspects of the solutions such as distinct modes~\cite{spanos1999probability}. Hence, rather than estimating only the moments of the distribution, the present study focuses on learning the full distribution $p(\mathbf{y}|\mathbf{x})$ as accurately as possible via a surrogate model parameterized by a set of learnable weights $\xi$ and trained on a dataset $\mathbf{d}$, i.e., to learn $p_\xi(\mathbf{y}|\mathbf{x},\mathbf{d}) \approx p(\mathbf{y}|\mathbf{x})$ without having access to the parameters $\psi$.

It is worth noting that the formulation above is applicable to a wide range of problems in scientific computing. E.g., for super-resolution flows ~\cite{SHU2023111972,Xie2018} or flow reconstructions from sparse observations~\cite{Druault2007, Franz_2021_CVPR, Maulik2020}, the uncertainty is introduced due to the sparse nature of the constraints. Likewise, the temporal evolution of turbulent flows from a given discrete state contains uncertainty in terms of the unresolved scales below the mesh spacing~\cite{PEI2021107513,Perot2007}. Hence, obtaining the full distribution of solutions for a given numerical problem is a fundamental challenge. In this context, we investigate the capabilities of DDPM to infer distributions of RANS simulations.

\subsection{Constructing the Distribution of Solutions $\mathbf{\psi}$ 
\label{sec:choice_of_parameter}}

RANS simulations are pivotal in iterative aerodynamic shape optimization, where the fluid dynamics performance of a given shape is accessed through a flow snapshot 
of a converged RANS simulation. 
However, the inherent flow instability around bodies poses a challenge to RANS simulations for shape optimization. For instance, high transient features like vortex shedding in the flow around airfoils will occur when certain Reynolds number and angle of attack are reached~\cite{Kiran2012}. Steady-state RANS simulations are inadequate in capturing these highly transient flows, inducing oscillation in number of simulation iterations. While alternative transient simulation methods are available for such unsteady flows, assessing flow steadiness adds challenges to the shift between steady and transient methods. This difficulty is pronounced during shape optimization iterations, where the shapes of airfoils could be highly flexible. Moreover, well-acknowledged limitations of RANS methods in capturing critical flow phenomena, such as separation~\cite{Spalart2000}, further compound the challenges. Temporal averaging inherent in RANS proves insufficient in accounting for turbulence energy input from dominant periodic wake components~\cite{Alfonsi2009,tucker2012}. Additionally, deficiencies in addressing Reynolds shear stress anisotropy and the impact of streamline curvature in separated flows are common among many RANS models~\cite{Ansys2011,Chunhua2020}. These inherent limitations of RANS methods introduce convergence difficulties, particularly when faced with separations. As a result, all the challenges from highly transient flow and separated flow result in the \textit{oscillating snapshots} in steady state RANS simulations, introducing uncertainties to the simulation results and finally adversely affecting shape optimization~\cite{Dicholkar2022,Shenren2015}. While Detached-Eddy Simulation (DES) and Large-Eddy Simulation (LES) offer more accurate alternatives for critical flows, steady-state RANS simulations retain favorability in engineering design due to their computational efficiency and reliable performance in the converged regime~\cite{lyu2013rans,lyu2014rans,chauhan2021rans}.  Notably, the research community has recognized this uncertainty in the RANS simulation and has undertaken numerous initiatives to mitigate its adverse implications~\cite {Dicholkar2022,aerospace10030230}.

Meanwhile, neural networks become more popular to serve as surrogate models for aerodynamic shape optimization. Most of these neural networks typically utilize only one snapshot or an average of snapshots of RANS simulation as training data. In such cases, the inherent uncertainty in the simulation tends to be overlooked. This neglect can result in suboptimal performance during the optimization process, as the network may struggle to accurately predict the correct flow dynamics. The present study employs the number of solver iterations, $\tau$, as a simple yet representative instance of the parameters $\psi$ for RANS simulation of airfoil flows. 
$\tau$ is used to draw samples from the target distribution $p(\mathbf{y}|\mathbf{x})$ in order to establish a dataset with multiple solutions for the flow field. This dataset, with the inherent uncertainty of RANS simulations being explicitly considered through multiple snapshots, allows us to construct an "uncertainty-aware surrogate model". Simultaneously, the dataset enables a quantitative comparison between different methods, utilizing the ground truth uncertainty information provided by these multiple snapshots. Through this configuration, we aim to enhance the reliability and effectiveness of the deep surrogate models in capturing the true uncertainties present in aerodynamic simulations. This, in turn, can improve their potential for applications such as aerodynamic shape optimization.

\subsection{Data Generation and Preprocessing\label{sec:data_generation}}

The data generation process in the present study follows an existing benchmark for learning RANS simulations of airfoil flows~\cite{thuerey2020}. All simulations are performed using the open-source code \textit{OpenFOAM}~\cite{Jasak1996,Weller1998} with SA one equation turbulence model~\cite{SA1992}. There are 1417 different airfoils from the UIUC database~\cite{UIUC} used to generate 5000 two-dimensional simulation cases ($M=5000$). The range of $Re$ and $\alpha$ are $(10^6,10^7)$ and $(-22.5^\circ, 22.5^\circ)$, respectively. 
A set of 30 airfoils not used in the training dataset are used to generate a test dataset with 130 simulation cases. 100 of these samples use the ($Re$,$\alpha$) domain of the training dataset. We denote these samples as the \textit{interpolation region}. The remaining 30 samples use the same previously unseen airfoils and additionally use parameters outside of the original distribution ($Re \in (5\times10^5,10^6) \cup (10^7,1.1\times10^7) $, $\alpha \in (-25^\circ, -22.5^\circ) \cup (22.5^\circ, 25^\circ) $. These tests in the \textit{extrapolation region} will be used to evaluate the generalization in terms of shape as well as flow condition. More detailed information on the $(\alpha, Re)$ distribution can be found in Appendix~\ref{sec:app:parameterdistribution}.


\begin{figure}[tbh]
    \centering
    \includegraphics[scale=0.33]{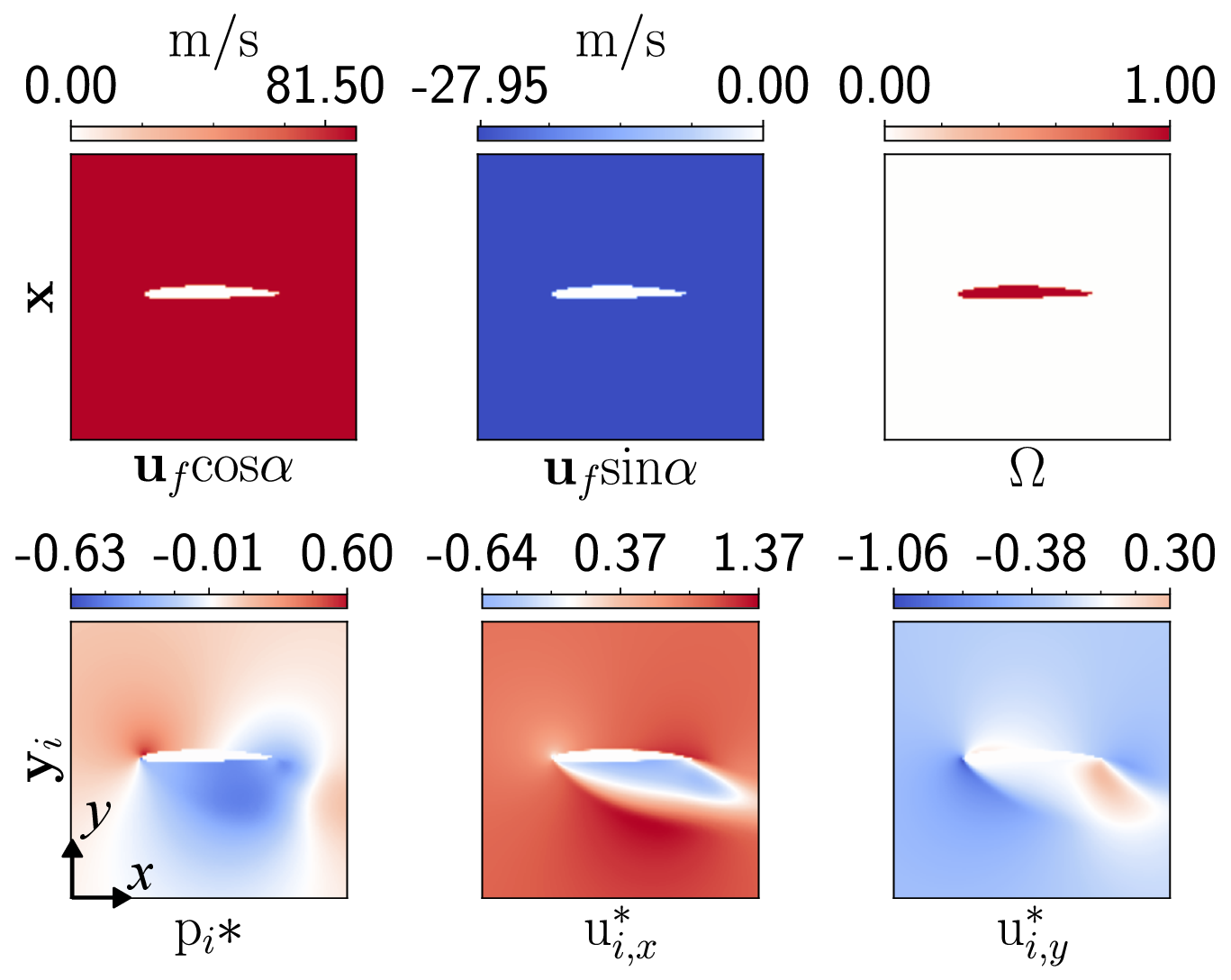}
    \caption{One instance of the encoded input and output simulation data (ah21-7 airfoil, $Re=8.616\times10^3$, and $\alpha=-18.93^\circ$).}
    \label{fig:dataset_sample}
\end{figure}

The simulation data is pre-processed to be normalized and nondimensionalized for training and inference. The $Re$ will be encoded as the freestream velocity and then embedded with $\Omega$ and $\alpha$ as a three-channel tensor $[|\mathbf{u_f}|\mathrm{cos}~\alpha, |\mathbf{u_f}|\mathrm{sin}~\alpha,\Omega]$. The decision to encode the input as a three-channel field was carefully considered and motivated by several factors. Related discussion can be found in Appendix~\ref{sec:app:discussion_input}. The simulation outputs are also encoded as a three-channel tensor where the first channel corresponds to the dimensionless pressure field $\mathrm{p}_{i}{*}=(\mathrm{p}_{i}-\mathrm{p}_{i,  a})/|\mathbf{u_f}|^2$ and the latter two are the dimensionless $x$ and $y$ components of the output velocity, i.e., $(\mathrm{u}_{x,i}^{*},\mathrm{u}_{y,i}^{*})=(\mathrm{u}_{x,i}/|\mathbf{u_f}|,\mathrm{u}_{y,i}/|\mathbf{u_f}|)$, respectively. 
Fig.~\ref{fig:dataset_sample} shows an instance of the encoded input and output simulation data. Finally, all input and output quantities are rescaled to $[-1,1]$ 
over the entire training dataset. The tensor resolution in each channel of the input and output data is interpolated to a square field of $s \times s$ values, for which we use $s \in \{32,64,128\}$ in the experiments below. The other preprocessing and data generation steps follow the previous benchmark~\cite{thuerey2020}. 

\begin{figure}[tbh]
    \centering
   \sidesubfloat[]{\includegraphics[scale=0.33]{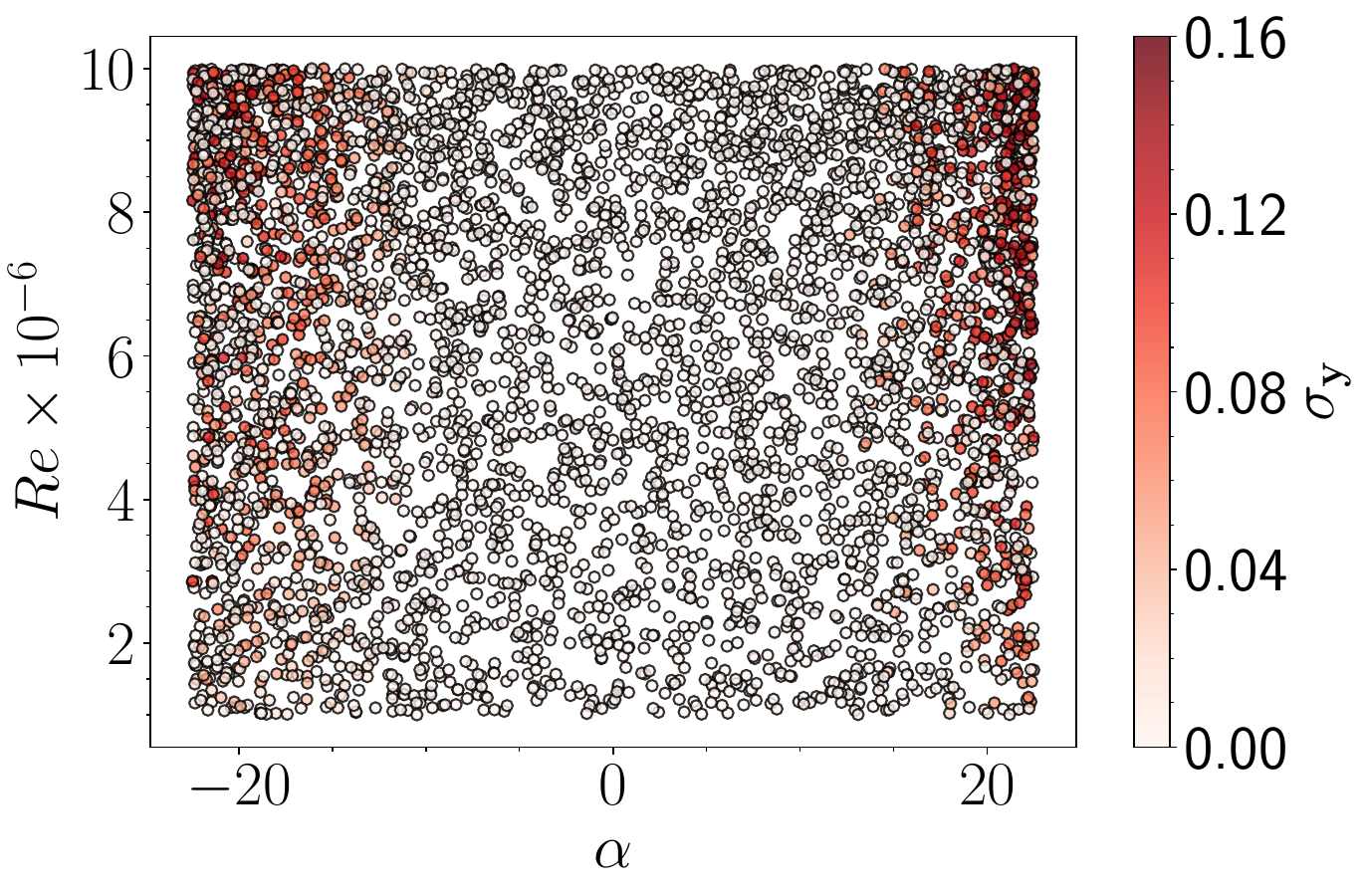}\label{fig:train_dataset}}
   \sidesubfloat[]{\includegraphics[scale=0.33]{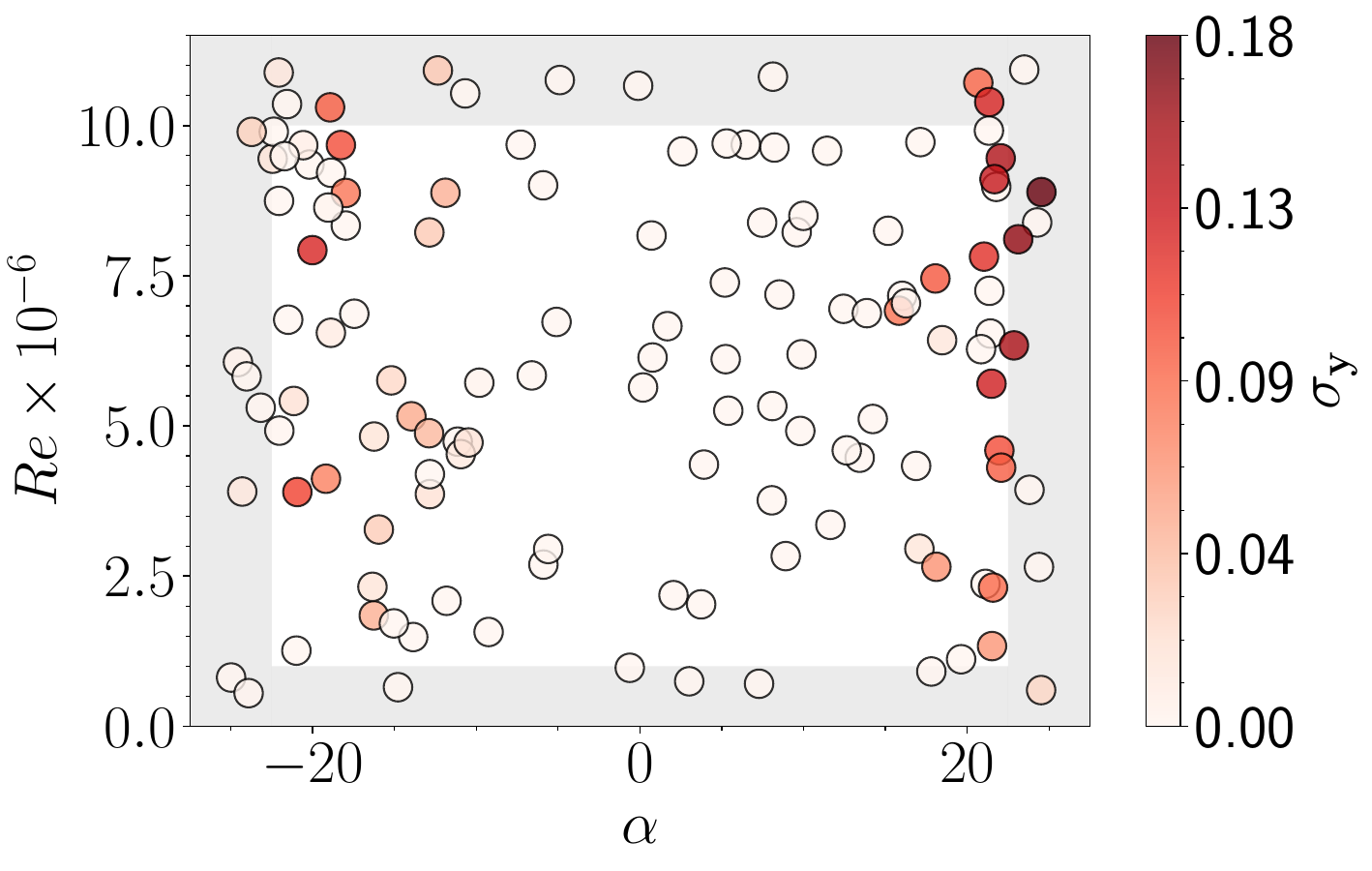}\label{fig:test_dataset}}
    \caption{The uncertainty distribution in the datasets. a) The training dataset. b) The test dataset. The shaded area shows the extrapolation region. }
    \label{fig:dataset}
\end{figure}

In training dataset, we draw $N$ samples of $\tau$ from a uniform distribution $\Tau\sim U(2500,3500)$ to obtain $N$ snapshots of flow fields, $\left \{ \mathbf{y}_{1}, \mathbf{y}_{2}, \cdots \mathbf{y}_{N} \right \}=\left\{ \mathcal{S}(\mathbf{x},\mathbf{\tau}_1), \mathcal{S}(\mathbf{x},\mathbf{\tau}_2), \cdots, \mathcal{S}(\mathbf{x},\mathbf{\tau}_{N})\right\}$, as a representation of the target distribution. Unless specified otherwise, the number of snapshots in the test dataset, $\widehat{N}$, is the same as the number of snapshots in the training dataset, i.e., $\widehat{N}=N=25$. Figure~\ref{fig:dataset} showcases the distribution of standard deviation among these 25 samples in the training and test datasets, serving as a quantification of uncertainty. The increase of $|\alpha|$ and $Re$ exacerbates the instability inherent in the flow, resulting in high uncertainty of the target distribution, particularly evident in the high ($\alpha$,$Re$) region at the top-left and top-right corners of Fig.~\ref{fig:dataset}. The shape of the airfoil also plays a crucial role in influencing the development of flow instability. Certain airfoils are meticulously designed to mitigate flow separation, as such phenomena can be detrimental to engineering design. This intricacy results in the low uncertainty points in the high ($\alpha$,$Re$) region, adding further complexity to the distribution of uncertainty and presenting heightened challenges for network predictions.



\section{Methodology\label{sec:method}}
This section introduces the basic theory of DDPMs, BNNs, and heteroscedastic models. The performance of the latter two methods will be compared with the performance of DDPM in the next section. The main target of this section is to introduce the loss formulation for each approach, $\mathcal{L}_{NN}$, and the target distribution of solutions, $p_\xi(\mathbf{y}|\mathbf{x},\mathbf{d})$, of each method. DDPMs and BNNs allude to the distribution of solutions through the posterior distribution of the surrogate models' parameters $\xi$.
Thus the distribution of predictions is represented by the following marginal distribution:
\begin{equation}
    p_\xi(\mathbf{y}|\mathbf{x},\mathbf{d})=\int p(\mathbf{y}|\mathbf{x},\xi) p(\xi|\mathbf{d})  d\xi 
    ,
    \label{eq:marginal_surrogate}
\end{equation}
where $ p(\mathbf{y}|\mathbf{x},\xi)$ represents the inference of the surrogate model, and $p(\xi|\mathbf{d})$ is the posterior distribution of the parameters of the surrogate model. To draw samples from the  solutions with Eq.~\eqref{eq:marginal_surrogate} we need to infer the predictions by the surrogate model with its parameters sampled from the posterior distribution. On the other hand, heteroscedastic models assume a certain type of distribution for the solution and directly predict its parameters. 
Thus, prediction samples of heteroscedastic models are directly drawn from the resulting distribution.

\subsection{Denoising Diffusion Probabilistic Model}
\begin{figure*}[tbh]
    \centering
    \includegraphics[scale=0.33]{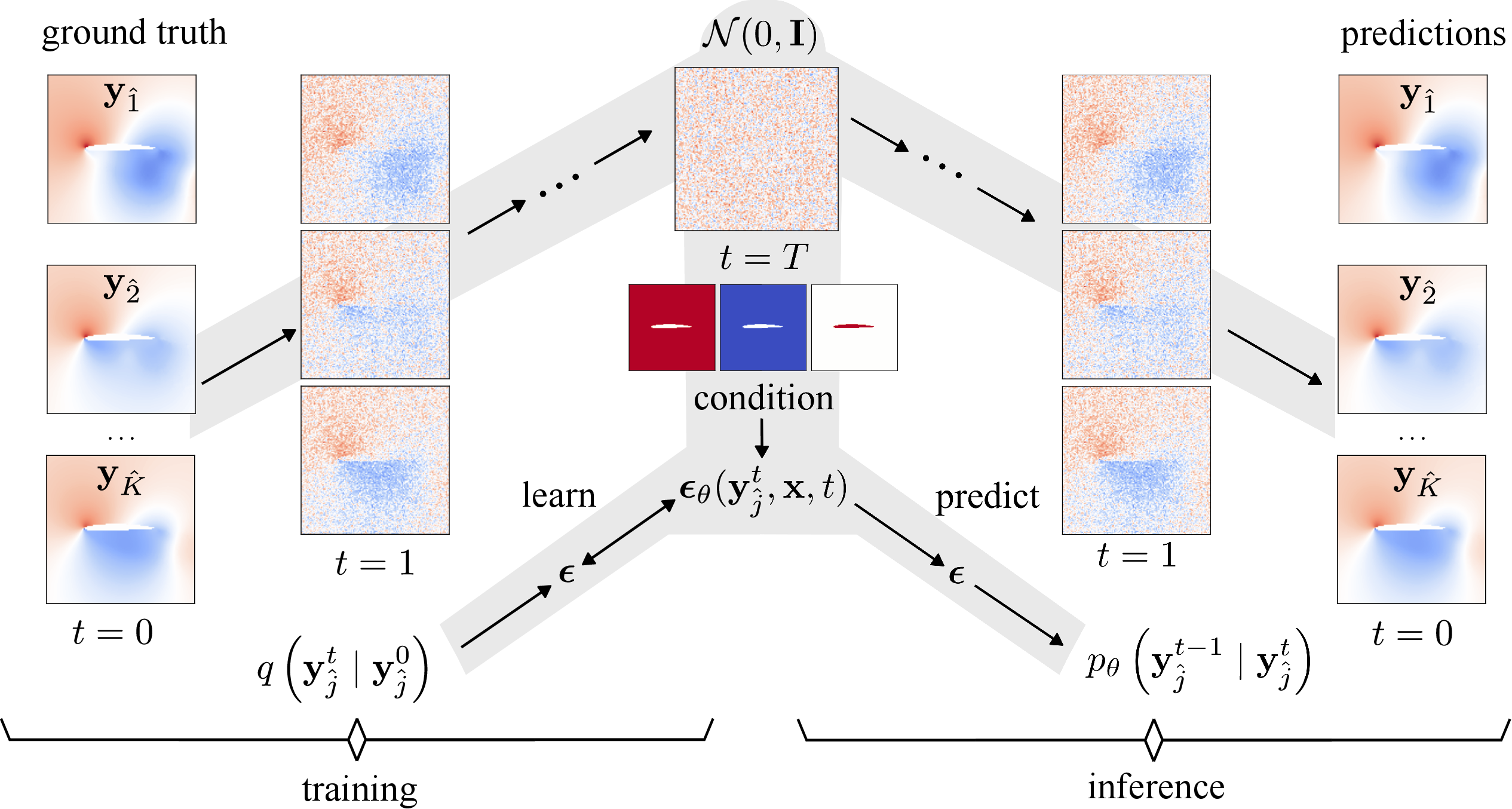}
    \caption{The sketch of uncertainty prediction process using the DDPM.}
    \label{fig:sketch_diffusion}
\end{figure*}
To train a DDPM-based surrogate model, we follow the canonical procedure established for DDPM~\cite{Sohl2015, Ho2020}, as illustrated in Fig.~\ref{fig:sketch_diffusion}. 
In the training process, a forward Markov chain is introduced to gradually distort the initial data distribution, $\mathbf{y}_{i}^0=\mathbf{y}_{i}$, $\mathbf{y}_{i}^0\sim q(\mathbf{y}_{i}^0)$ to a standard Gaussian distribution $\mathcal{N}\left(\mathbf{0}, \mathbf{I}\right)$. The forward Markov chain is defined as
\begin{equation}
    q\left(\mathbf{y}_{i}^{0:T}\right)
    =
    q(\mathbf{y}_{i}^0)
    \prod_{t=1}^{T} 
    q\left(\mathbf{y}_{i}^{t} \mid \mathbf{y}_{i}^{t-1}\right)
    ,
    \label{eq:dif:forward_diffusion}
\end{equation}
where 
\begin{equation}
    q\left(\mathbf{y}_{i}^{t} \mid \mathbf{y}_{i}^{t-1}\right)
    =
    \mathcal{N}
    \left(\mathbf{y}_{i}^{t} ; \sqrt{1-\beta^{t}} \mathbf{y}_{i}^{t-1}, \beta^{t} \mathbf{I}\right)
    .
    \label{eq:dif:forward_step}
\end{equation}
The hyperparameter $\beta^t \in (0,1)$ controls the noise schedule in the forward chain. The value of $\beta^T$ should be close to 1 to make $q\left(\mathbf{y}_{i}^{T}\right) \approx \mathcal{N}\left(\mathbf{y}_{i}^{T} ; \mathbf{0}, \mathbf{I}\right)$. In the present study, $\beta^t$ follows a cosine schedule~\cite{Nichol2021}.

A nice property of Eq.~\eqref{eq:dif:forward_step} is that we can use it to marginalize Eq.~\eqref{eq:dif:forward_diffusion}, which results in
\begin{equation}
    q\left(\mathbf{y}_{i}^{t} \mid \mathbf{y}_{i}^{0}\right)
    =
    \mathcal{N}
    \left(\mathbf{y}_{i}^{t} ; \sqrt{\bar{\gamma}^t} \mathbf{y}_{i}^{0}, \left(1-\bar{\gamma}^t\right) \mathbf{I}\right)
    ,
    \label{eq:dif:forward_step_x0}
\end{equation}
where $\gamma^t = 1 - \beta^t$ and $\bar{\gamma}^t = \prod_{i=1}^t \gamma^i$. Via the reparameterization trick~\cite{Kingma2015}, $\mathbf{y}_{i}^t$ can then be sampled from a standard Gaussian distribution $\boldsymbol{\epsilon }\sim\mathcal{N}(\mathbf{0},\mathbf{I})$: 
\begin{equation}
    \mathbf{y}_{i}^{t}
    =
    \sqrt{\bar{\gamma}^t} \mathbf{y}_{i}^{0}
    +\sqrt{\left(1-\bar{\gamma}^t\right)}\boldsymbol{\epsilon}
    .
    \label{eq:dif:reparameterization}
\end{equation}

In the inference process of DDPM, another Markov chain is used to recover the data from the added Gaussian noise step by step. The reverse Markov chain is built with a learned transition parameterized by $\theta$:
\begin{equation}
    p_\theta\left(\mathbf{y}_{i}^{0:T}\right) 
    =
    p(\mathbf{y}_{i}^T)
    \prod_{t=1}^{T}
    p_\theta \left(\mathbf{y}_{i}^{t-1} \mid \mathbf{y}_{i}^{t}\right)
    .
    \label{eq:dif:reverse_diffusion}
\end{equation}
To approximate the forward chain with the reverse chain, we can minimize the Kullback-Leibler (KL) divergence $\mathrm{KL}\left( q\left(\mathbf{y}_{i}^{0:T}\right) \parallel p_\theta\left(\mathbf{y}_{i}^{0:T}\right)\right)$ between these two distributions. Ho et al.~\cite{Ho2020} gives a specific form of $p_\theta \left(\mathbf{y}_{i}^{t-1} \mid \mathbf{y}_{i}^{t}\right)$ as

\begin{equation}
    p_\theta \left(\mathbf{y}_{i}^{t-1} \mid \mathbf{y}_{i}^{t}\right)
    =
    \mathcal{N}
    \left(\mathbf{y}_{i}^{t-1} 
    ; 
    \boldsymbol{\mu}_{\theta,\mathbf{y^{t-1}_{i}}}
    , 
    \frac{1-\bar{\gamma}^{t-1}}
    {1-\bar{\gamma}^{t}}\beta^{t} \mathbf{I}\right)
    ,
    \label{eq:dif:reverse_diffusion_step}
\end{equation}
and
\begin{equation}
    \boldsymbol{\mu}_{\theta,\mathbf{y^{t-1}_{i}}}
    =
    \frac{1}{\sqrt{\gamma^t}} 
    \Big( 
        \mathbf{y}_{i}^t 
        - 
        \frac{\beta^t}{\sqrt{1 - \bar{\gamma}^t}} \boldsymbol{\epsilon}_\theta(\mathbf{y}_{i}^t, t) 
    \Big)
    .
    \label{eq:dif:reverse_diffusion_step_mean}
\end{equation}

Then, minimizing the KL divergence $\mathrm{KL}\left( q\left(\mathbf{y}_{i}^{0:T}\right) \parallel p_\theta\left(\mathbf{y}_{i}^{0:T}\right)\right)$ is equivalent to the following simple loss function for training~\cite{Ho2020}:
\begin{equation}
    \mathcal{L}_{NN}(\theta)
    =
    \mathbb{E}_{\mathbf{x},\boldsymbol{\epsilon }\sim\mathcal{N}(\mathbf{0},\mathbf{I}),t  } 
    \left[
    \parallel
        \boldsymbol{\epsilon}
        -
        \boldsymbol{\epsilon_\theta}(\mathbf{y}_{i}^t,t)
    \parallel^2
    \right]
    ,
    \label{eq:dif:loss_function}
\end{equation}
where $\boldsymbol{\epsilon}$ is the Gaussian noise used to compute $\mathbf{y}_{i}^t$ through Eq.~\eqref{eq:dif:reparameterization} and $\boldsymbol{\epsilon}_\theta$ is the neural network predicting $\boldsymbol{\epsilon}$. 

In the present study, DDPM is used to synthesize a specific $\mathbf{y}_{i}$ 
that corresponds to the chosen physical parameters, provided as condition $\mathbf{x}$. Following the conditioning approach of Lyu et al.~\cite{lyu2022,Saharia2022,Saharia2023, Dule2023}, the condition $\mathbf{x}$ is added as the input for the network via $\boldsymbol{\epsilon_\theta}(\mathbf{y}_{i}^t,\mathbf{x},t)$, which
gives the loss function of
\begin{equation}
    \mathcal{L}_{NN}(\theta)
    =
    \mathbb{E}_{\mathbf{x},\boldsymbol{\epsilon }\sim\mathcal{N}(\mathbf{0},\mathbf{I}),t  } 
    \left[
    \parallel
        \boldsymbol{\epsilon}
        -
        \boldsymbol{\epsilon_\theta}(\mathbf{y}_{i}^t,\mathbf{x},t)
    \parallel^2
    \right]
    .
    \label{eq:dif:loss_function_final}
\end{equation}

Finally, using the reverse chain shown in Eq.~\eqref{eq:dif:reverse_diffusion}-\eqref{eq:dif:reverse_diffusion_step_mean}, the distribution of solutions is obtained as

\begin{align}
\begin{split}
p_\xi\left(\mathbf{y}|\mathbf{x},\mathbf{d}\right)
    &=
    \int
\mathcal{N}
\left(
\mathbf{y}_{i}^{T}
;
\mathbf{0},
\mathbf{I}  
\right)
    \prod_{t=1}^{T}
    p_\theta \left(\mathbf{y}_{i}^{t-1} \mid \mathbf{y}_{i}^{t}\right)
d \mathbf{y}_{i}^1\cdots d \mathbf{y}_{i}^{T-1}    
d \mathbf{y}_{i}^T
\\
    &=
    \int
\mathcal{N}
\left(
\mathbf{y}_{i}^{T}
;
\mathbf{0},
\mathbf{I}  
\right)
    \prod_{t=1}^{T}
    \mathcal{N}
    \left(\mathbf{y}_{i}^{t-1} 
    ; 
    \frac{1}{\sqrt{\gamma^t}} 
    \Big( 
        \mathbf{y}_{i}^t 
        - 
        \frac{\beta^t}{\sqrt{1 - \bar{\gamma}^t}} \boldsymbol{\epsilon}_\theta(\mathbf{y}_{i}^t,\mathbf{x}, t) 
    \Big)
    , 
    \frac{1-\bar{\gamma}^{t-1}}
    {1-\bar{\gamma}^{t}}\beta^{t} \mathbf{I}\right)
d \mathbf{y}_{i}^1\cdots d \mathbf{y}_{i}^{T-1}    
d \mathbf{y}_{i}^T
\end{split}
\label{eq:prediction_DDPM}
\end{align}
Eq.~\eqref{eq:prediction_DDPM} shows that the surrogate model is the whole reverse Markov chain for DDPM, and the network only works as a component in the surrogate model. The posterior distribution of the parameters is the distribution of $\mathbf{y}_{i}^t|^{T}_{t=1}$ while the parameters of the neural network are deterministic after training. The posterior sampling is achieved by sampling $\mathbf{y}_{i}^t|^{T}_{t=1}$ from Gaussian distribution parameterized by the given expectation and standard deviation value.

\subsection{Alternatives for Uncertainty Estimation with Deep Learning}
Working with uncertainties has  been an active topic in both fluid mechanics and deep learning communities~\cite{Najm2009,Moloud2021}. This subsection introduces two established methods from this area which will serve as baseline methods in the experiments section.

\subsubsection{Bayesian Neural Networks\label{sec:method:BNN}}
In contrast to DDPMs, BNNs directly use the network as the surrogate model~\cite{neal1996, Wang2020}. Then the posterior distribution of the surrogate model, $p(\xi|\mathbf{d})$, turns into the posterior distribution of the network $p(\theta|\mathbf{d})$. Theoretically, this posterior could be computed with the Bayes rule as 
\begin{equation}
    p({\theta}|\mathbf{d}) = 
    \frac{p(\mathbf{d}|{\theta})p({\theta})}{p(\mathbf{d})}
    ,
    \label{eq:bnn:analytical}
\end{equation}
where $p({\theta})$ is a pre-defined prior distribution representing our prior knowledge of network parameters. 
A standard Gaussian distribution is often a reasonable choice since the network parameters are typically small and can be positive or negative~\cite{neal1996, Wang2020,Moloud2021}. Nonetheless, a direct calculation of $p(\theta\mid \mathbf{d})$ via Eq.~\eqref{eq:bnn:analytical} is often intractable, and thus methods like Monte Carlo (MC) dropout~\cite{Nitish2014,Gal2016}, Markov chain Monte Carlo (MCMC)~\cite{Matthew2003,Chen2014}, and variational inference (VI)~\cite{Hinton1993,Graves2011,Ranganath2014} have been proposed to solve this problem.  VI uses a parameterized variational distribution $q_{\phi}(\theta)$ to approximate the posterior of model parameters and then minimize the KL divergence $\mathrm{KL}\left( q_{\phi}(\theta) \parallel p(\theta\mid \mathbf{d})\right)$ between them, which leads to the following negative evidence lower bound (ELBO) as loss function~\cite{Neal1998,Blundell2015}
\begin{equation}
    \mathcal{L}_{\mathrm{NN}}(\phi)= \lambda\mathrm{KL}(q_{\phi}({\theta})||p({\theta}))-
    \mathbb{E}_{q_{\phi}}[\log(p(\mathbf{d}|{\theta}))]
    .
    \label{eq:bnn:loss_func}
\end{equation}
Here, the first and second terms in the loss function pose a trade-off for the network to approach the prior distribution and the ground truth data~\cite{Blundell2015,Moloud2021}. A scaling factor $\lambda < 1$ is introduced for the KL divergence term to adjust this balance, which was shown to turn $p({\theta}|\mathbf{d})$ into a cold posterior~\cite{Wenzel2020, Laurence2021}. On the one hand, a higher scaling factor makes the distribution of the model more like the prior Gaussian distribution, that is, more random. On the other hand, a smaller scaling factor forces the model to learn more from the dataset, and the whole model will degenerate into a deterministic model when the scaling factor becomes zero.

The distribution of solutions is finally obtained as
\begin{equation}
    p_\xi(\mathbf{y}|\mathbf{x},\mathbf{d})
    =
    \int p(\mathbf{y}|\mathbf{x},\theta)p(\theta|\mathbf{d}) d\theta
    =\int p(\mathbf{y}|\mathbf{x},\theta)q_\phi(\theta) d\theta
    .
\end{equation}
The posterior sampling is achieved by sampling network parameters from the learned variational distribution $q_\phi(\theta)$. 

\subsubsection{Heteroscedastic Models}

Aleatoric uncertainty can be further divided into homoscedastic and heteroscedastic uncertainty, where the former represents a constant for all data, while the latter varies w.r.t. different input data~\cite{Nix1994, Le2005}. Heteroscedastic uncertainty is usually more relevant since some data in the dataset typically have higher uncertainty than others (e.g., airfoil flow simulations with higher $Re$s as shown in Fig.~\ref{fig:dataset}). To model the heteroscedastic uncertainty of the airfoil flow simulations, we assume that an ideally-configured simulation with the physical parameter $\mathbf{x}$ converges to a single flow solution $\mathbf{y}_{g}$ where the subscript $g$ indicates ground truth data, and additional simulation results $\mathbf{y}=\left \{ \mathbf{y}_{1}, \mathbf{y}_{2}, \cdots \mathbf{y}_{N} \right \}$ that contain errors can be treated as a noisy set around the ground truth $\mathbf{y}_{g}$. By doing so, we are directly modeling the distribution of solutions rather than modeling the posterior distribution of the surrogate model’s parameters. 
For the heteroscedastic model, the surrogate simulator is then realized by sampling from the modeled distribution.
Under the assumption of normally distributed noise, a network parameterized by $\theta$ can be trained to predict the standard deviation $\boldsymbol{\sigma}_{\mathbf{y}}$ and the expectation $\boldsymbol{\mu}_{\mathbf{y}}$ of $\mathbf{y}$ through the maximum a posterior probability inference as~\cite{Nix1994, Kendall2017} 
\begin{equation}
    \mathcal{L}_{\mathrm{NN}}(\theta)=\frac{1}{N}\sum_{i=1}^{N}\left[ \frac{1}{2 [\boldsymbol{\sigma}_{\theta,\mathbf{y}}\left(\mathbf{x}\right)]^{2}}\left\|\mathbf{y}_{i}-\boldsymbol{\mu}_{\theta,\mathbf{y}}\left(\mathbf{x}\right)\right\|^{2}+\frac{1}{2} \log [\boldsymbol{\sigma}_{\theta,\mathbf{y}}\left(\mathbf{x}\right)]^{2}\right]
    .
    \label{eq:het:loss_func}
\end{equation}
Note that the network actually predicts $\mathrm{log} \boldsymbol({\sigma}_{\theta,\mathbf{y}_{i}})^2$ rather than $\boldsymbol{\sigma}_{\theta,\mathbf{y}_{i}}$ in practice since the latter may result in a negative standard deviation and induce numerical instability ~\cite{Kendall2017}. The limitation of the normally distributed noise can be mitigated by introducing mixture density networks that use Gaussian mixture distributions to model potentially more complex distribution for the noise~\cite{bishop1994mixture}. However, the simple assumption of a single Gaussian distribution is the easiest and most commonly used one~\cite{Nix1994,Kendall2017,Maulik2020}.

Due to the assumed normally distributed noise, the distribution of solutions can be written as a Gaussian distribution parameterized by the predicted expectation and standard deviation:
\begin{equation}
    p_\xi\left(\mathbf{y}|\mathbf{x},\mathbf{d}\right)
    =
    \mathcal{N}\left(
    \mathbf{y};
    \boldsymbol{\mu}_\theta(\mathbf{x}),\boldsymbol{\sigma}_\theta(\mathbf{x})
    \right)
    .
\end{equation}

\section{Experiments\label{sec:experiment}}

In this section, we assess the performance of DDPM in inferring distributions of RANS solutions for airfoil flows. However, prior to evaluating the method's performance in detail, we conduct an analysis of the origins of uncertainty. By studying the varying patterns of uncertainty and flow fields with different angles of attack and Reynolds numbers, our  aim is to deepen the understanding of the underlying uncertainty in the RANS simulations. A reduced single-parameter study is then carried out to assess the accuracy of predictions by DDPM w.r.t. their distribution of uncertainties. Afterward, we broaden the scope of the experiments to investigate the prediction accuracy with multi-parameter experiments using an enlarged parameter space and higher data resolution. Predictions of corresponding BNNs and heteroscedastic models are evaluated for comparison.
The network architectures and training procedures of these 3 models, which are detailed in Appendix~\ref{sec:app:training}, are kept as identical as possible in each experiment to ensure fair comparisons. For each model, we draw 500 samples from the distribution of predictions with the same condition $\mathbf{x}$ as a practical estimation of the ground truth distribution. Unless otherwise specified, all the accuracy metrics and line plots for each model are obtained with 3 networks trained with different initial random seeds, while the field plots are drawn from a randomly chosen network.

\subsection{An analysis of the aleatoric uncertainty of the dataset}

\begin{figure}[tbh]
    \centering
    \includegraphics[scale=0.33]{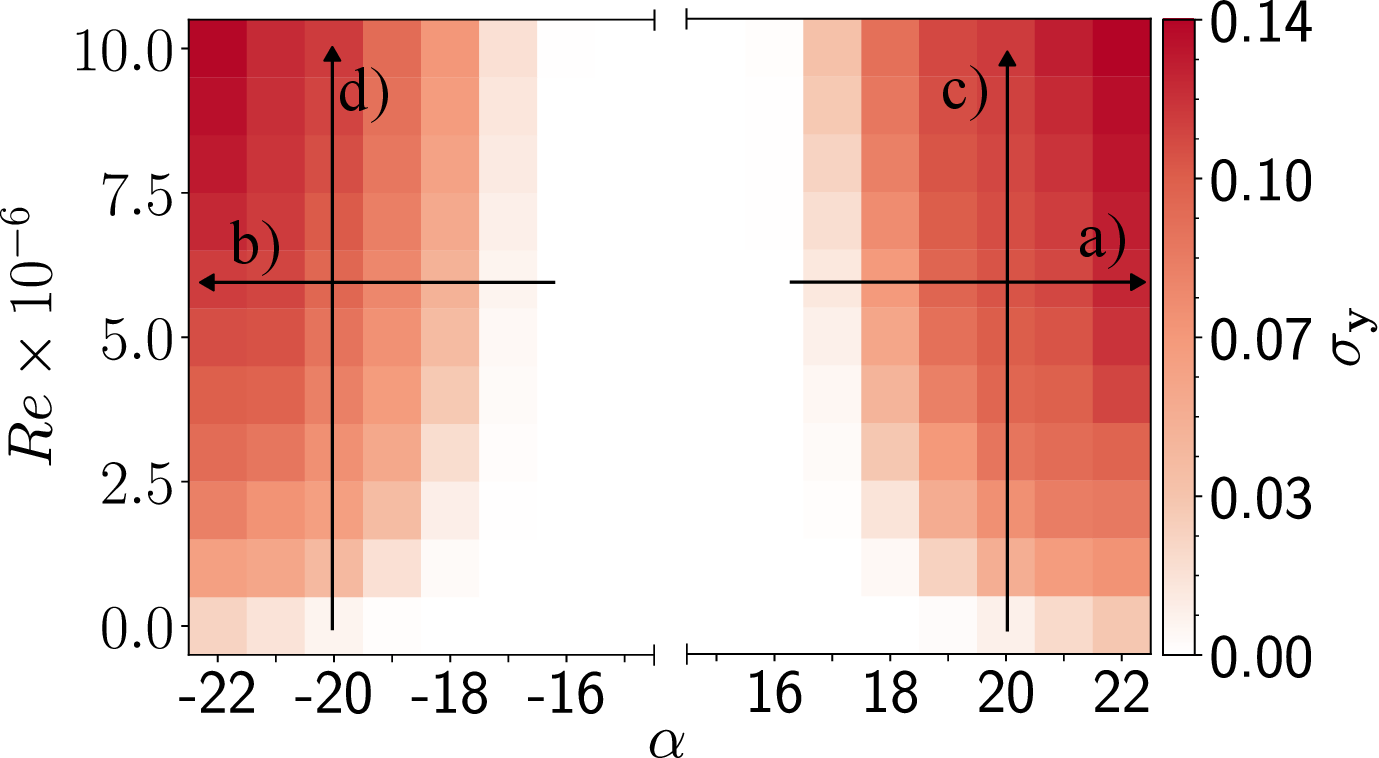}
    \caption{Uncertainty transitions of the RANS simulation for raf30 airfoil. Labeled arrows denote parameter regions as in Fig.~\ref{fig:flow_std}.}
   \label{fig:raf30}
\end{figure}

\begin{figure}[tbh]
    \centering
    \sidesubfloat[]{\includegraphics[scale=0.33]{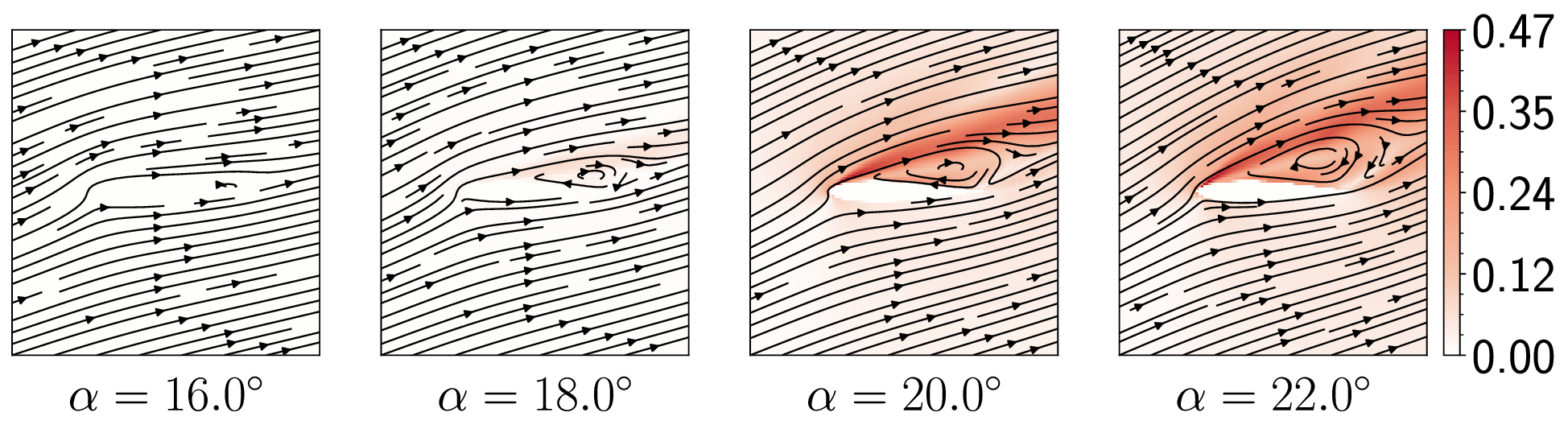}\label{fig:flow_std_aoa_positive}}\\
    \sidesubfloat[]{\includegraphics[scale=0.33]{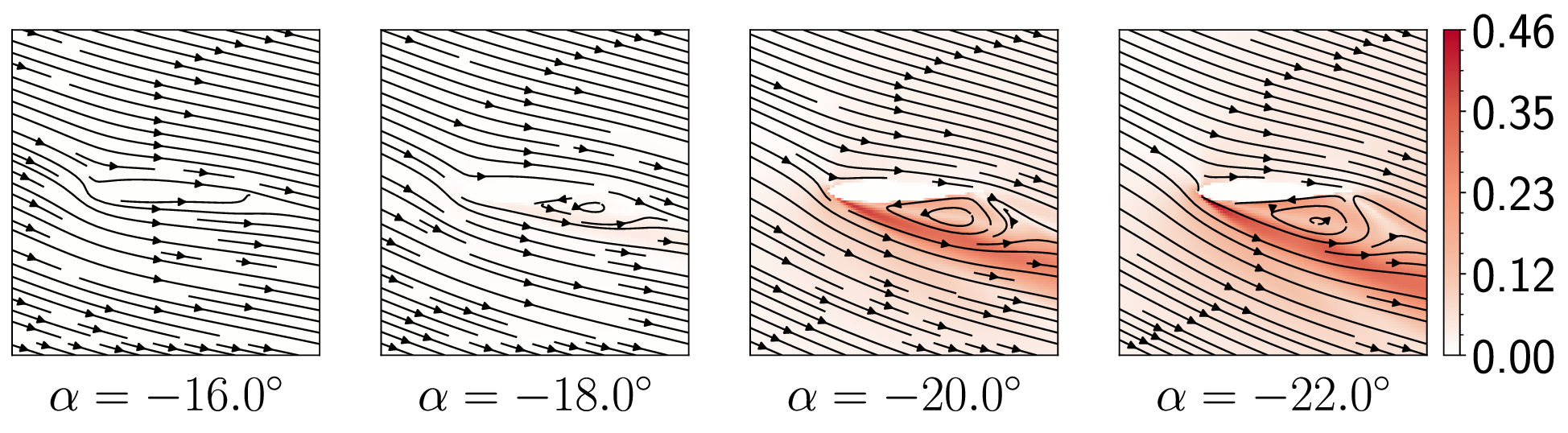}\label{fig:flow_std_aoa_negative}}\\
    \sidesubfloat[]{\includegraphics[scale=0.33]{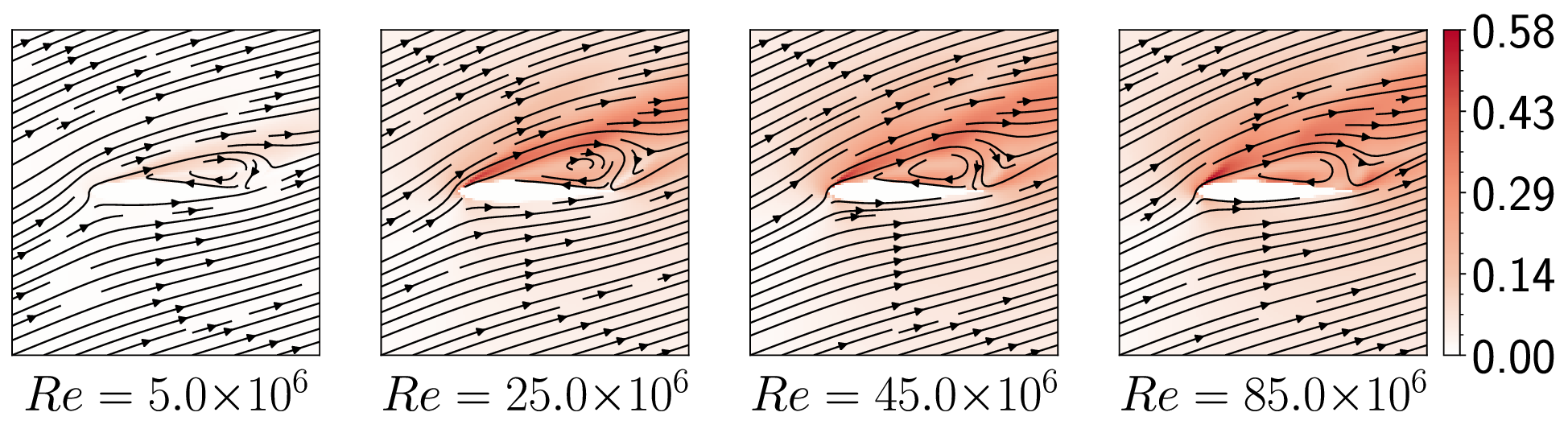}\label{fig:flow_std_re_positive}}\\
    \sidesubfloat[]{\includegraphics[scale=0.33]{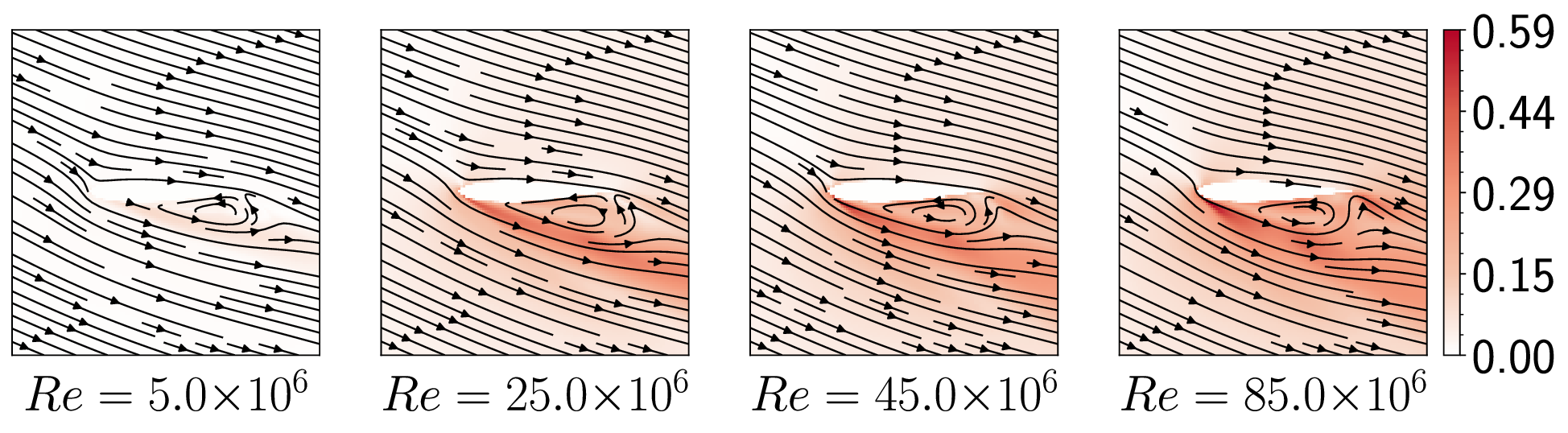}\label{fig:flow_std_re_negative}}
    \caption{Mean streamlines and the uncertainty distribution of velocity magnitude $\sigma_{|\mathbf{u^*}|}$ (raf30 airfoil). a,b) $Re=6.5 \times 10^6$ with varying $\alpha$. c) $\alpha=20^{\circ}$, and d) $\alpha=-20^{\circ}$, both with varying $Re$.}
   \label{fig:flow_std}
\end{figure}

Before evaluating the different methods, a comprehensive exploration of the discussed uncertainty is imperative. Fig.~\ref{fig:raf30} provides an exhaustive depiction of the uncertainty transition in the simulation of the raf30 airfoil across the $(\alpha,Re)$ parameter space. Corresponding to the trend in Fig.~\ref{fig:dataset}, higher uncertainty is observed in the high angle of attack region, escalating with the increase in $Re$. A closer look at the flow field shows that the boundary of the uncertainty transition is also where the flow separation occurs. 
Fig.~\ref{fig:flow_std} provides a more detailed investigation, which presents streamlines of the mean flow field and the uncertainty distribution of velocity magnitude.

Analyzing Fig.~\ref{fig:flow_std_aoa_positive} and Fig.~\ref{fig:flow_std_aoa_negative} reveals a clear correlation between the progression of flow separation and an increasing angle of attack. The growth of the angle of attack augments the radius of curvature of the streamline and enhances the adverse pressure gradient on the upper surface of the airfoil, resulting in an expanding separation bubble and a gradual forward shift in the separation point. Meanwhile, increasing the Reynolds number leads to a gradual transition to turbulence in the flow. 
Despite the significant increase in flow chaos with the development of turbulence, vortices within the turbulent regime enhance momentum transfer perpendicular to the flow direction, bringing streamlines closer to the airfoil. This proximity mitigates adverse pressure gradients, impeding the advancement of flow separation. The separation bubble structures in Fig.~\ref{fig:flow_std_re_positive} and Fig.~\ref{fig:flow_std_re_negative} illustrate limited growth with rising Reynolds numbers, highlighting the stabilizing effect of turbulence on separation. The uncertainty distribution in velocity closely corresponds to the presence of separation, primarily concentrated around the separation bubble. It increases together with the expansion of the separation bubble and the intensity of turbulence near the airfoil. Beyond regions directly influenced by separation, farther away from the airfoil, the impact on uncertainty gradually diminishes. This intricate relationship highlights the interplay between observed uncertainty and the inherent challenges of RANS simulation in capturing critical flows.

\subsection{Single-parameter Experiments}\label{sec:single}

It is instructive to evaluate the accuracy of the different approaches in terms of a reduced setting with a single parameter before turning to the full dataset.
Here, we consider a problem with a resolution of $32\times32$, where the airfoil shape $\Omega$ (raf30) and the angle of attack $\alpha$ ($20^\circ$) are kept constant, i.e., line c) in Fig.~\ref{fig:raf30}. The $Re$ is the only independent variable, and the training dataset encompasses $Re \in$ 
\{1.5$\times 10^6$, 3.5$\times 10^6$, 5.5$\times 10^6$, 7.5$\times 10^6$, 9.5$\times 10^6$\}. As the test dataset, the $Re$ is interpolated and extrapolated into \{2.5$\times 10^6$, 4.5$\times 10^6$, 6.5$\times 10^6$, 8.5$\times 10^6$\} and \{5$\times 10^5$, 10.5$\times 10^6$\}, respectively.

\begin{figure}[tbh]
    \centering
    \includegraphics[scale=0.33]{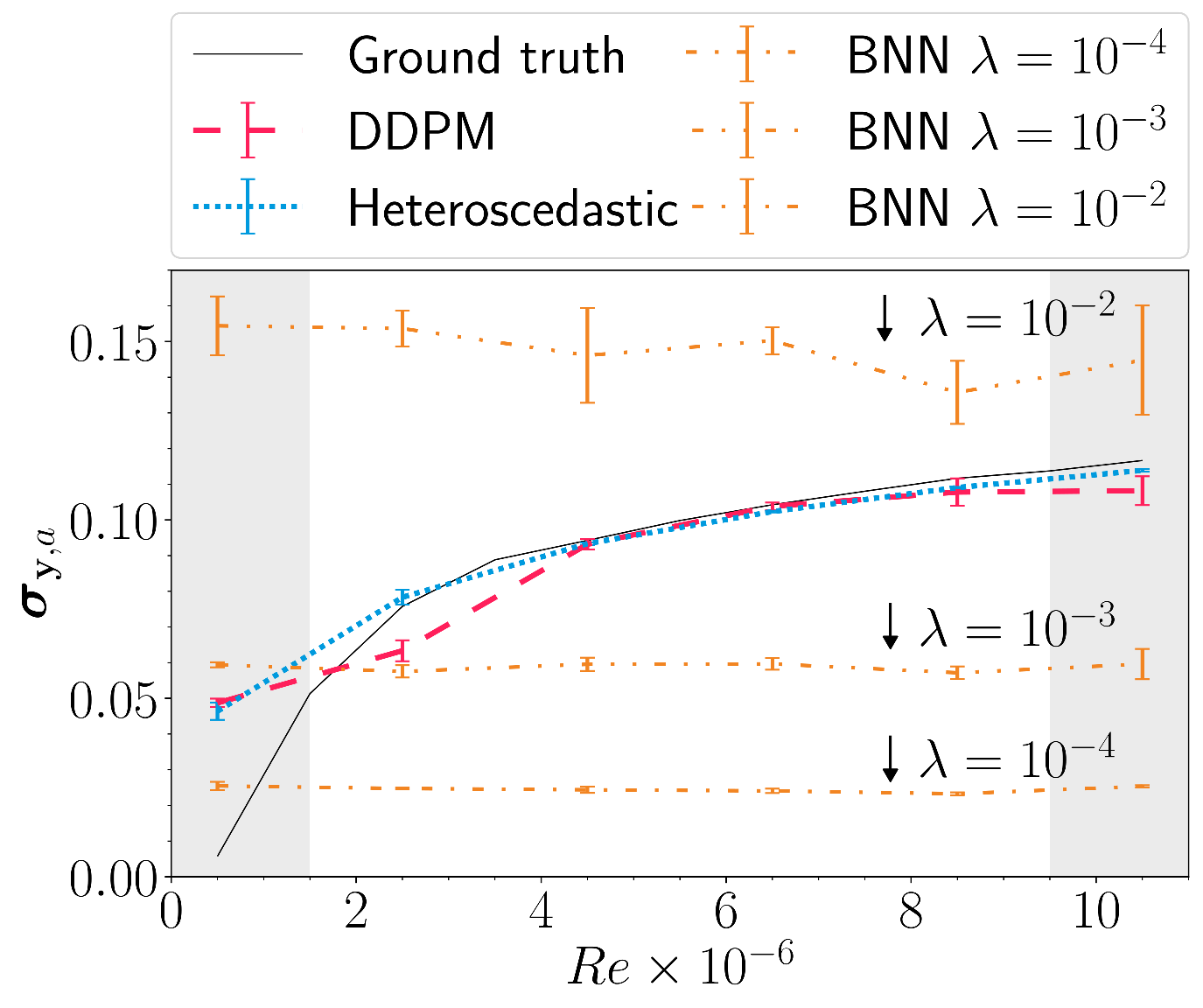}
    \caption{The predicted average standard deviation with increasing $Re$ (raf30 airfoil, $\alpha=20.00^\circ$). The shaded area indicates the extrapolation region of the test dataset. }
    \label{fig:1D_std_linecompare_veall}
\end{figure}

\paragraph{Accuracy.} Fig.~\ref{fig:1D_std_linecompare_veall} plots the average standard deviation of the field predicted by the DDPM, heteroscedastic model, and 3 BNNs with different scaling factors. Compared with BNNs, the DDPM and heteroscedastic model predictions agree well with the ground truth in the interpolation region. For the extrapolation region, the predictions of DDPM and the heteroscedastic model are still adequate for high $Re$, while the standard deviation is over-estimated in the low $Re$ region. 
This is not completely unexpected since the cases in the low $Re$ region substantially differ from those in the training dataset and exhibit essentially constant fields. For BNNs, all predictions of the standard deviation  are far from the ground truth. The higher the scaling factor of the BNN, the higher the standard deviation it predicts. The differences between different networks initialized with different random seeds likewise increase, implying a more random distribution of network parameters as discussed in Sec.~\ref{sec:method:BNN}.

\begin{figure}[tbh]
    \centering
    \sidesubfloat[]{\includegraphics[scale=0.33]{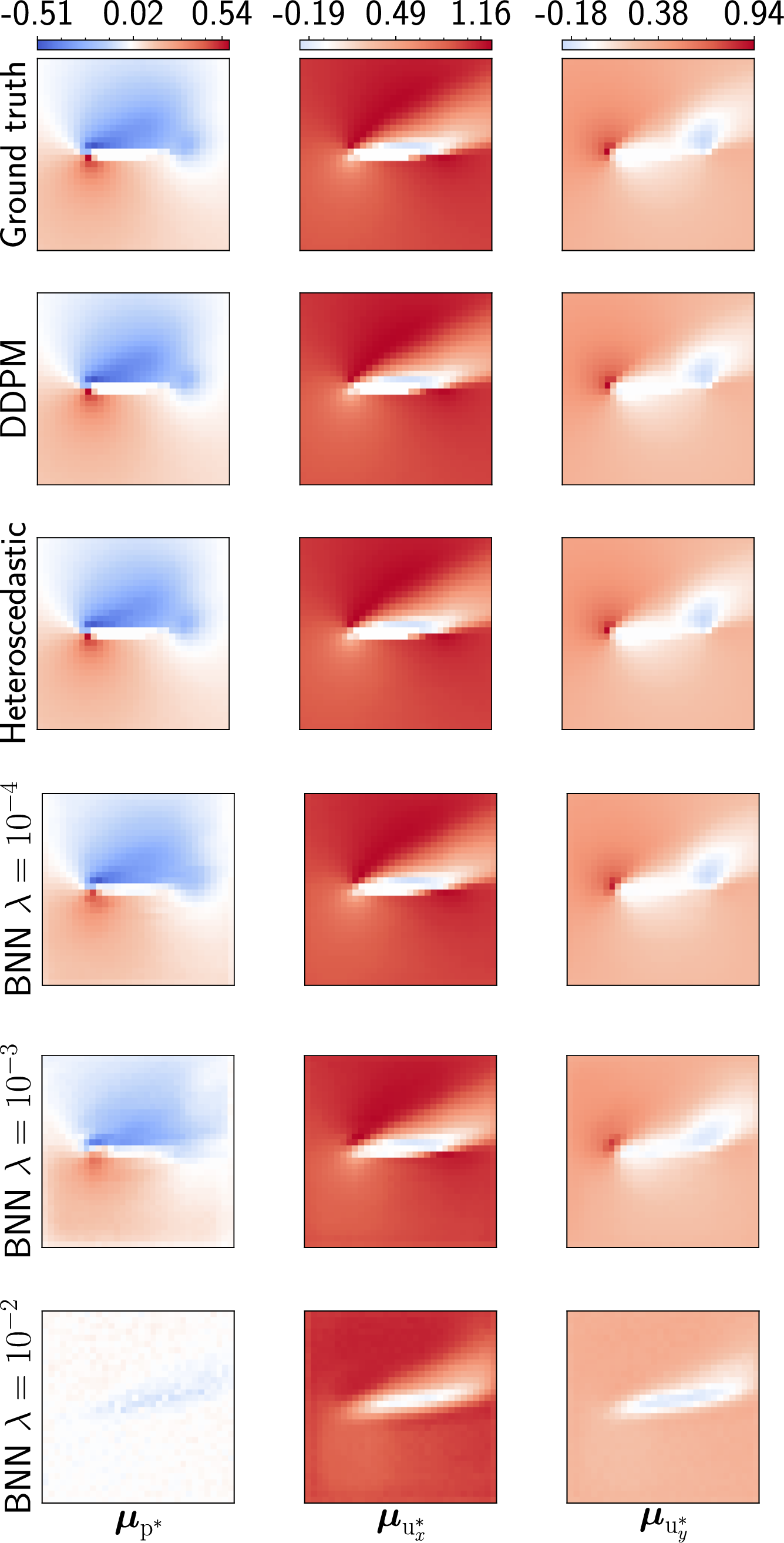}\label{fig:1D_cloudcompare_ve65_mean}}
    \sidesubfloat[]{\includegraphics[scale=0.33]{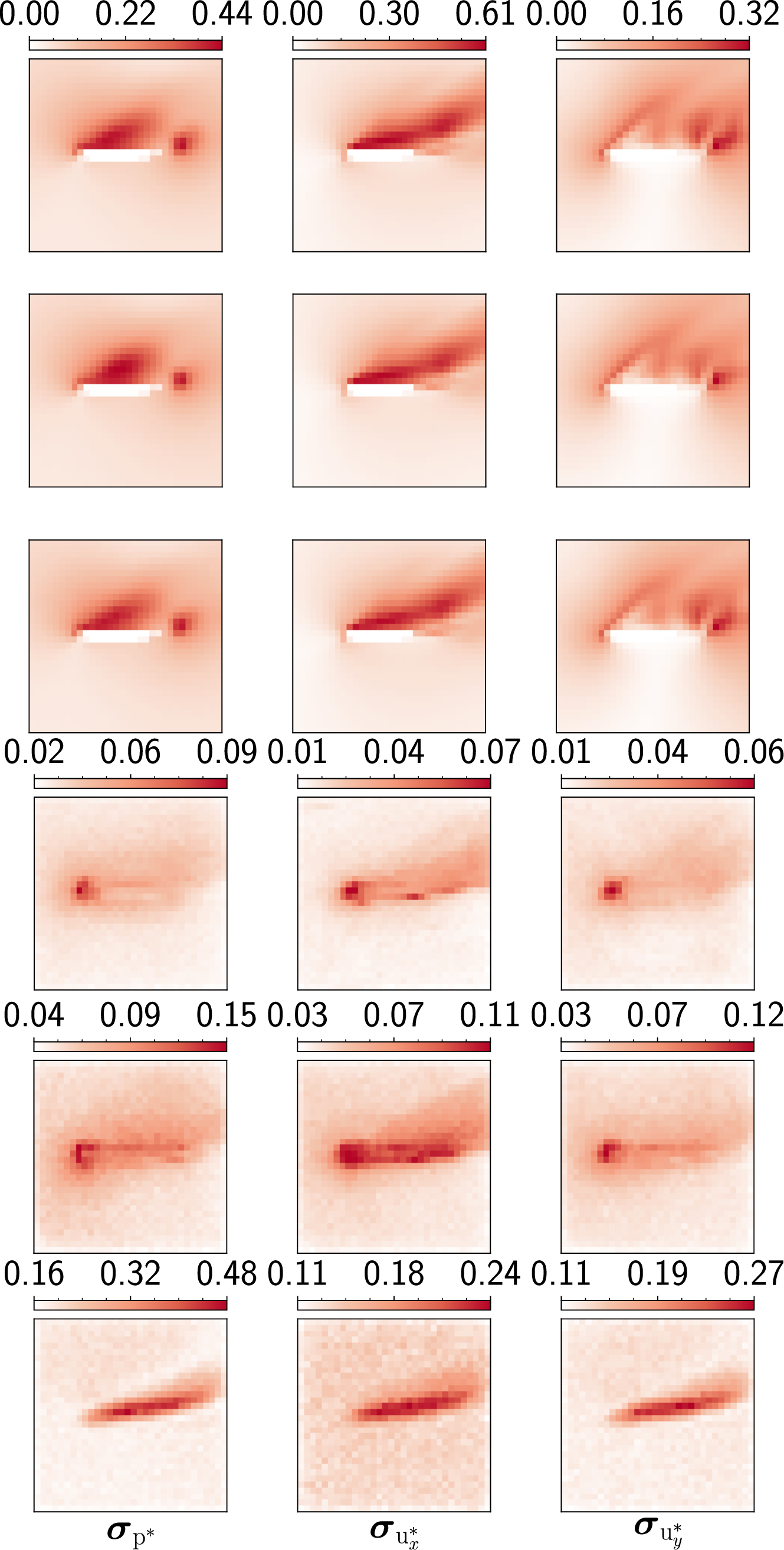}\label{fig:1D_cloudcompare_ve65_std}}
    \caption{The expectation a) and standard deviation b) distribution of the flow field (raf30 airfoil, $Re=6.5\times 10^6$, $\alpha=20.00^\circ$).}
   \label{fig:1D_cloudcompare_ve65}
\end{figure}

Meanwhile, the pattern of the predicted expectation and standard deviations predicted by DDPM and heteroscedastic model agree well with the ground truth, as shown in Fig.~\ref{fig:1D_cloudcompare_ve65}. The performance of the BNNs strongly depends on the scaling factor $\lambda$ instead, where a more deterministic BNN with a smaller $\lambda$ can predict a more accurate expectation field as shown in Fig.~\ref{fig:1D_cloudcompare_ve65}. Nonetheless, it is worth noting that the pattern of the standard deviation predicted by BNN is far from the ground truth even when $\lambda$ is manually adjusted to match the magnitudes of the standard deviation of the ground truth distribution. An extended discussion on the reason behind the BNNs' performance can be found in Appendix~\ref{sec:app:discusssion_bnn}.

\begin{figure}[tbh]
    \centering
    \sidesubfloat[]{\includegraphics[scale=0.33]{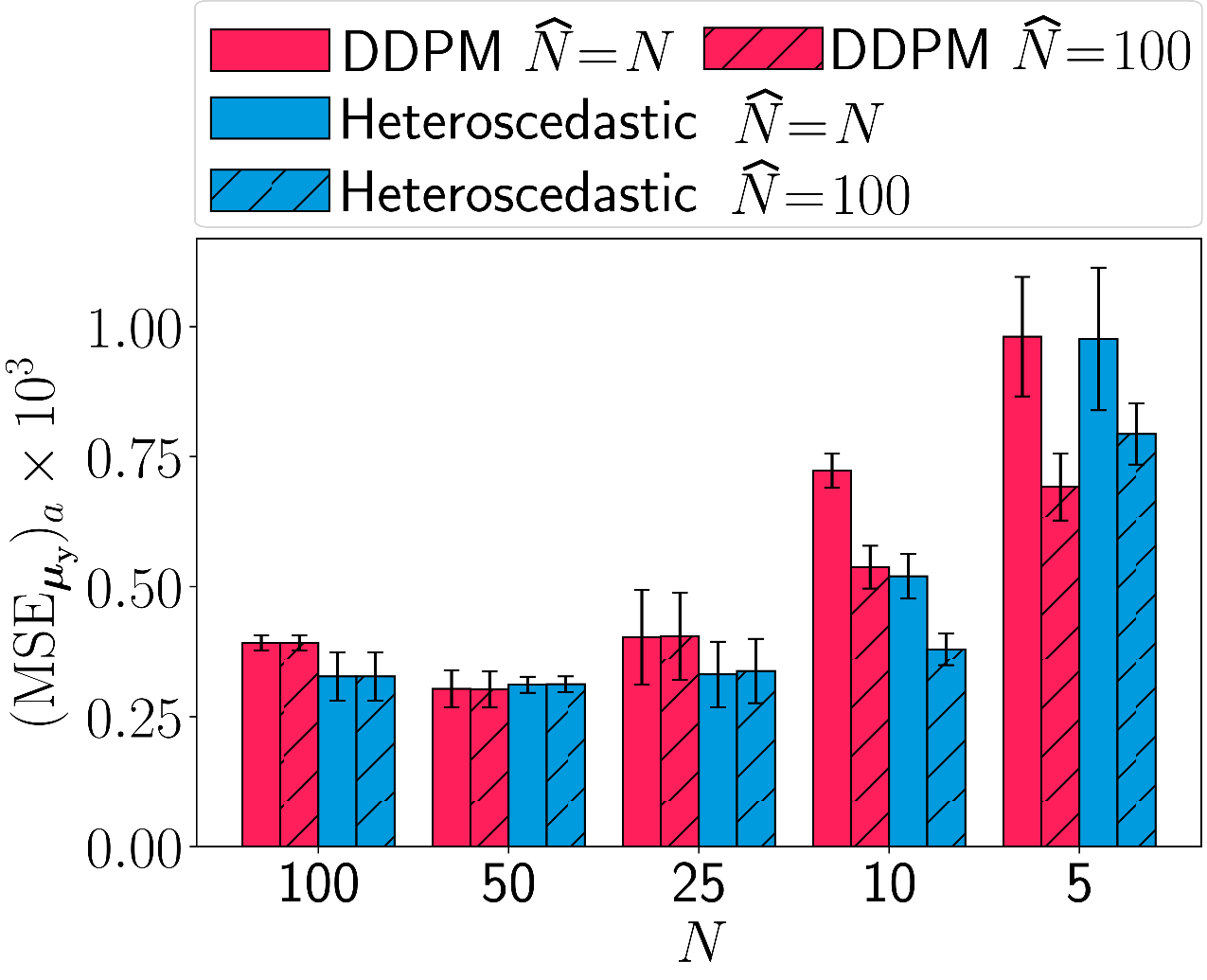}\label{fig:different_sample_mean}}
    \sidesubfloat[]{\includegraphics[scale=0.33]{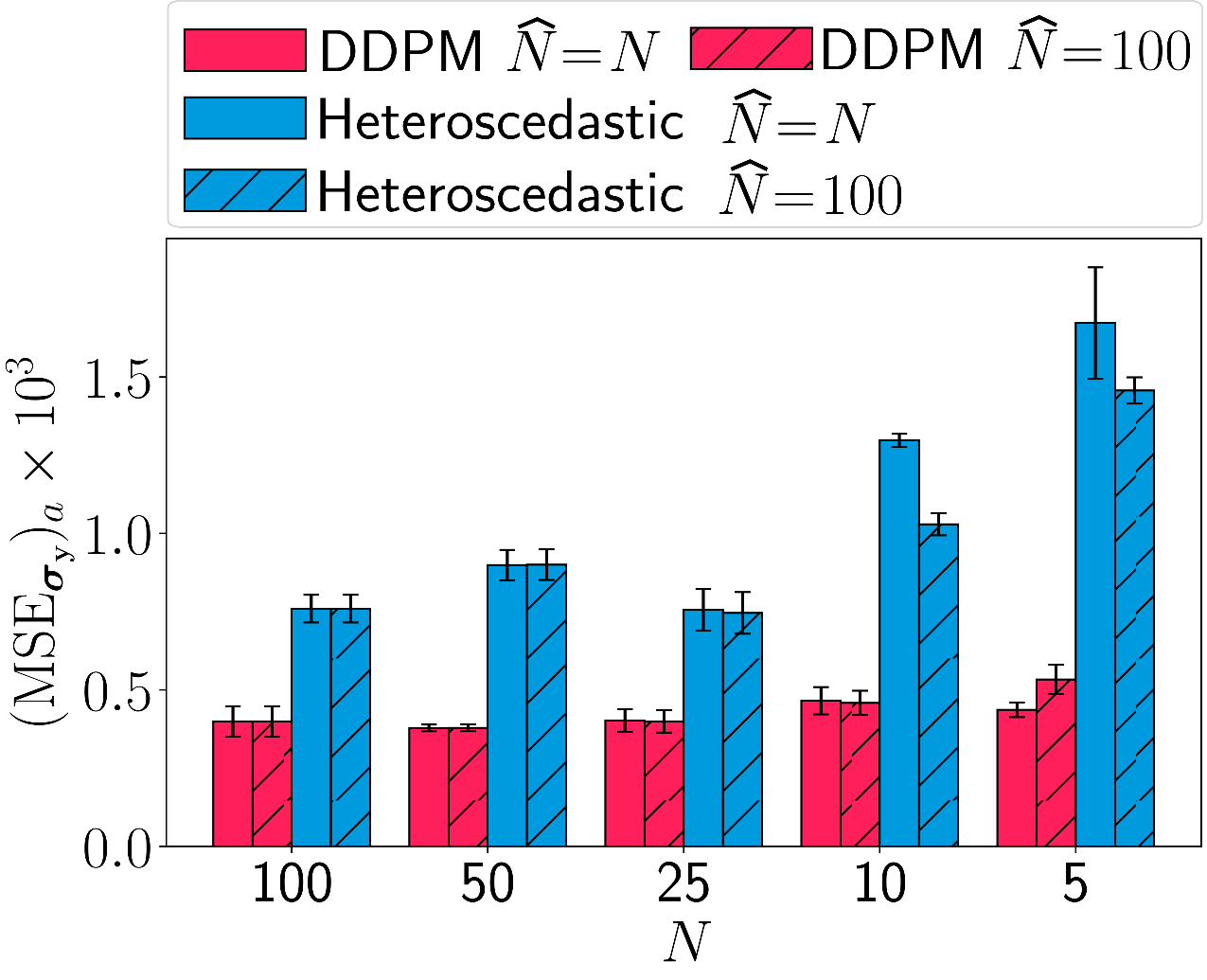}\label{fig:different_sample_std}}
    \caption{The prediction error of expectation a) and standard deviation b) of DDPM and heteroscedastic model trained with different number of snapshot samples.}
    \label{fig:different_sample}
\end{figure}

\paragraph{Number of snapshots samples.} The influence of the number of snapshot samples in the training dataset on DDPM's performance is examined in Fig.~\ref{fig:different_sample}. The comparison involves the performance of DDPM and the heteroscedastic model trained with varying numbers of snapshot samples. 
BNNs are omitted due to their suboptimal accuracy, as shown above. Predictions are evaluated against two reference test dataset: one calculated using the same number of training samples ($\widehat{N}=N$), and a second one using 100 snapshot samples ($\widehat{N}=100$). DDPM and the heteroscedastic model exhibit similar performance for predicting the expectation field, with a noticeable reduction when the number of samples drops below 25. Notably, DDPM consistently outperforms the heteroscedastic model in standard deviation prediction. Although Fig.~\ref{fig:1D_std_linecompare_veall} suggests that the heteroscedastic model has a lower error in absolute uncertainty values, Fig.~\ref{fig:different_sample_std} establishes that DDPM excels in accurately predicting uncertainty distribution. The uncertainty prediction accuracy of the heteroscedastic model experiences a comparatively larger decline when $N<25$ compared with DDPM. Furthermore, Fig.~\ref{fig:different_sample_std} demonstrates that the Mean Squared Error (MSE) calculated based on the $\widehat{N}=N$ dataset and $N=100$ dataset begins to diverge only after $N<25$, validating our decision to train the networks with 25 samples. Fig.~\ref{fig:1D_number_of_samples} in Appendix~\ref{sec:app:generalization} additionally provides plots depicting distributions of ground truth and predictions generated using different numbers of snapshot samples.

\begin{figure}[htb]
    \centering
    \includegraphics[scale=0.33]{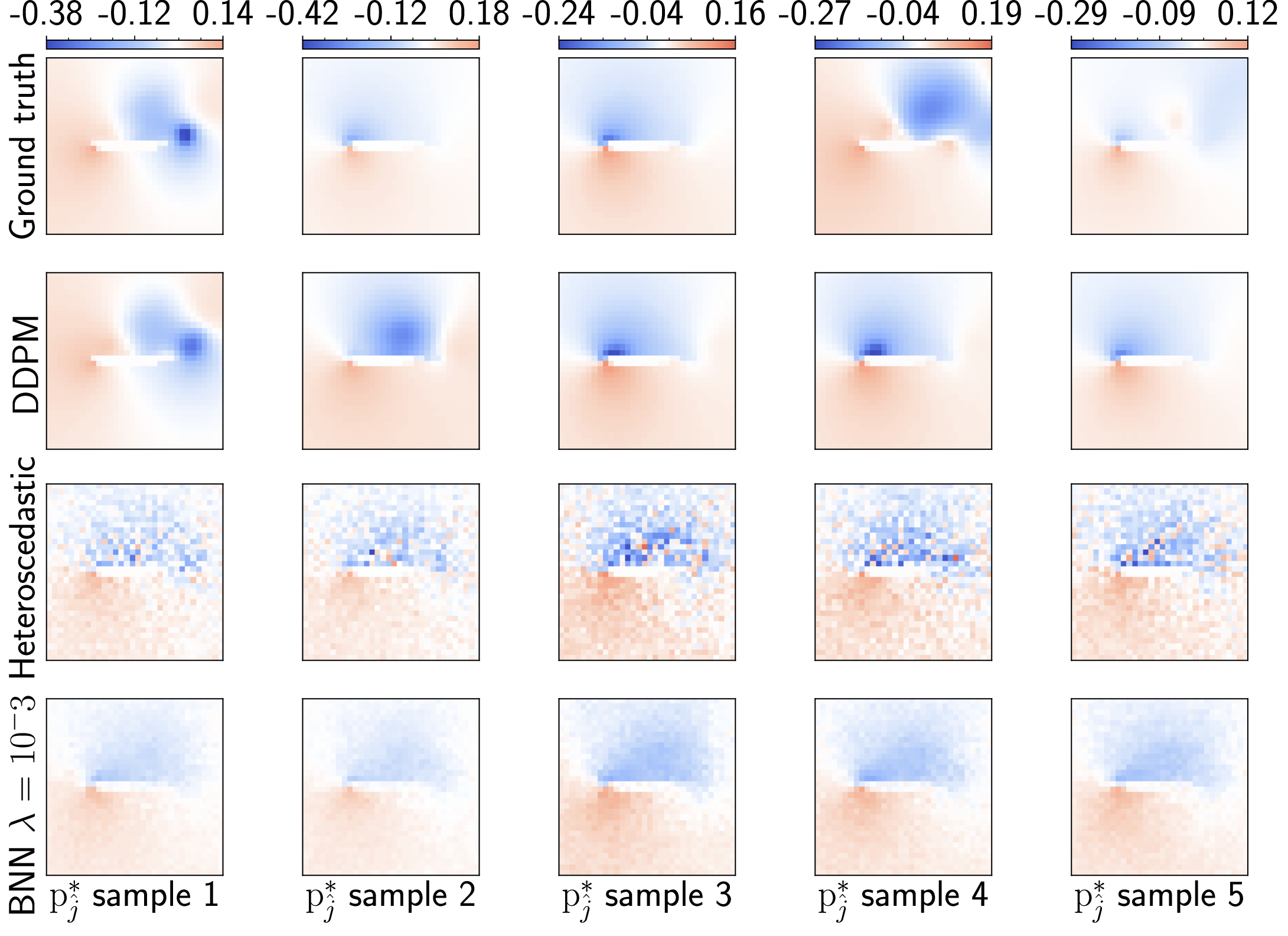}
    \caption{Samples from the distribution of solutions predicted by different models and the ground truth distribution (raf30 airfoil, $Re=6.5\times 10^6$, $\alpha=20.00^\circ$).}
    \label{fig:1D_samples}
\end{figure}

\paragraph{Individual samples.} Fig.~\ref{fig:1D_samples} shows 5 samples from the distribution of solutions predicted by different models and the ground truth distribution for $Re=6.5\times10^6$. It is apparent that the DDPM gives meaningful samples, while the sampling processes for the other two learned methods result in incoherent and noisy fields. Each sample of the distribution of solutions predicted by the DDPM is obtained by iteratively transforming the Gaussian noise field into a prediction, taking neighborhoods and the global state into account via the neural network. In contrast, the sampling of the heteroscedastic model is a series of independent Gaussian sampling steps at each data point in the flow field, which only depends on the local parameters predicted for the distribution and is independent of adjacent points. Hence, the obtained flow field sample is fundamentally unsmooth. For the BNN, each sample is predicted by the network with differently sampled parameters. While this could theoretically, like the DDPM, take neighborhoods and the global state into account, the noisy samples illustrate the shortcomings of the BNN training and inference process. 

\subsection{Multi-parameter Experiments}
In this section, we train the networks on the full dataset where the airfoil shape $\Omega$, $Re$, and $\alpha$ are all independent variables. Since BNNs do not predict acceptable results in the simpler single-parameter case, we focus on the DDPM and the heteroscedastic model for the following learning tasks with increased difficulty. 

\paragraph{Accuracy.} 
We first evaluate the accuracy 
with a dataset of targets with a resolution of $32\times32$. The test dataset is further divided into low-uncertainty($\boldsymbol{\sigma}_{\mathbf{y},a} < 5\times10^{-3}$) and high-uncertainty($\boldsymbol{\sigma}_{\mathbf{y},a} \geq 5\times10^{-3}$) cases to evaluate the model predictions separately. 

\begin{table}[tbh]
\caption{The average MSE of the model prediction on the test dataset. Cases where DDPM outperforms the heteroscedastic model are shown bolded.}
\label{tab:evaluation_on_test}
\begin{tabular}{cccccc}
\hline
\multirow{2}{*}{Dataset region} & Uncertainty         & \multicolumn{2}{c}{$(\mathrm{MSE}_{\boldsymbol{\mu}_\mathbf{y}})_a \times 10^3$} & \multicolumn{2}{c}{$(\mathrm{MSE}_{\boldsymbol{\sigma}_\mathbf{y}})_a \times 10^3$} \\ \cline{3-6} 
                                & categories          & Heteroscedastic                        & DDPM                                  & Heteroscedastic                         & DDPM                                   \\ \hline
                                & low $\boldsymbol{\sigma}_y$ cases  & 0.834$\pm$0.043                        & \textbf{0.320$\pm$0.037}              & 1.329$\pm$0.293                         & \textbf{0.384$\pm$0.112}               \\
Interpolation region            & high $\boldsymbol{\sigma}_y$ cases & 1.029$\pm$0.041                        & \textbf{1.014$\pm$0.144}              & 1.240$\pm$0.388                         & \textbf{0.885$\pm$0.059}               \\
                                & All cases           & 0.900$\pm$0.038                        & \textbf{0.556$\pm$0.033}              & 1.299$\pm$0.325                         & \textbf{0.555$\pm$0.056}               \\
                                & low $\boldsymbol{\sigma}_y$ cases  & 1.465$\pm$0.132                        & \textbf{0.837$\pm$0.050}              & 0.363$\pm$0.206                         & \textbf{0.027$\pm$0.020}               \\
Extrapolation region            & high $\boldsymbol{\sigma}_y$ cases & \textbf{2.284$\pm$0.169}               & 2.838$\pm$0.249                       & \textbf{1.744$\pm$0.191}                & 3.196$\pm$0.916                        \\
                                & All cases           & 1.765$\pm$0.127               & \textbf{1.571$\pm$0.118}                       & \textbf{0.869$\pm$0.124}                & 1.189$\pm$0.333                        \\ \hline
\end{tabular}
\end{table}

\begin{figure}[tbh]
    \centering
    \sidesubfloat[]{\includegraphics[scale=0.33]{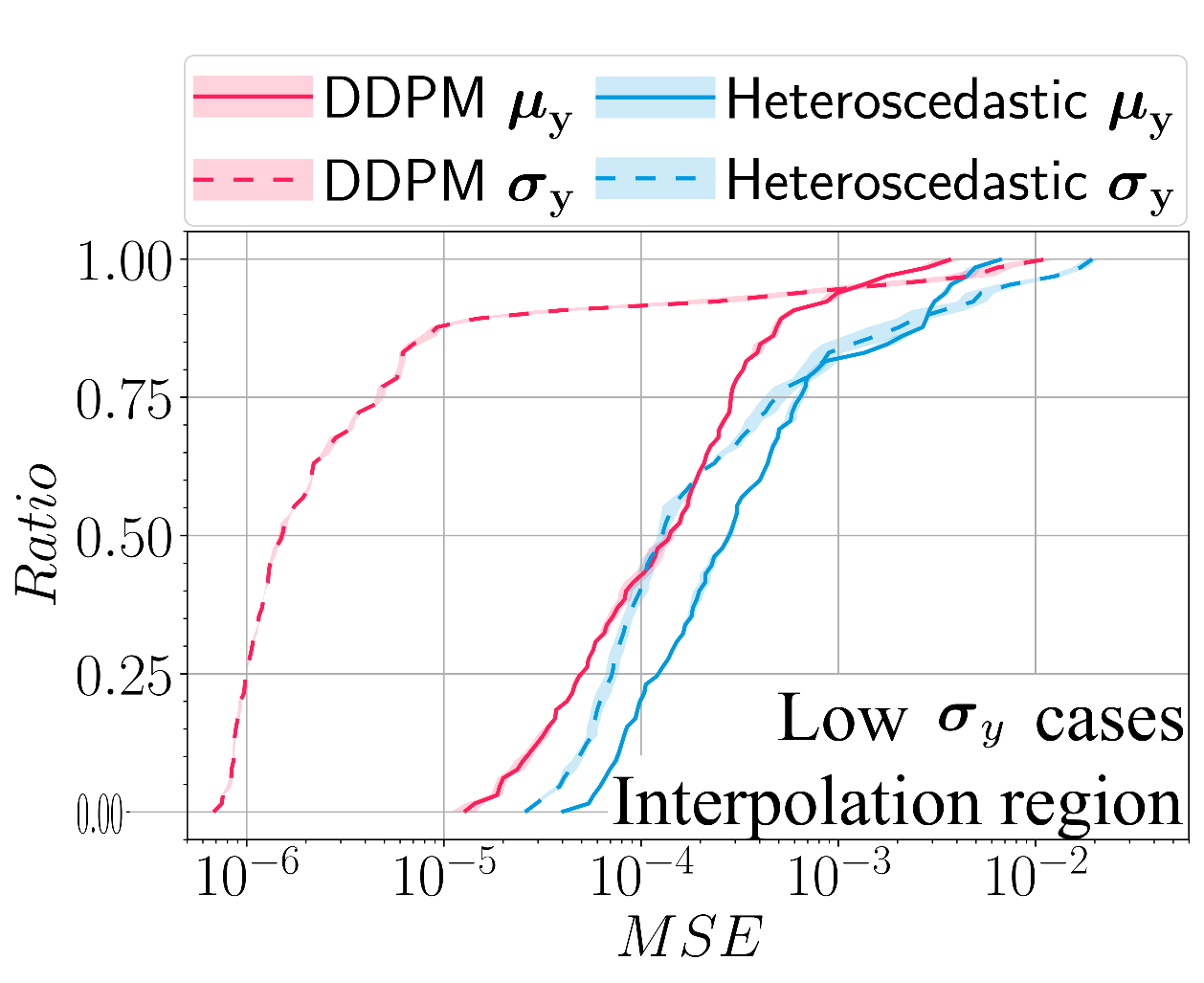}\label{fig:error_distribution_diffusion_he_lowstd_in}}
    \sidesubfloat[]{\includegraphics[scale=0.33]{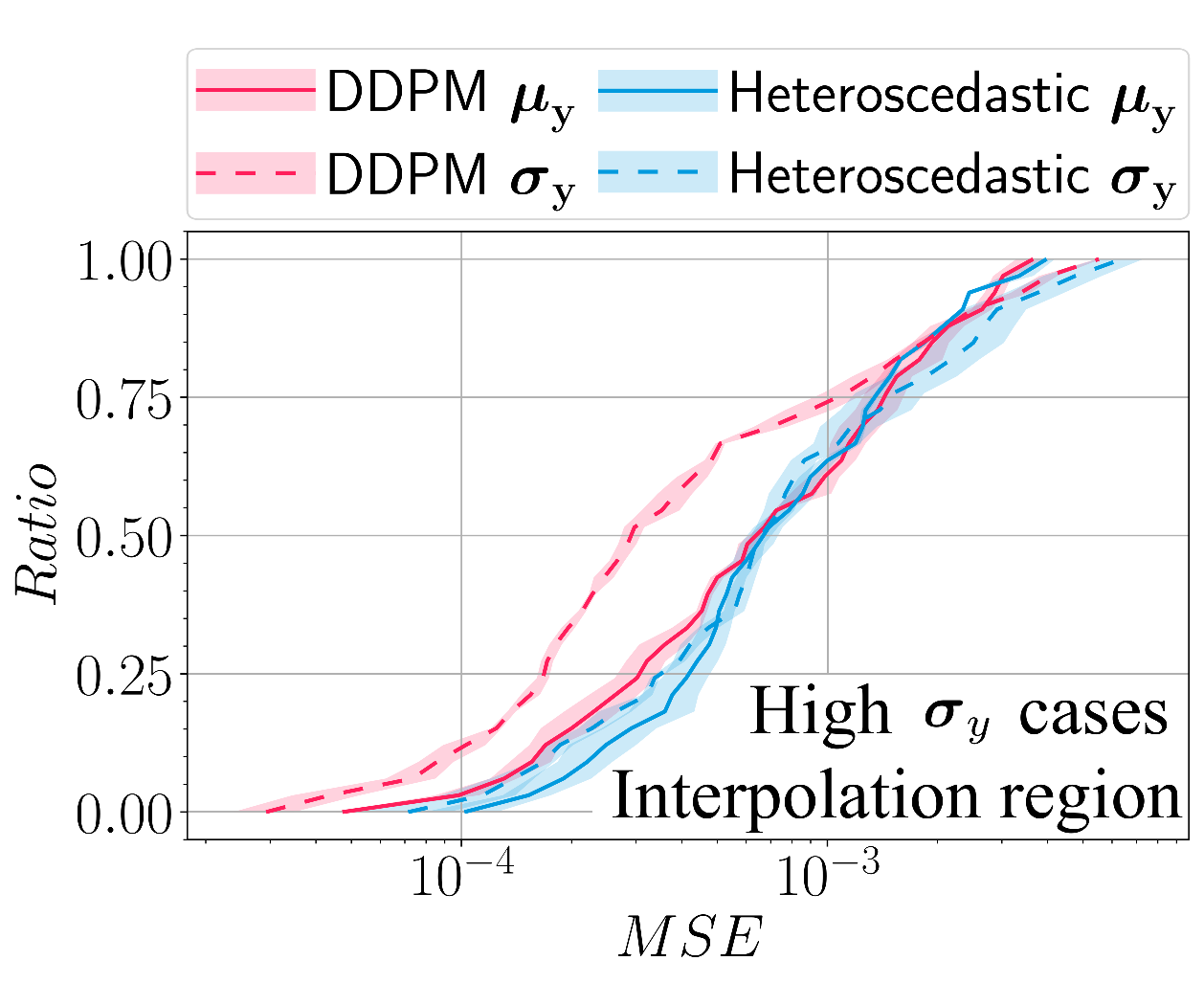}\label{fig:error_distribution_diffusion_high_lowstd_in}}
    \\
    \sidesubfloat[]{\includegraphics[scale=0.33]{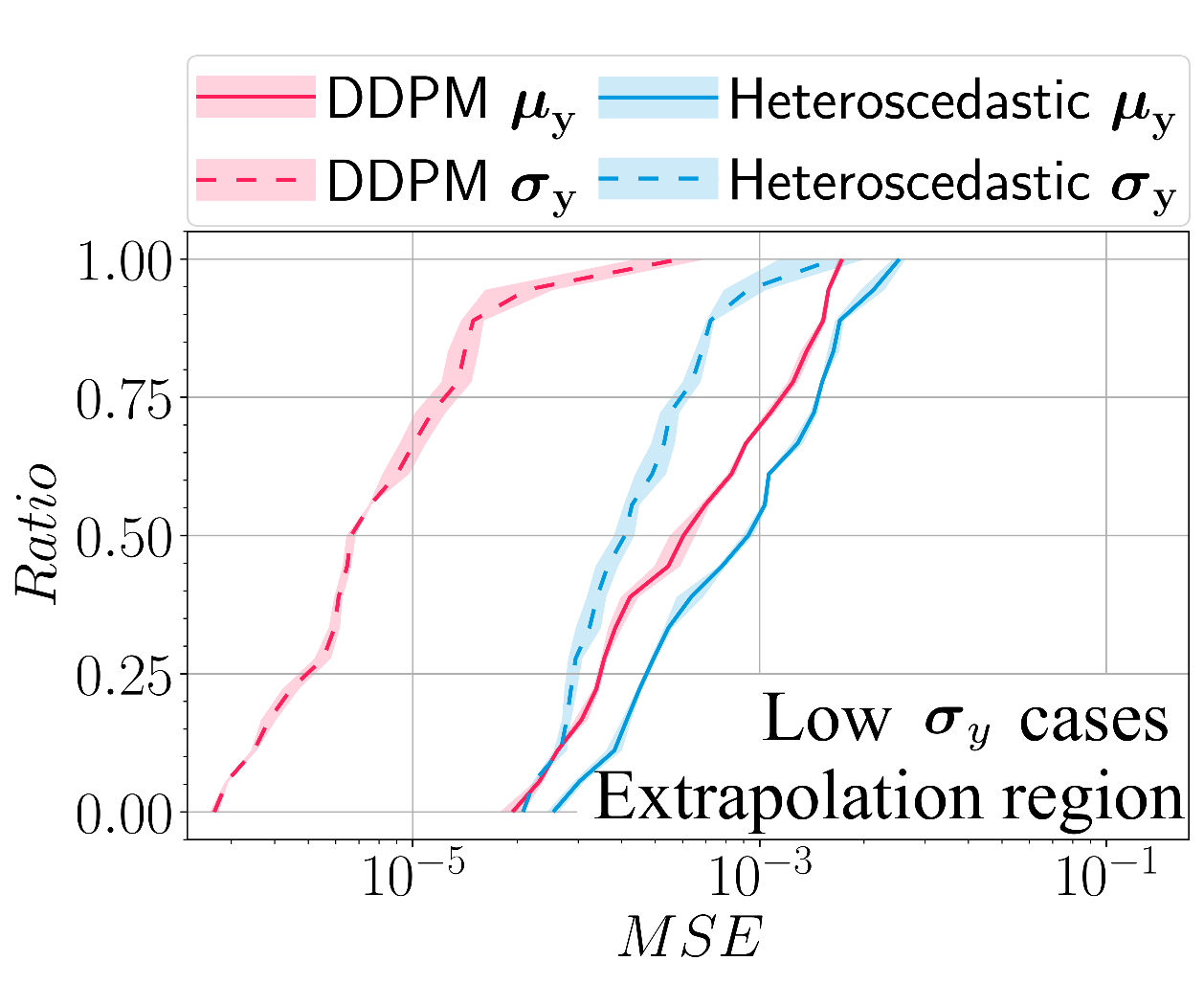}\label{fig:error_distribution_diffusion_he_lowstd_out}}
    \sidesubfloat[]{\includegraphics[scale=0.33]{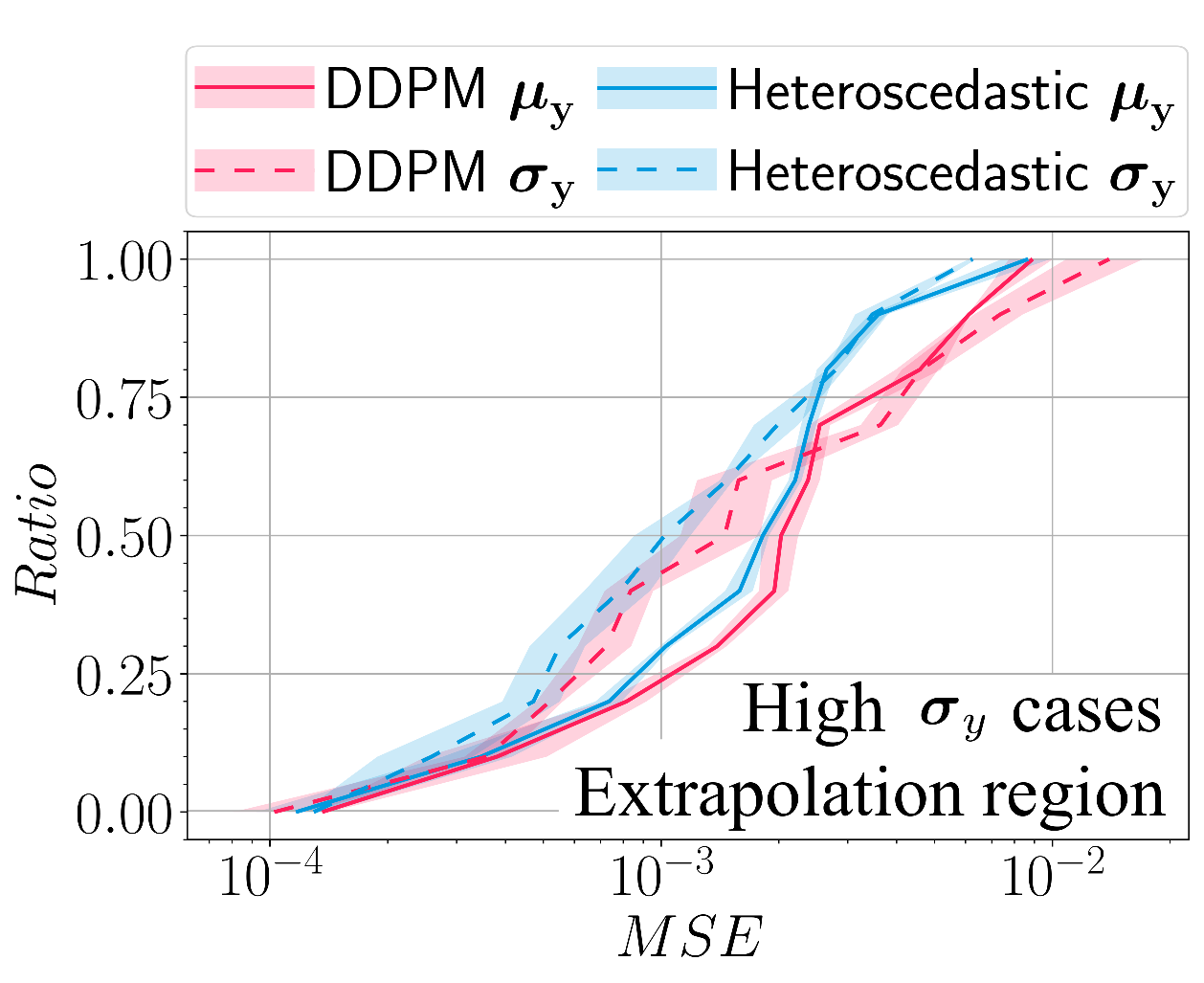}\label{fig:error_distribution_diffusion_high_lowstd_out}}
    \caption{Prediction error distributions of DDPM and heteroscedastic model. a), b): interpolation region; c), d): extrapolation region. a), c): low $\boldsymbol{\sigma}_y$ cases ; b), d): high $\boldsymbol{\sigma}_y$ cases.}
    \label{fig:error_distribution_diffusion_he}
\end{figure}

\begin{figure}[tbh]
	\centering
	\sidesubfloat[]{\includegraphics[scale=0.33]{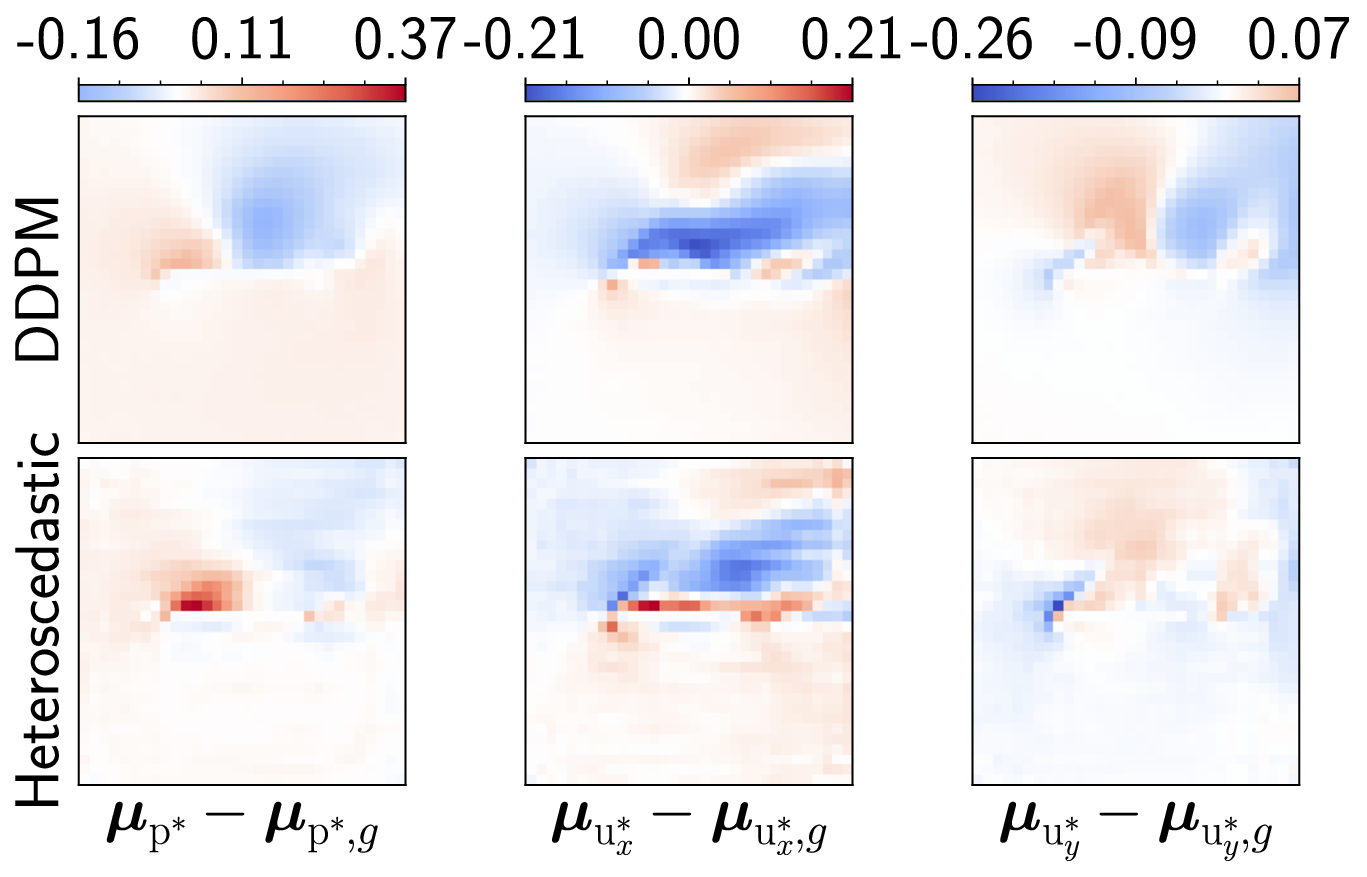}\label{fig:error_mean_fx84w097_7818_2103}}
	\sidesubfloat[]{\includegraphics[scale=0.33]{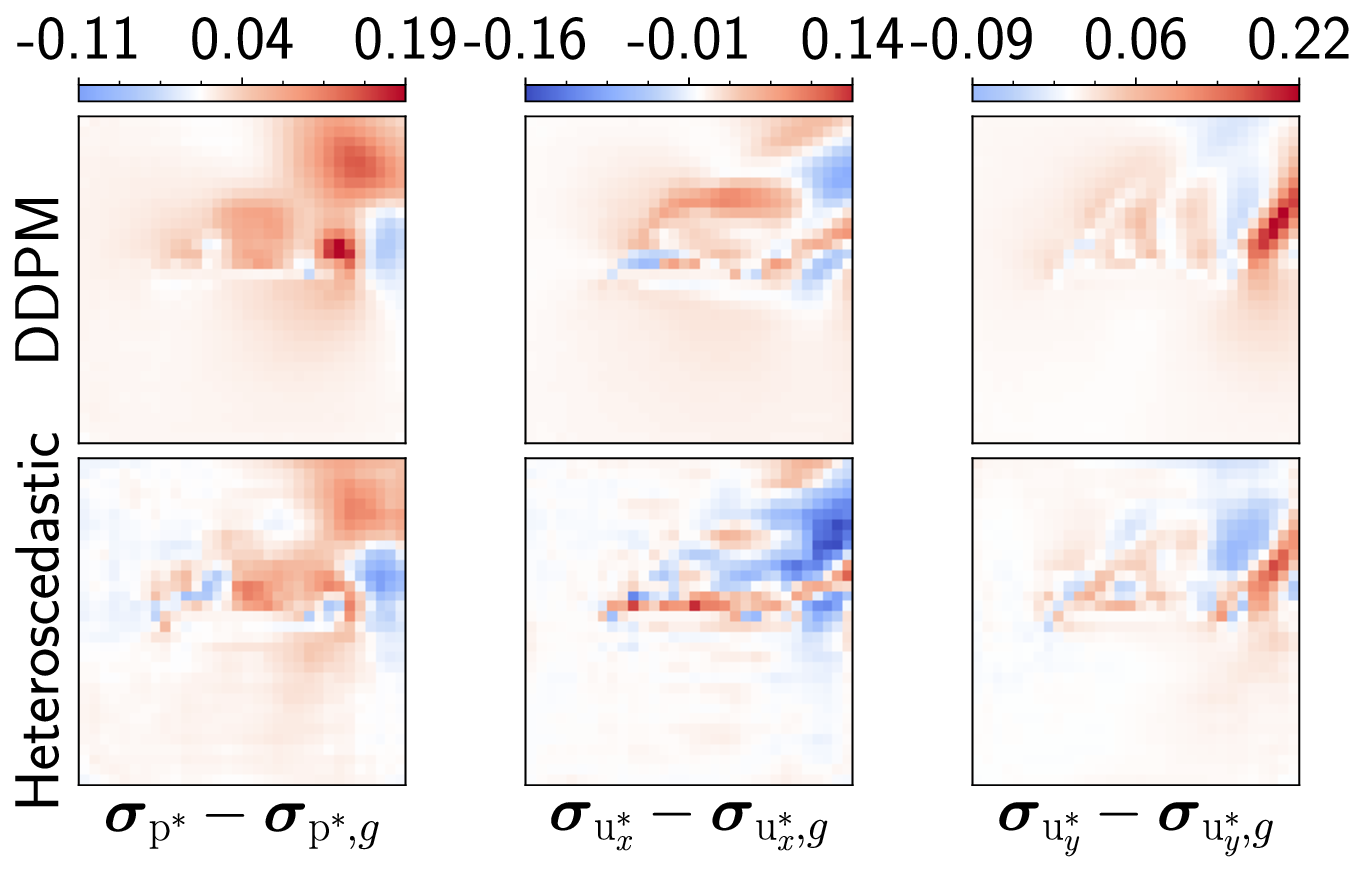}\label{fig:error_std_fx84w097_7818_2103}}
	\caption{The deviation of the expectation a) and standard deviation b) field predicted by DDPM and heteroscedastic model(fx84w097 airfoil $Re=78.18\times 10^6$, $\alpha=21.03^\circ$).}
	\label{fig:error_fx84w097_7818_2103}
\end{figure}

The MSE of the predicted expectation fields and standard deviation fields are summarized in Table.~\ref{tab:evaluation_on_test}. Among the cases from the interpolation region, the DDPM outperforms the heteroscedastic model in all metrics. 
While the heteroscedastic model seems to perform slightly better for predictions of high-uncertainty cases in the extrapolation region, Fig.~\ref{fig:error_distribution_diffusion_he} provides a more detailed evaluation of the error distributions. 
Here, values on the y-axis represent the ratio of predictions whose MSE is less than the corresponding value on the x-axis. Fig.~\ref{fig:error_distribution_diffusion_he_lowstd_in} and Fig.~\ref{fig:error_distribution_diffusion_he_lowstd_out} clearly show that DDPM surpasses the heteroscedastic model in terms of predicting expectation and standard deviation in the low-uncertainty test cases. Excellent performance is shown in the predictions of the standard deviation, where more than 80$\% $ and 60$\% $ of the DDPM predictions have an error less than $10^{-5}$ for interpolation and extrapolation region, respectively. In contrast, the minimal error of the heteroscedastic model prediction is greater than $10^{-5}$ in both cases. Meanwhile, Fig.~\ref{fig:error_distribution_diffusion_high_lowstd_in} indicates that the DDPM still gives a better prediction of standard deviations  for high-uncertainty cases in the interpolation region, while the difference between the expectation predictions of these two models is not significant. Similarly, Fig.~\ref{fig:error_distribution_diffusion_high_lowstd_out} also demonstrates that although the DDPM seems to produce cases with higher maximum error than the heteroscedastic model, the accuracy of DDPM's predictions is nonetheless mostly on-par with the heteroscedastic model for high uncertainty cases in the extrapolation region. Fig.~\ref{fig:error_fx84w097_7818_2103} shows the deviation of the expectation and standard deviation field predicted by DDPM and the heteroscedastic model for a specific example. The distribution of deviations of DDPM is smoother and more meaningful than the predictions of the heteroscedastic model,
but the visible regions with larger errors indicate an over- or under-representation of specific flow patterns in the predictions.  

As illustrated in Fig.5, one of the major advantages of the DDPM is that it can produce meaningful target samples. To characterize the distribution of solutions predicted by the DDPM in more detail, we compare the distribution of the drag coefficient $C_d$ computed from the sampled flow fields with the \replaced{distribution obtained from simulation data with the same resolution}{ground truth} in Fig.~\ref{fig:Cd_distribution}. Details of the $C_d$ calculation can be found in Appendix~\ref{sec:app:drag_coef}. The $C_d$ distribution obtained from the samples of the heteroscedastic model is also shown for comparison. As illustrated in the figure, the Gaussian hypothesis of the heteroscedastic model does not capture the distribution of the ground truth drag coefficients. DDPM model, on the other hand, infers samples that closely align with the ground truth distribution, accurately capturing the peak around $C_d=0.1$.

\begin{figure}[tbh]
    \centering
    \includegraphics[scale=0.33]{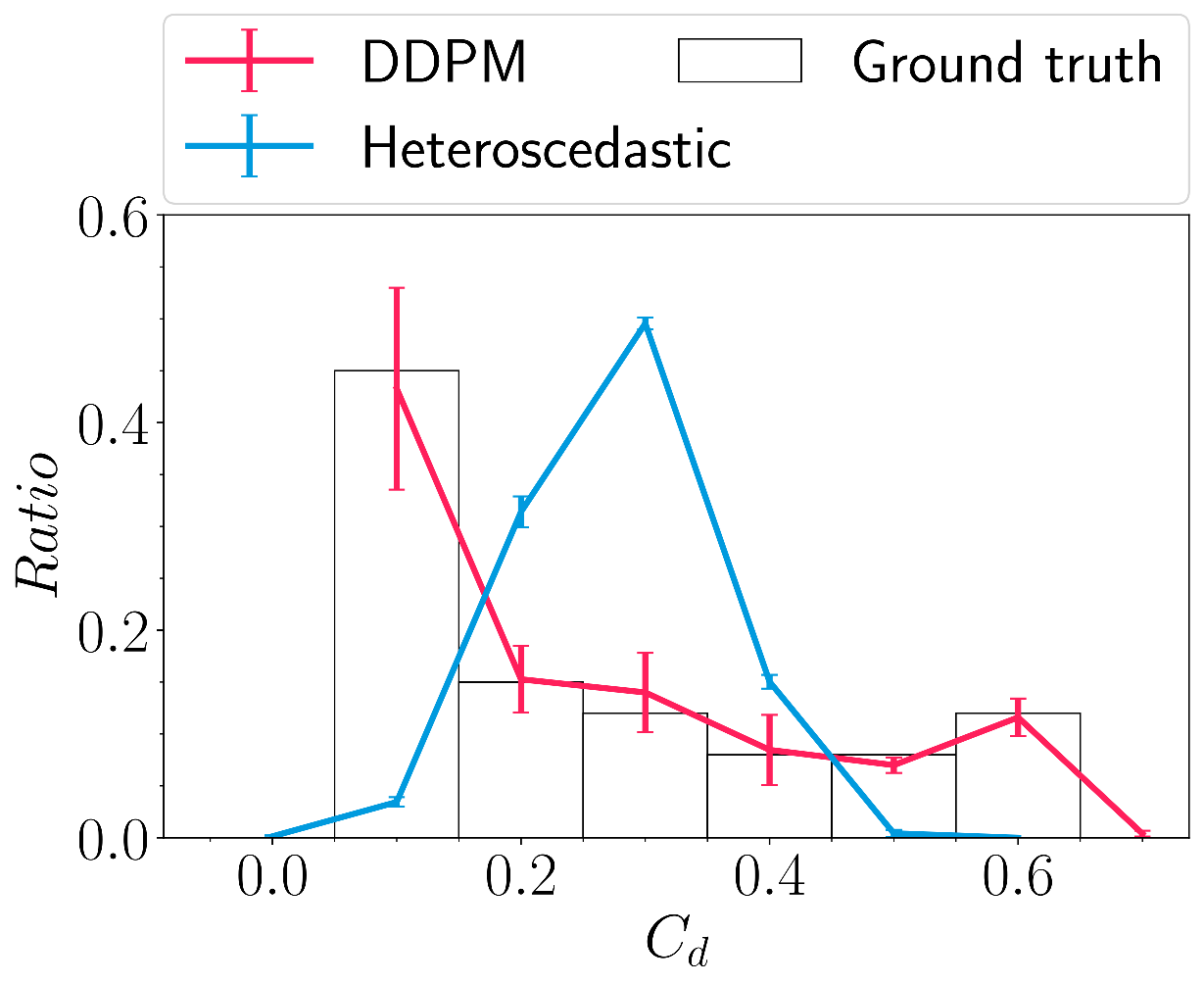}
    \caption{The drag coefficient distribution predicted by the heteroscedastic model and DDPM (ag09 airfoil, $Re=6.918\times 10^6$, $\alpha=15.83^\circ$)}
    \label{fig:Cd_distribution}
\end{figure}

\begin{figure}[tbh]
    \centering
    \sidesubfloat[]{\includegraphics[scale=0.33]{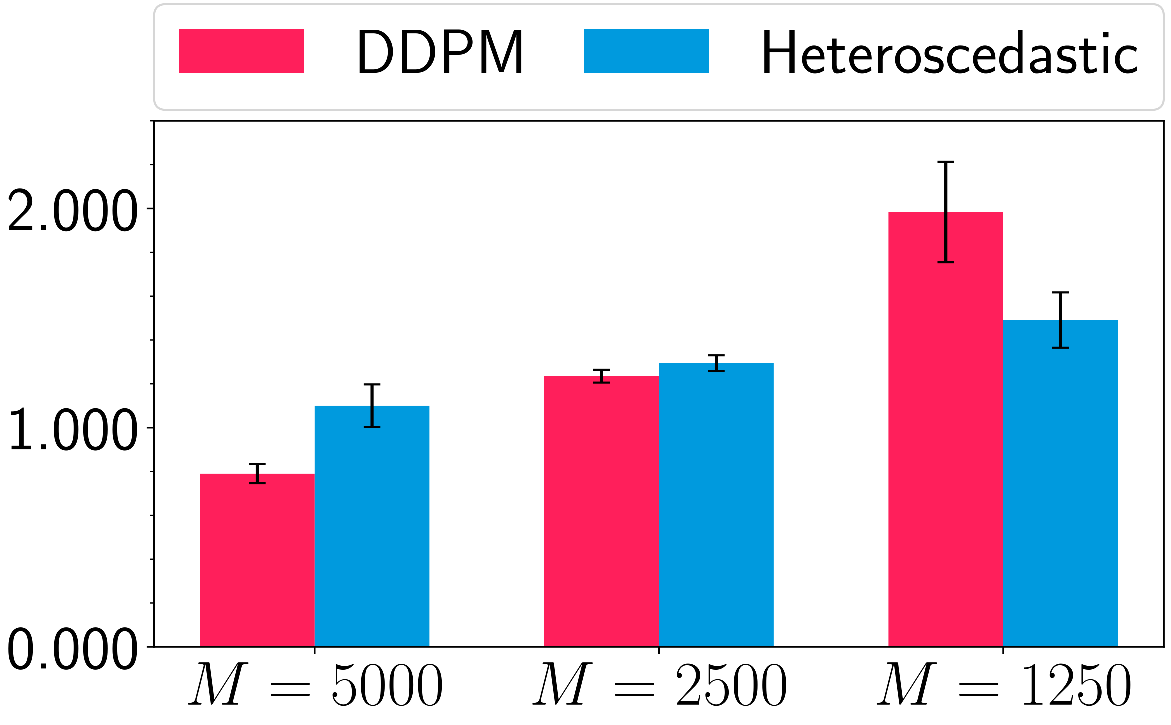}\label{fig:different_num_cases_mean}}
    \sidesubfloat[]{\includegraphics[scale=0.33]{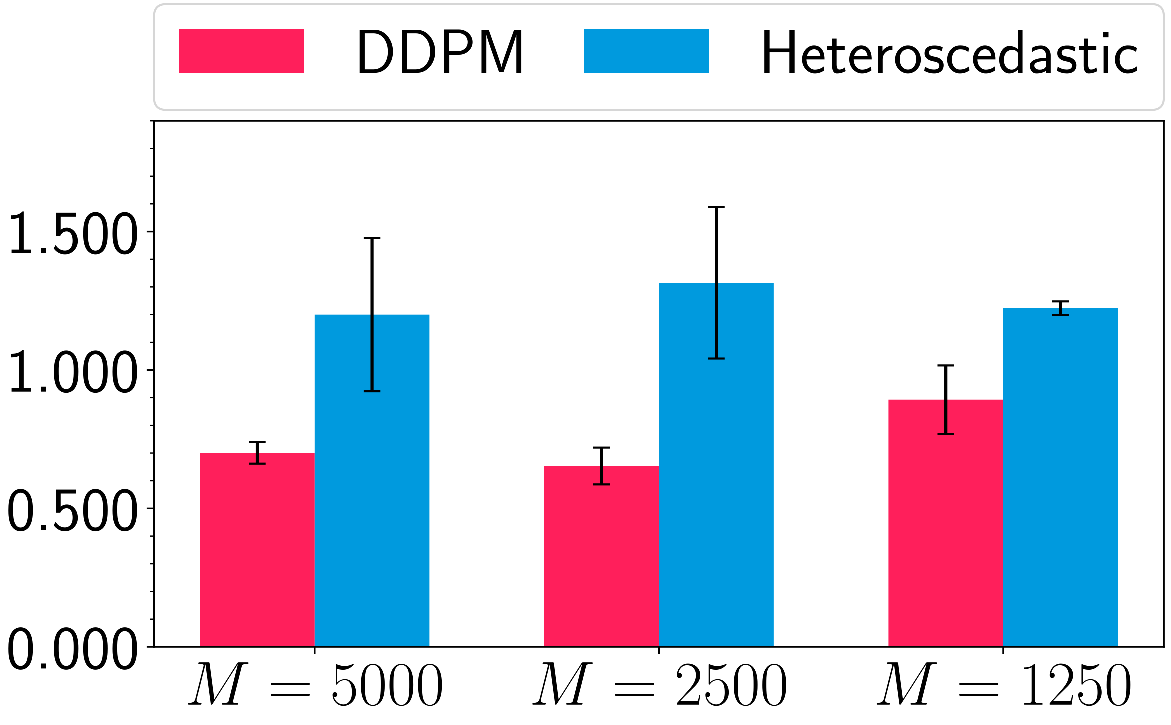}\label{fig:different_num_cases_std}}
    \caption{The prediction error of expectation a) and standard deviation b) of DDPM and heteroscedastic model trained with different number of simulation cases.}
    \label{fig:different_num_cases}
\end{figure}

\paragraph{Number of training cases.}
In Fig.~\ref{fig:different_sample}, we explore the impact of the size of the training dataset on the model performance by adjusting the number of snapshots in each simulation case. Beyond the number of snapshots, the total number of simulation cases is also a critical aspect of the dataset size. Fig.~\ref{fig:different_num_cases} compares the prediction accuracy between DDPM and heteroscedastic models trained with varying numbers of simulation cases. The effect of the number of training cases on the predicted expectations is more pronounced than on the predictions of the  standard deviation.  A discernible trend of decreasing performance in expectation prediction is observed compared to standard deviation prediction. Despite the apparent faster decay in the accuracy of expectation prediction with the reduction of training data for DDPM compared to the heteroscedastic model, DDPM demonstrates superior predictive performance in both expectation and standard deviation when the number of simulation cases is set at $M=5000$ and $M=2500$. Even with a reduced dataset size of $M=1250$, DDPM maintains a considerable advantage in standard deviation prediction.

\begin{figure}[tbh]
    \centering
    \sidesubfloat[]{\includegraphics[scale=0.33]{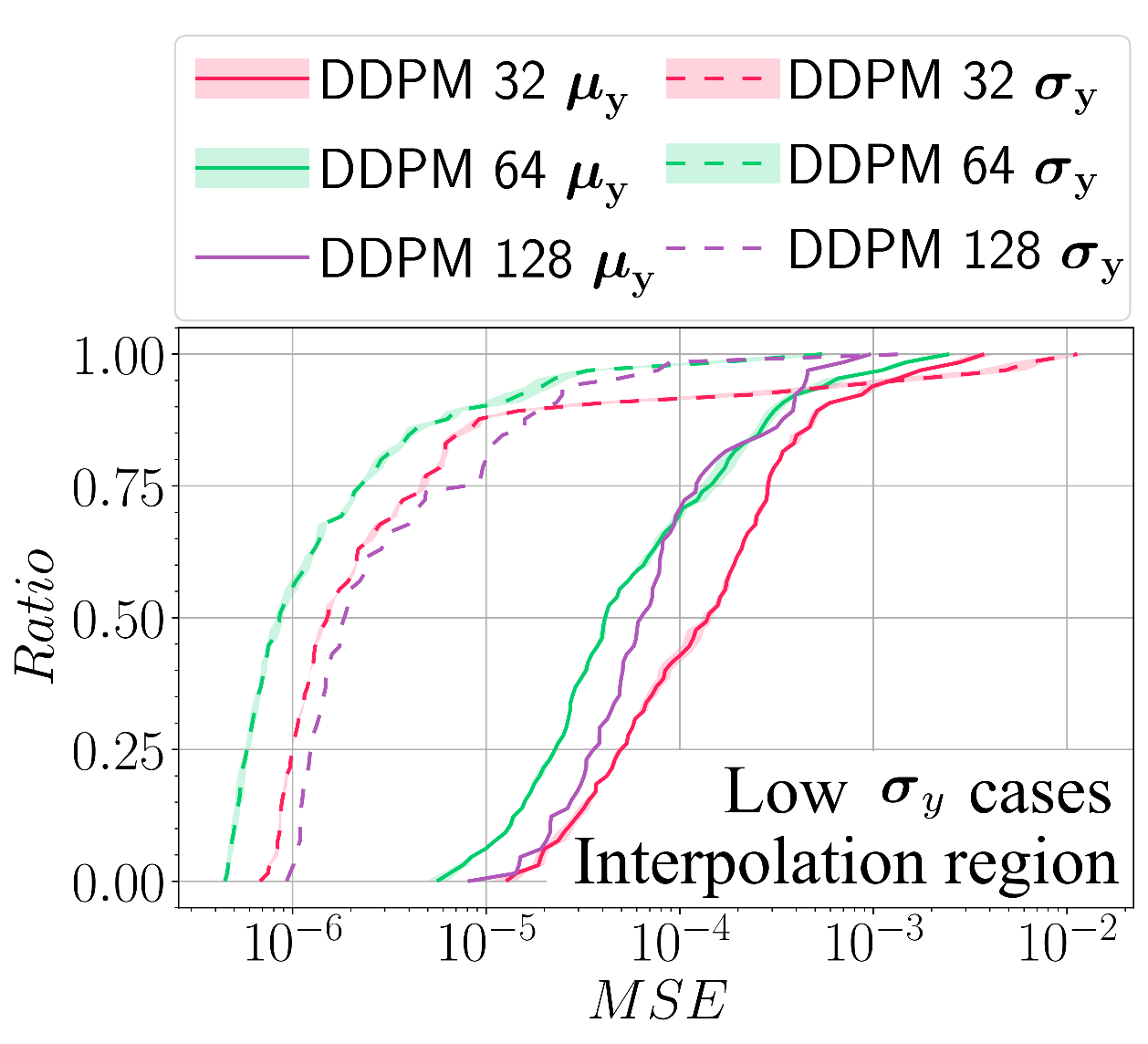}\label{fig:error_distribution_diffusion_64_lowstd_in}}
    \sidesubfloat[]{\includegraphics[scale=0.33]{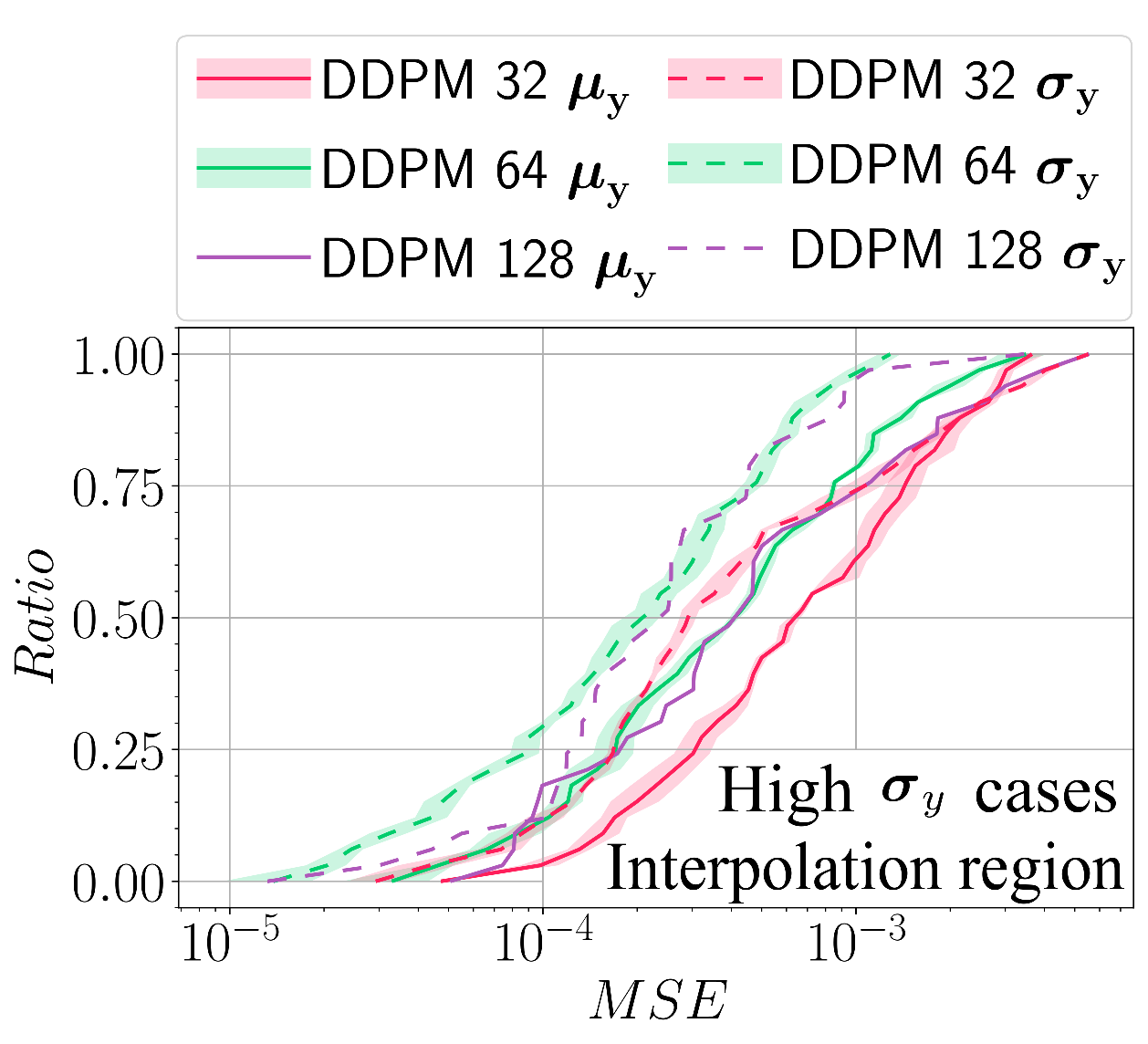}\label{fig:error_distribution_diffusion_64_highstd_in}}
    \\
    \sidesubfloat[]{\includegraphics[scale=0.33]{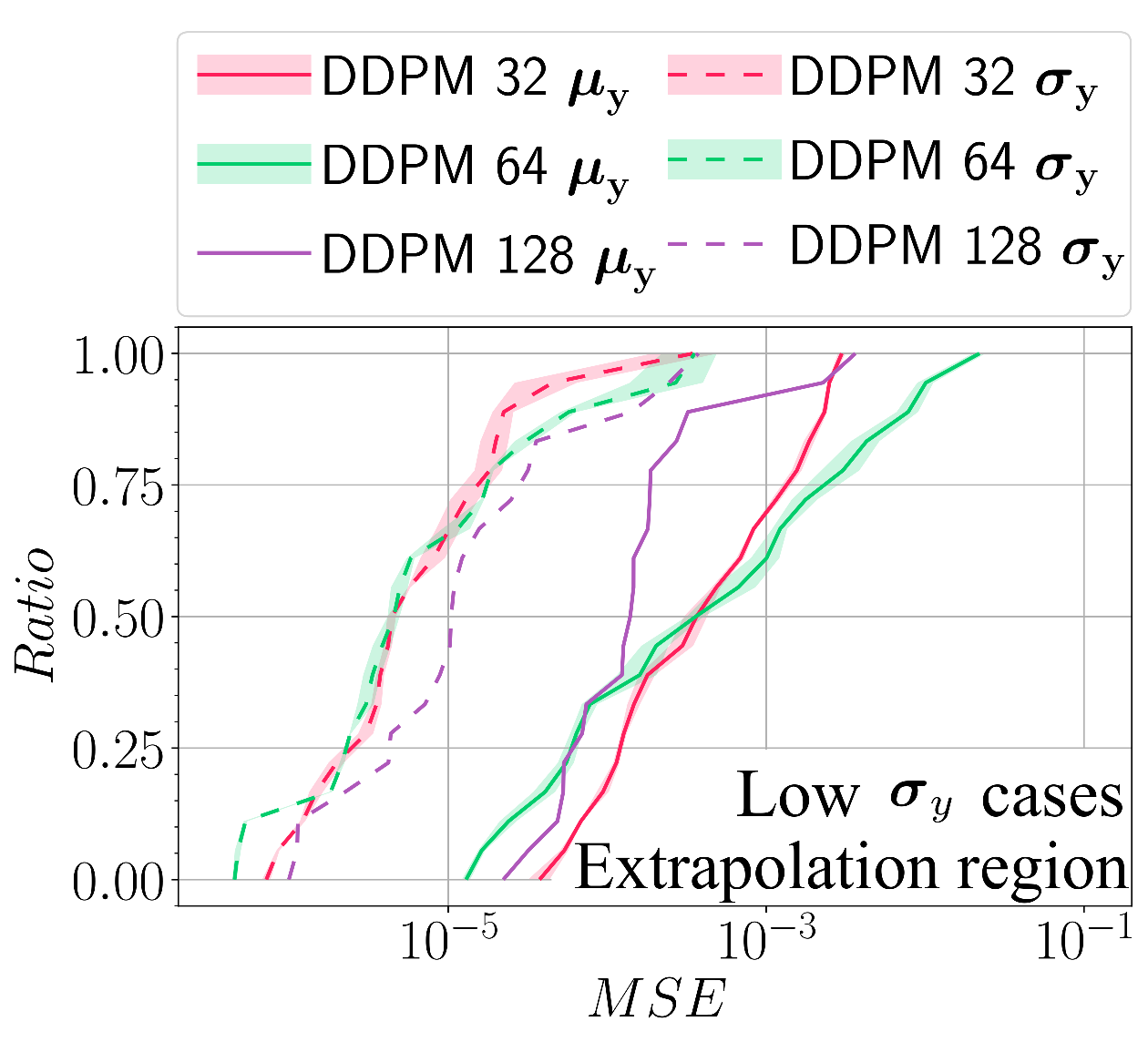}\label{fig:error_distribution_diffusion_64_lowstd_out}}
    \sidesubfloat[]{\includegraphics[scale=0.33]{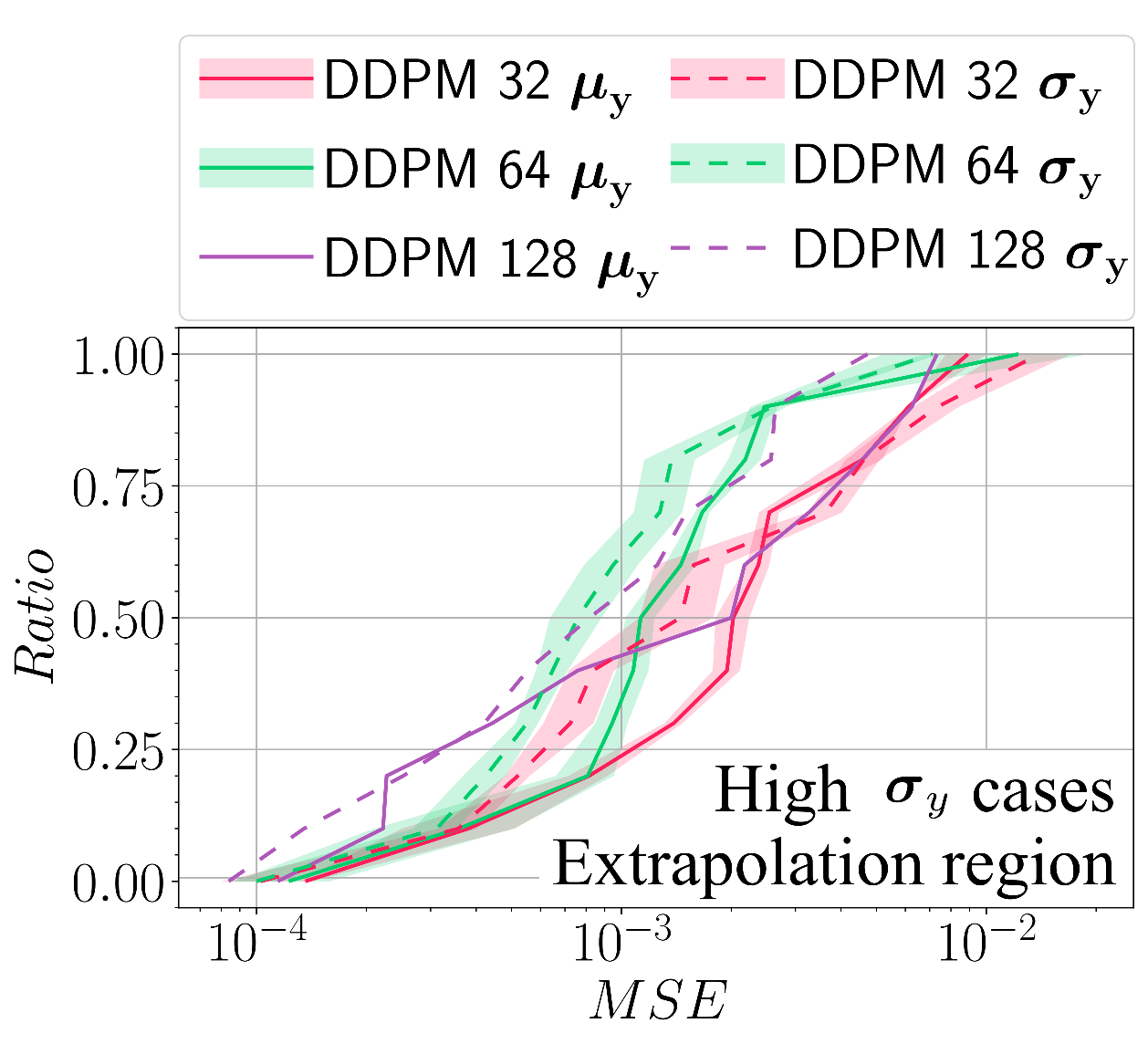}\label{fig:error_distribution_diffusion_64_highstd_out}}
    \caption{Prediction error distributions of  DDPM  with different data resolutions. a), b): interpolation region; c), d): extrapolation region. a), c): low $\boldsymbol{\sigma}_y$ cases ; b), d): high $\boldsymbol{\sigma}_y$ cases.}
    \label{fig:error_distribution_diffusion_64}
\end{figure}

\begin{figure}[tbh]
	\centering
	\sidesubfloat[]{\includegraphics[scale=0.33]{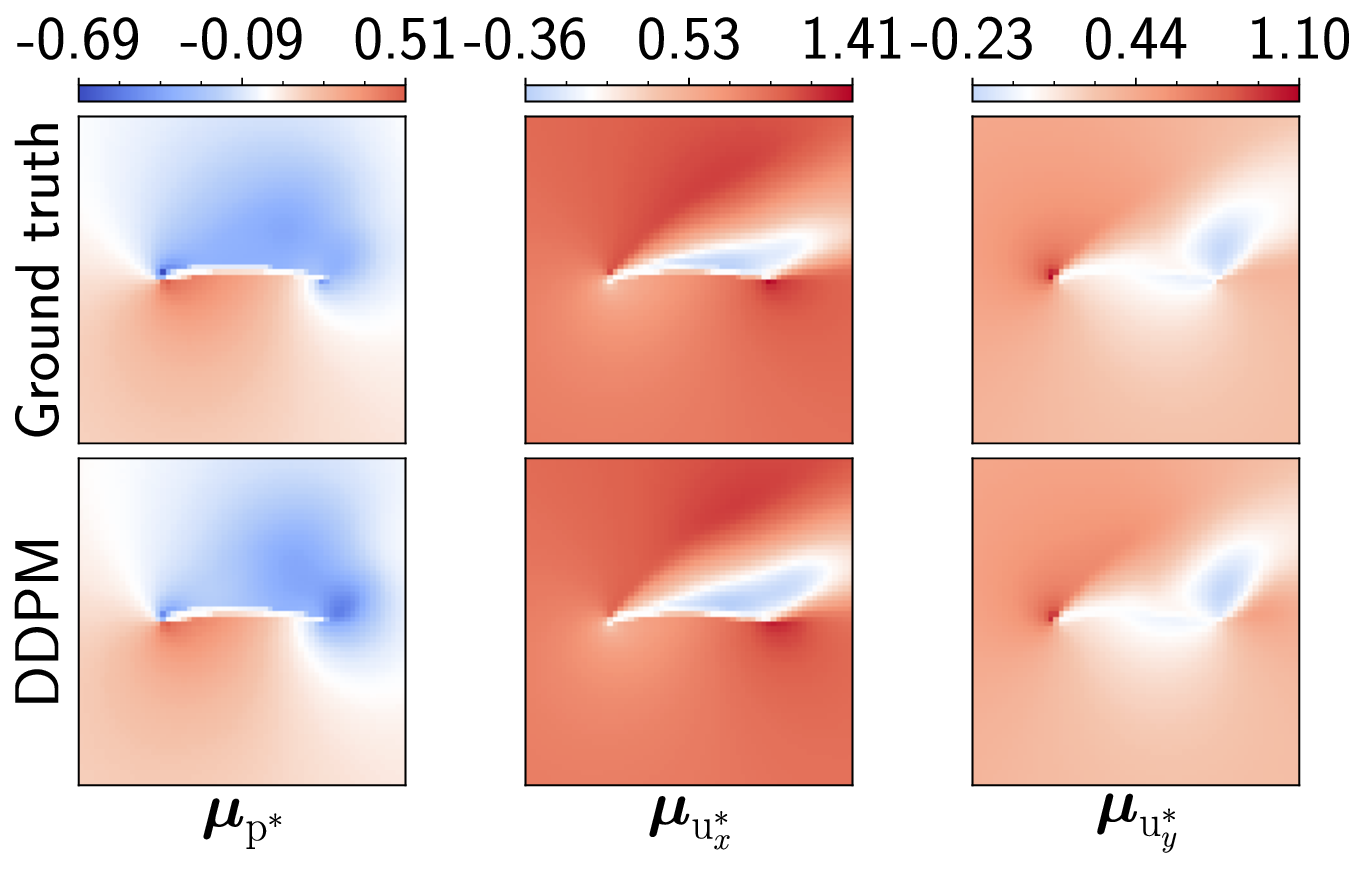}\label{fig:highesterror_mean_in_64}}
	\sidesubfloat[]{\includegraphics[scale=0.33]{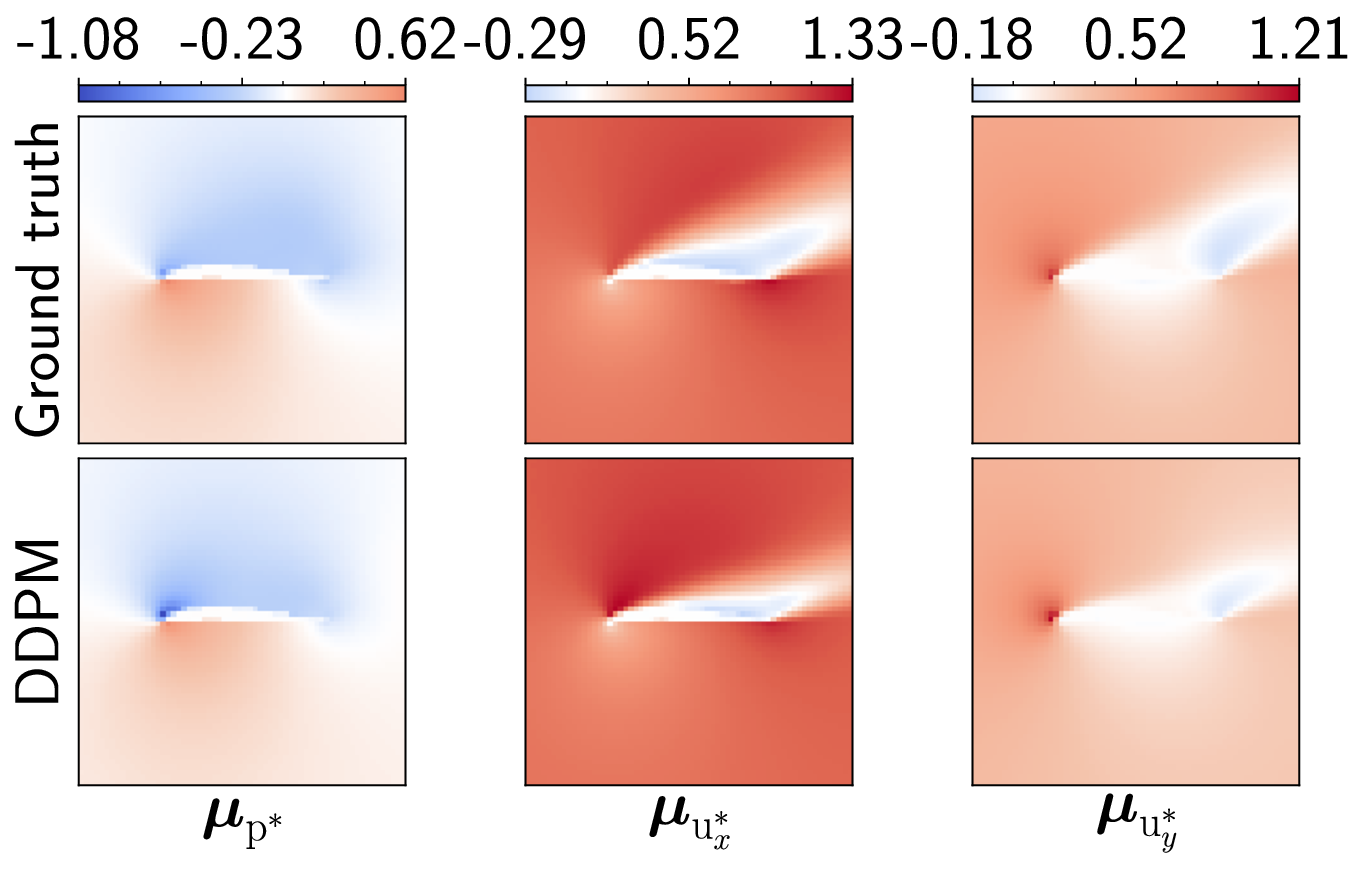}\label{fig:highesterror_mean_64_out}}
	\caption{DDPM's expectation distribution prediction with the largest error in $64\times64$ data. a) Interpolation region (bw3 airfoil, $Re=4.305\times 10^6$, $\alpha=22.08^\circ$). b) Extrapolation region (m17 airfoil, $Re=6.02\times 10^5$, $\alpha=24.52^\circ$).}
	\label{fig:highesterror_64_mean}
\end{figure}

\begin{figure}[tbh]
	\centering
	\sidesubfloat[]{\includegraphics[scale=0.33]{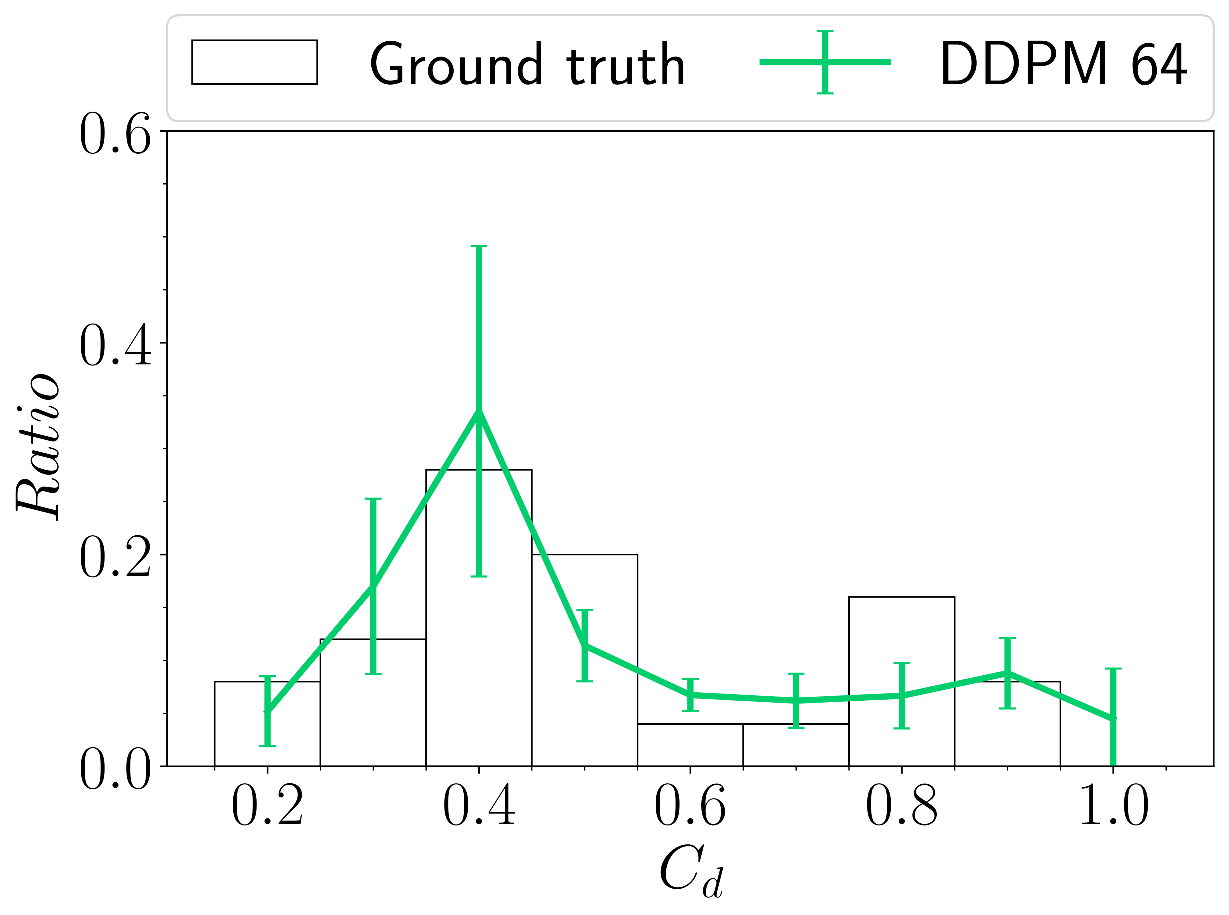}\label{fig:highesterror_std_64_in}}
	\sidesubfloat[]{\includegraphics[scale=0.33]{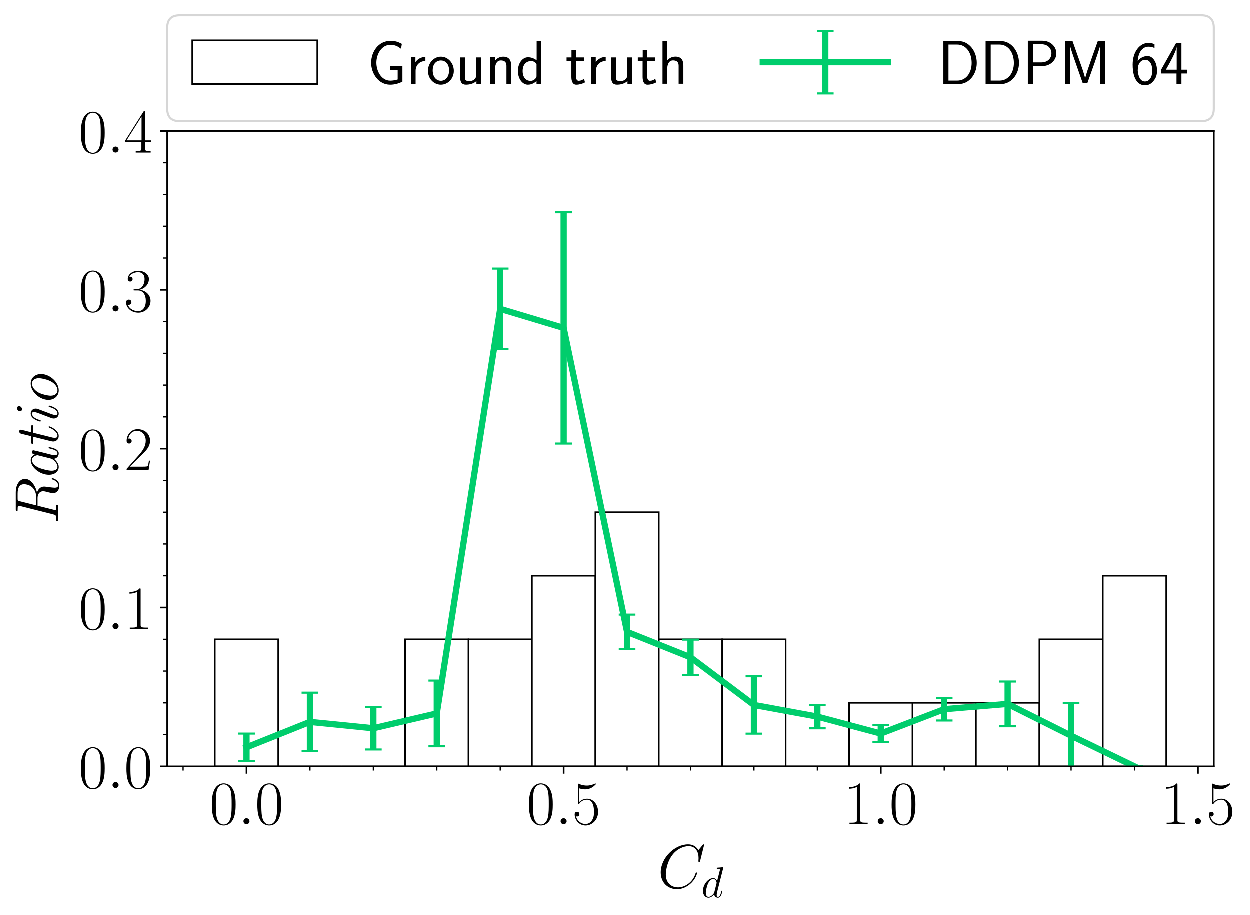}\label{fig:highesterror_std_64_out}}
	\caption{DDPM's drag coefficient distribution prediction with the largest error in $64\times64$ data. a) Interpolation region(bw3 airfoil, $Re=4.305\times 10^6$, $\alpha=22.08^\circ$). b) Extrapolation region (goe566 airfoil, $Re=8.893\times 10^6$, $\alpha=24.54^\circ$).}
	\label{fig:highesterror_64_std}
\end{figure}

\paragraph{Predictions with enlarged resolutions.\label{sec:experiment:large_resolution}}
Scaling the predictions to high resolutions is a crucial aspect of all practical applications of a learning algorithm.
To evaluate the capabilities of DDPM to scale to larger resolutions, we extend the depth and width of networks and retrain them on datasets with higher resolution. Fig.~\ref{fig:error_distribution_diffusion_64} compares DDPM's predictions with a resolution of $32 \times 32$ and $64 \times 64$, showing that the performance of the $64 \times 64$ predictions is in line with the accuracy of the $32 \times 32$ predictions despite the larger number of degrees of freedom. To illustrate the behavior of DDPM with high-resolution data in more detail, we show two cases with the highest prediction error for the $64\times64$ data in Fig.~\ref{fig:highesterror_64_mean} and Fig.~\ref{fig:highesterror_64_std}. \added{The reference drag coefficient distribution in Fig.~\ref{fig:highesterror_64_std} is also obtained from simulation data with the resolution of $64\times64$.} While the DDPM prediction of the sample from the interpolation region gives a lower pressure field at the rear of the airfoil and a larger reflux region in the $\mathrm{u_x}$ field, the predominant flow patterns are nonetheless well captured, as illustrated in Fig.~\ref{fig:highesterror_mean_in_64}.
Besides, Fig.~\ref{fig:highesterror_std_64_in} shows that flow patterns corresponding to $C_d$=0.5, and 0.8 are under-represented while others such as $C_d$=0.2 are over-represented in the predicted distribution. In the extrapolation region, the predicted pressure in the maximum expectation error case is lower at the front edge of the airfoil, and the reflux region is smaller both in the $\mathrm{u_x}$ and the $\mathrm{u_y}$ field as shown in Fig.~\ref{fig:highesterror_mean_64_out}. For the uncertainty distribution, a larger error is visible in Fig.~\ref{fig:highesterror_std_64_out} compared with the previous interpolation case. The flow patterns corresponding to $C_d$=0.4 and 0.5 are over-represented. Nonetheless, 
the range of distributions of the drag coefficient is generally well captured by the large majority of the samples produced by the DDPM.

As an outlook, we have additionally trained and evaluated a network that infers outputs with the resolution of $128\times128$ using a training dataset with a corresponding resolution. The accuracy of this network is analyzed in Fig.~\ref{fig:error_distribution_diffusion_64}, where its performance is on-par with the previously analyzed networks. A full set of DDPM predictions evaluated on the whole test dataset with the output resolution of $128\times 128$ is also available in Fig.~\ref{fig:test_all_mean} and Fig.~\ref{fig:test_all_std} of Appendix~\ref{sec:app:testset}. Fig.~\ref{fig:high_resolution} qualitatively compares predictions with different resolutions showing a zoomed region near the rear of the airfoil. The flow details in the high-resolution field are faithfully reconstructed by the $128\times128$ predictions, giving a noticeable gain in sharpness and generated flow structures. In Fig.~\ref{fig:ux_distribution}, we present a quantitative comparison between the predictions of DDPM and the ground truth, focusing on the velocity distribution in the freestream direction. This distribution assesses the development of flow separation. The results illustrate that DDPM adeptly captures both the flow separation and the associated uncertainty observed in the RANS simulation. Furthermore, Fig.~\ref{fig:ux_distribution} depicts two separation bubbles near the airfoil with reattachment occurring at the trailing edge. The obtained uncertainty is high at the center of the separation bubble, suggesting the shedding of the vortex in the simulation snapshots. 

To show the flexibility of the posterior sampling enabled by DDPM, we analyze a specific airfoil case in terms of a Proper Orthogonal Decomposition (POD) as a popular representative of tools for flow analysis~\cite{BerkoozPOD1993}. We perform POD on both the ground truth and a set of samples inferred by the pre-trained DDPM model for airfoil kc135d, $Re=5.702\times 10^6$, and $\alpha=21.49^\circ$ to investigate the potential for vortex shedding. Fig.~\ref{fig:pod} displays the first three modes along with their corresponding energy fractions. While the first mode, typically associated with the mean of the snapshots, dominates all modes with an energy fraction around 90\%, the second and third modes distinctly reflect the vortex shedding pattern present in the simulation. Importantly, the predictions of DDPM align well with the ground truth in the POD results, underscoring the position of DDPM as the only method capable of generating physical samples for POD analysis. 

These experiments demonstrate that DDPM networks, like regular neural networks, can be scaled up to produce more detailed outputs. At the same time, DDPM retains its capabilities to produce accurate samples from the distribution of solutions at enlarged resolutions.

The resolution of $128\times128$ additionally allows us to directly compare to the results from previous networks trained for the benchmark setup used above~\cite{thuerey2020}.
We use a pre-trained $128\times128$ neural network with 30.9M parameters, denoted as \textit{DFP} model in the following (see Appendix~\ref{sec:app:training} for details).
It is deterministic and trained with a dataset in a supervised manner assuming $\boldsymbol{\sigma}_y=0$. Thus, only the predictions on low uncertainty cases are meaningful for inference with the DFP model. 
Despite being 1.5 times larger than the DDPM model, the DFP model gives significantly lower accuracy. Over 40$\%$ of the DDPM predictions have lower errors than the best prediction error of the DFP model. This holds for the interpolation as well as the extrapolation region, as shown in Fig.~\ref{fig:error_distribution_diffusion_128_DFP_in} and Fig.~\ref{fig:error_distribution_diffusion_128_DFP_out}.
The supervised training of the DFP model forces it to learn averaged solutions for ambiguous inputs, which invariably lowers the quality of the inferred solutions for cases with high uncertainty. 
However, the evaluation above shows that the DDPM model still has an advantage for cases with low uncertainty, for which the DFP model could theoretically have learned a similarly accurate solution.

\begin{figure}[tbh]
    \centering
   	\sidesubfloat[]{\includegraphics[scale=0.33]{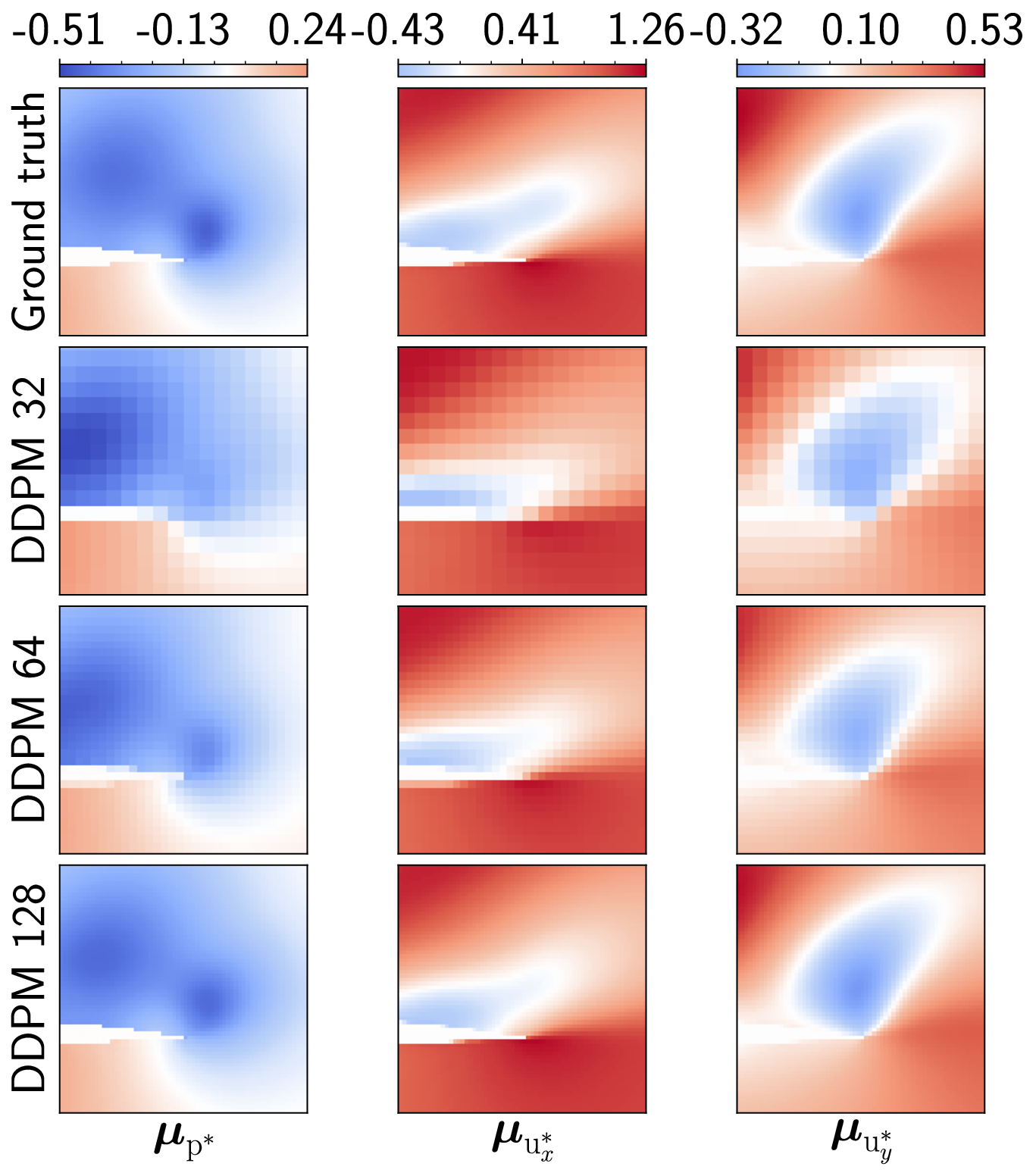}\label{fig:high_resolution_mean}}
    \sidesubfloat[]{\includegraphics[scale=0.33]{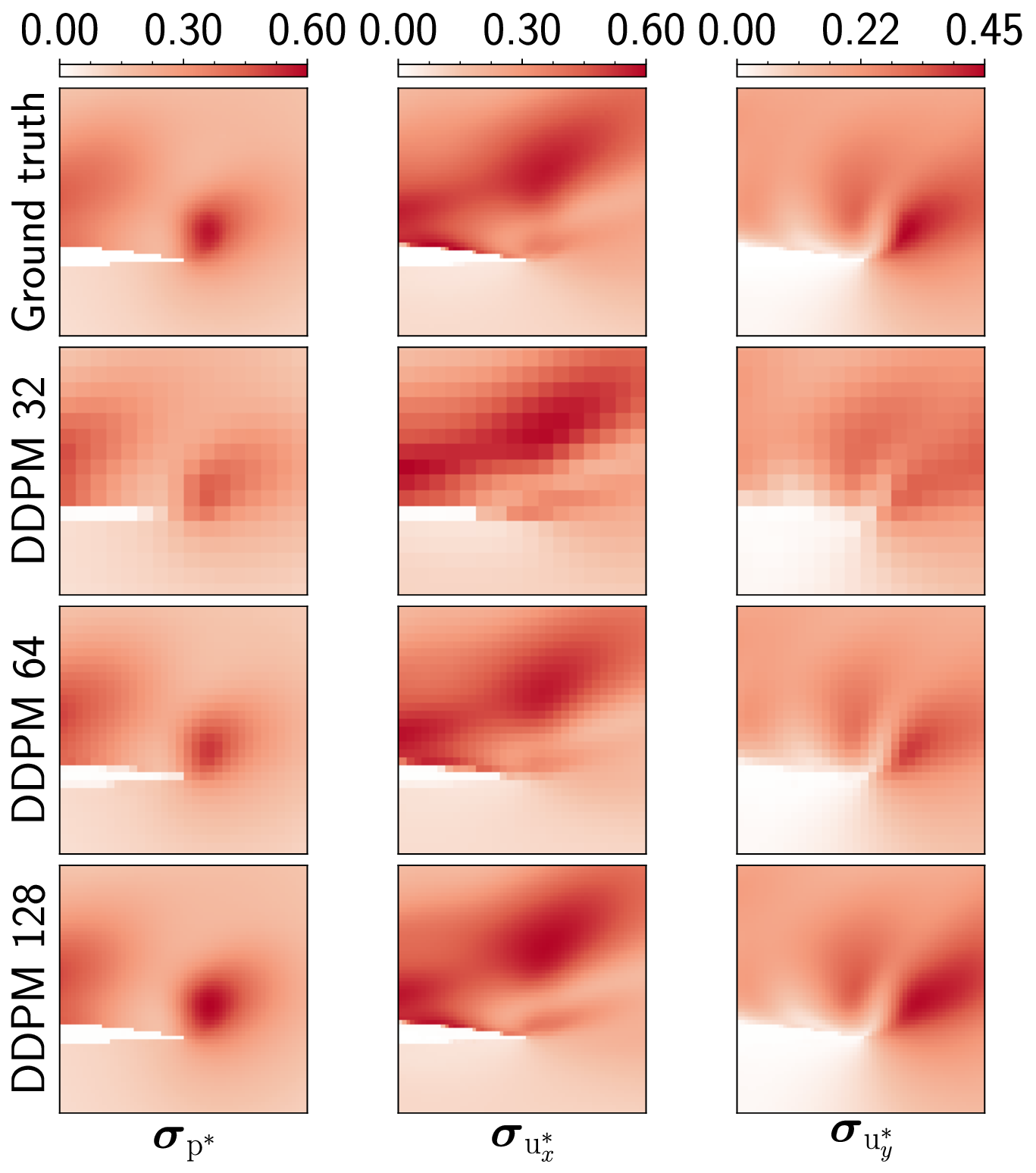}\label{fig:high_resolution_std}}
    \caption{Predictions of the DDPM with varying resolutions in terms of expectation a) and standard deviation b) (kc135d airfoil, $Re=5.702\times 10^6$, $\alpha=21.49^\circ$).}
    \label{fig:high_resolution}
\end{figure}

\begin{figure}[tbh]
    \centering
    \includegraphics[scale=0.33]{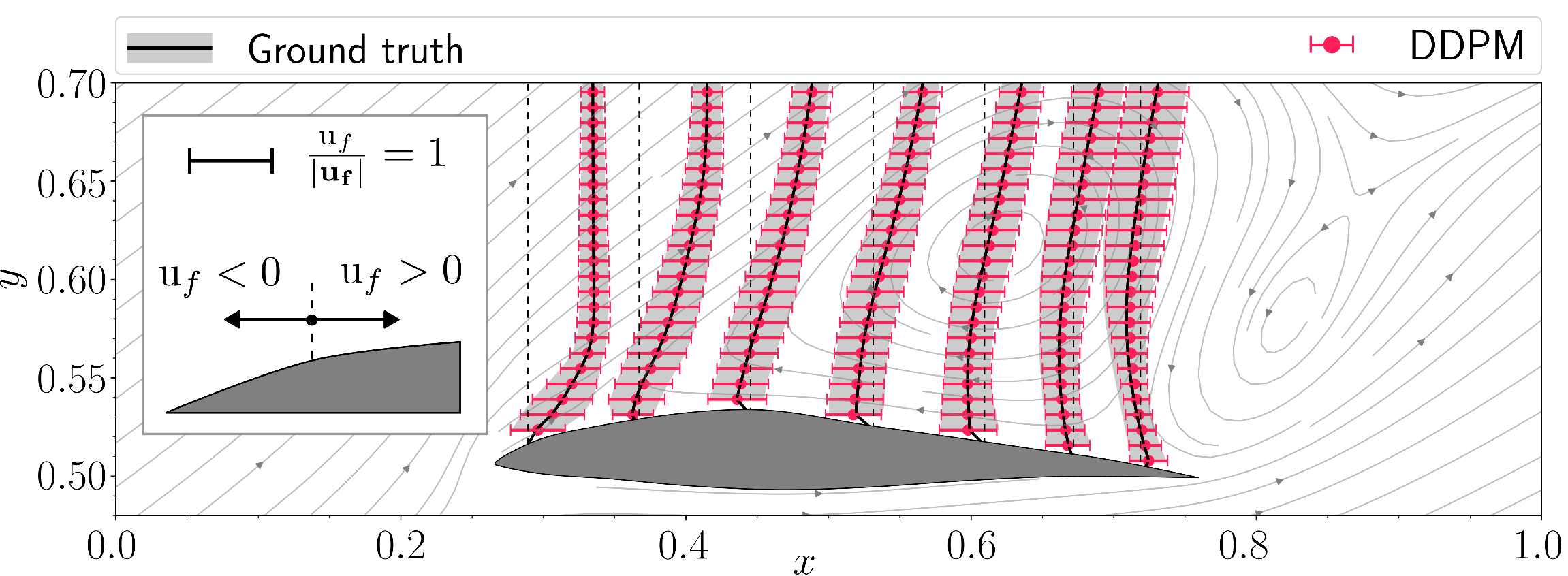}
    \caption{The distribution of the velocity component in the freestream direction with corresponding uncertainty (kc135d airfoil, $Re=5.702\times 10^6$, $\alpha=21.49^\circ$).
    }\label{fig:ux_distribution}
\end{figure}

\begin{figure}[tbh]
    \centering
    \includegraphics[scale=0.33]{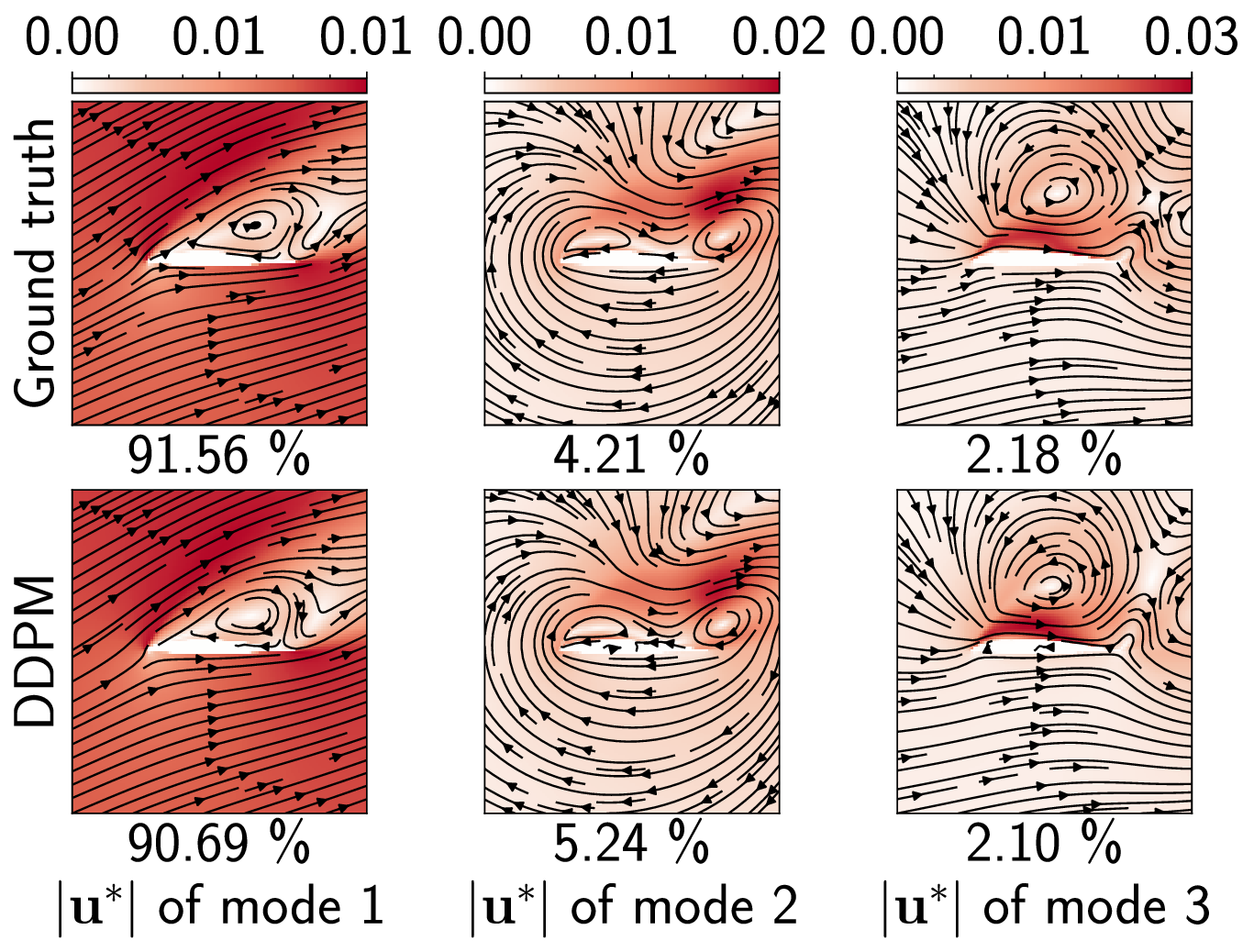}
    \caption{The first 3 POD modes of the ground truth and DDPM predictions with corresponding energy fractions. (kc135d airfoil, $Re=5.702\times 10^6$, $\alpha=21.49^\circ$).
    }\label{fig:pod}
\end{figure}

\begin{figure}[tbh]
    \centering
    \sidesubfloat[]{\includegraphics[scale=0.33]{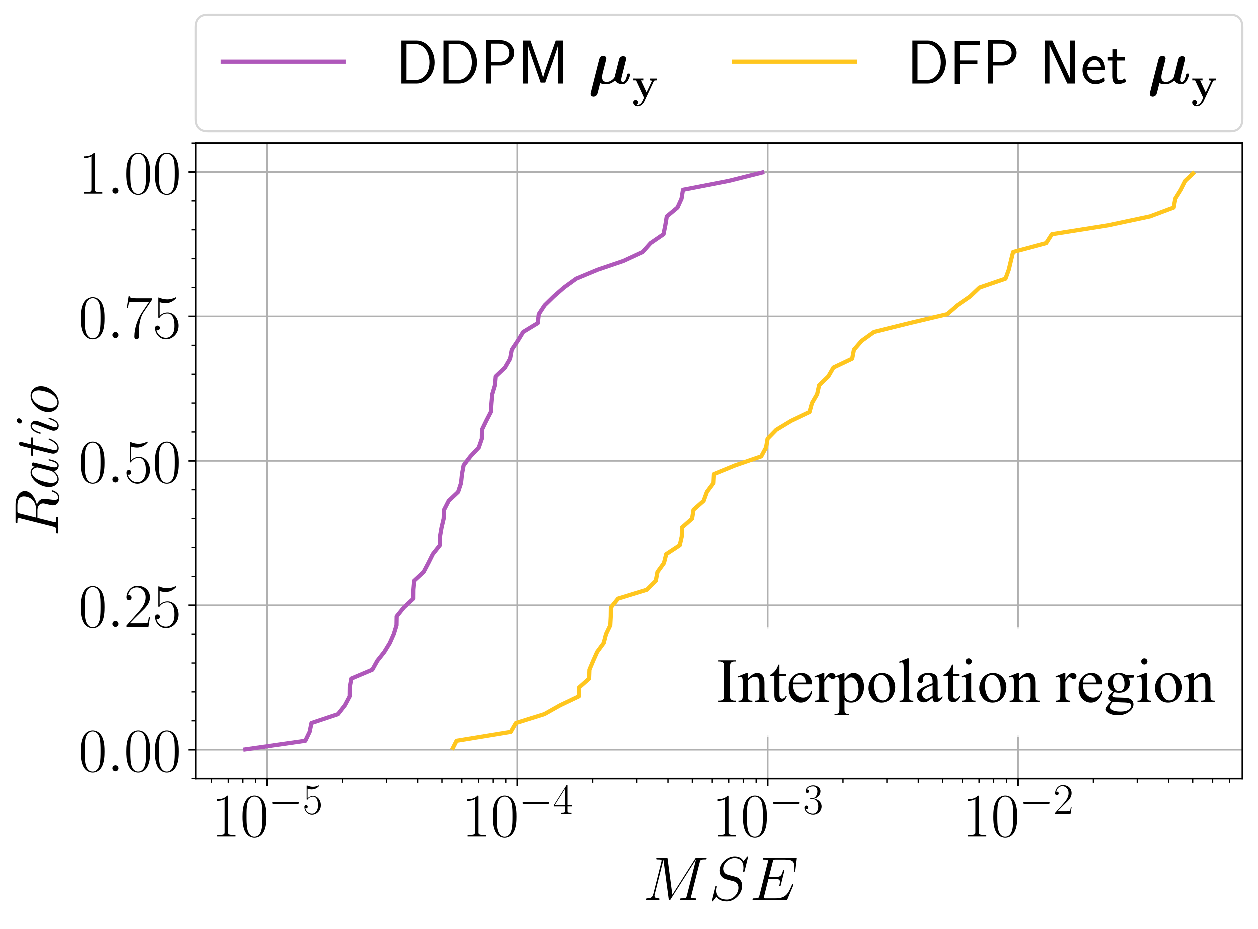}\label{fig:error_distribution_diffusion_128_DFP_in}}
    \sidesubfloat[]{\includegraphics[scale=0.33]{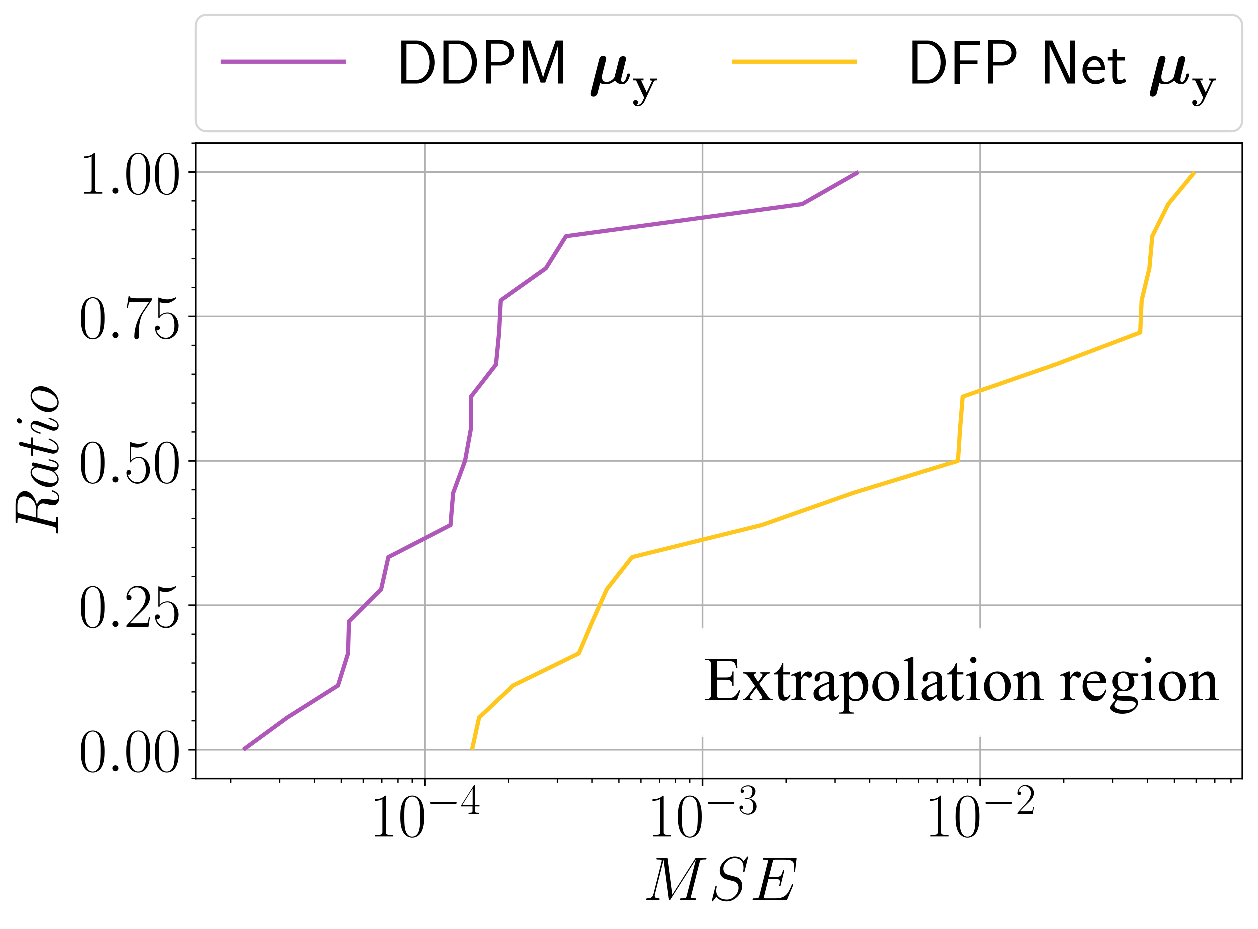}\label{fig:error_distribution_diffusion_128_DFP_out}}
    \caption{Prediction error distributions of DDPM and DFP model~\cite{thuerey2020} on low $\boldsymbol{\sigma}_y$ cases with the resolution of $128\times 128$. a) Interpolation region.  b) Extrapolation region.}
    \label{fig:error_distribution_diffusion_128_DFP}
\end{figure}

\paragraph{Acceleration.}

As the DDPM approach incurs an enlarged computational cost due to its iterative nature, it is important to evaluate whether the trained models retain an advantage over regular simulations in terms of resources required for producing an output.
Table.~\ref{tab:acceleration} summarizes the inference times of the diffusion models with different resolutions on both GPU and CPU. As a reference, we compare with the OpenFOAM simulations that were used to compute the training dataset samples.
Our measurements show that the DDPM can provide acceleration by a factor of 4.5 or 25 to generate a $128 \times 128$ sample using CPU or GPU, respectively. The GPU support offers the potential to increase the sample resolution without strongly impacting the runtime: generating 10 samples with $128 \times 128$ resolution only results in a 3.2 times longer runtime, while generating 10 $64\times64$ samples requires only 1.3 times longer than a single sample. On the other hand, a significant number of samples is required in practice to obtain stable statistics for a given input condition and airfoil (typically $N=25$ simulation samples are used in the present study). However, the DDPM runtime to generate 25 $128\times 128$ samples on a GPU is still less than half of the original simulation runtime.

\begin{table}[]
\caption{The inference time (s) of DDPM on the Intel® Core™ i9-11900K CPU and NVIDIA GeForce RTX 3060 GPU. Cases where DDPM outperforms the simulation are shown bolded.}
\label{tab:acceleration}
\begin{tabular}{lccccc}
\hline
\multirow{2}{*}{Device} & \multirow{2}{*}{Batch size} & \multicolumn{3}{c}{DDPM}                                                          & Simulation                            \\ \cline{3-6} 
                        &                             & $32\times32$              & $64\times64$              & $128\times128$            & 30.032$\pm$1.504 K cells              \\ \hline
\multirow{5}{*}{CPU}    & 1                           & \textbf{1.398$\pm$0.008}  & \textbf{2.439$\pm$0.056}  & \textbf{16.242$\pm$0.218} & \multirow{5}{*}{72.998$\pm$4.786}     \\
                        & 5                           & \textbf{2.836$\pm$0.027}  & \textbf{6.766$\pm$0.135}  & 87.822$\pm$0.618          &                                       \\
                        & 10                          & \textbf{4.418$\pm$0.038}  & \textbf{12.950$\pm$0.073} & 187.751$\pm$0.742         &                                       \\
                        & 25                          & \textbf{10.150$\pm$0.019} & \textbf{33.821$\pm$0.201} & 494.848$\pm$8.174         &                                       \\
                        & 50                          & \textbf{23.109$\pm$0.197} & 83.508$\pm$1.655          & 942.307$\pm$21.951        &                                       \\
\multirow{5}{*}{GPU}    & 1                           & \textbf{1.294$\pm$0.471}  & \textbf{1.488$\pm$0.466}  & \textbf{2.845$\pm$0.494}  & \multicolumn{1}{l}{\multirow{5}{*}{}} \\
                        & 5                           & \textbf{1.086$\pm$0.006}  & \textbf{1.616$\pm$0.012}  & \textbf{9.071$\pm$0.119}  & \multicolumn{1}{l}{}                  \\
                        & 10                          & \textbf{1.154$\pm$0.014}  & \textbf{2.103$\pm$0.002}  & \textbf{17.392$\pm$0.145} & \multicolumn{1}{l}{}                  \\
                        & 25                          & \textbf{1.658$\pm$0.010}  & \textbf{4.543$\pm$0.023}  & \textbf{41.966$\pm$0.256} & \multicolumn{1}{l}{}                  \\
                        & 50                          & \textbf{2.861$\pm$0.005}  & \textbf{8.777$\pm$0.005}  & 81.255$\pm$1.198          & \multicolumn{1}{l}{}                  \\ \hline
\end{tabular}
\end{table}


\paragraph{\textcolor{darkgreen}{ Performance evaluation with coarse configurations $^\star$}}

Although the simulation is slower in evaluating uncertainty compared to diffusion models, it is possible to accelerate the simulation at the cost of reduced accuracy. Hence, while performance comparisons are notoriously difficult, we aim to ensure a fairer comparison in the following: we increase the relative tolerance of the linear solver in the simulation and evaluate the predicted uncertainty value. This makes the simulator  comparable with the diffusion models. Similarly, we only use the CPU for the diffusion model, and assess the performance of the model with reduced / different numbers of samples: their inference speed improves linearly when generating fewer samples. For this comparison, we focus on the one-dimensional test from Section~\ref{sec:single}.

\begin{figure}[tbh]
    \centering
    \sidesubfloat[]{\includegraphics[scale=0.33]{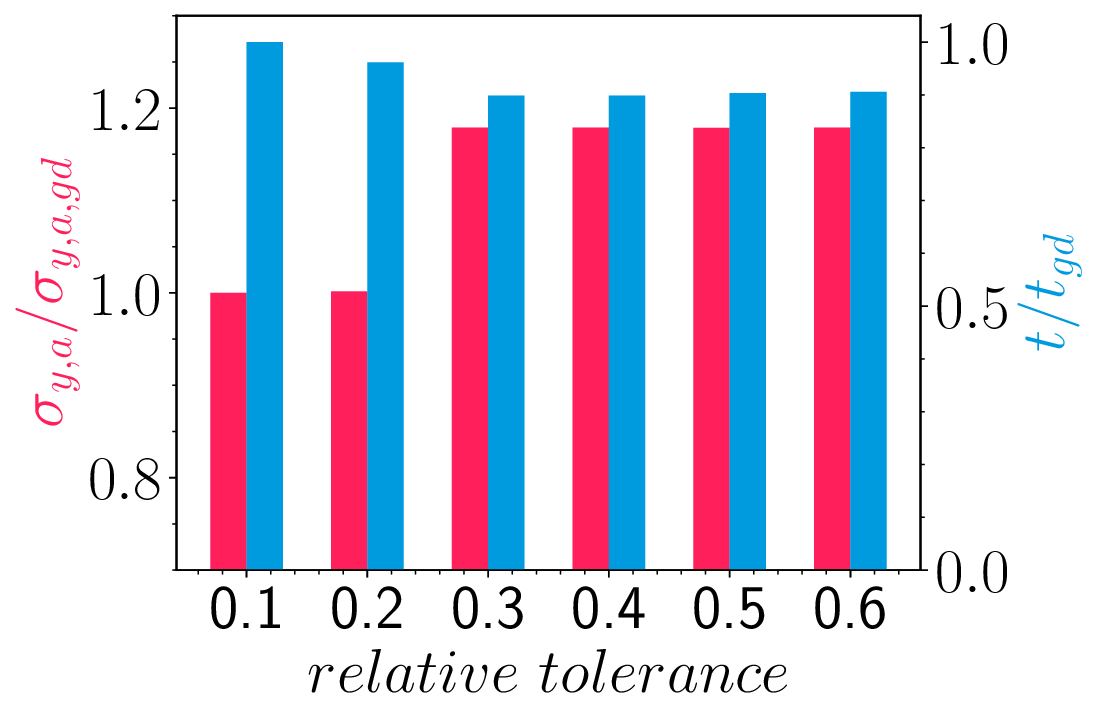}\label{fig:coarse_sim_of}}
    \sidesubfloat[]{\includegraphics[scale=0.33]{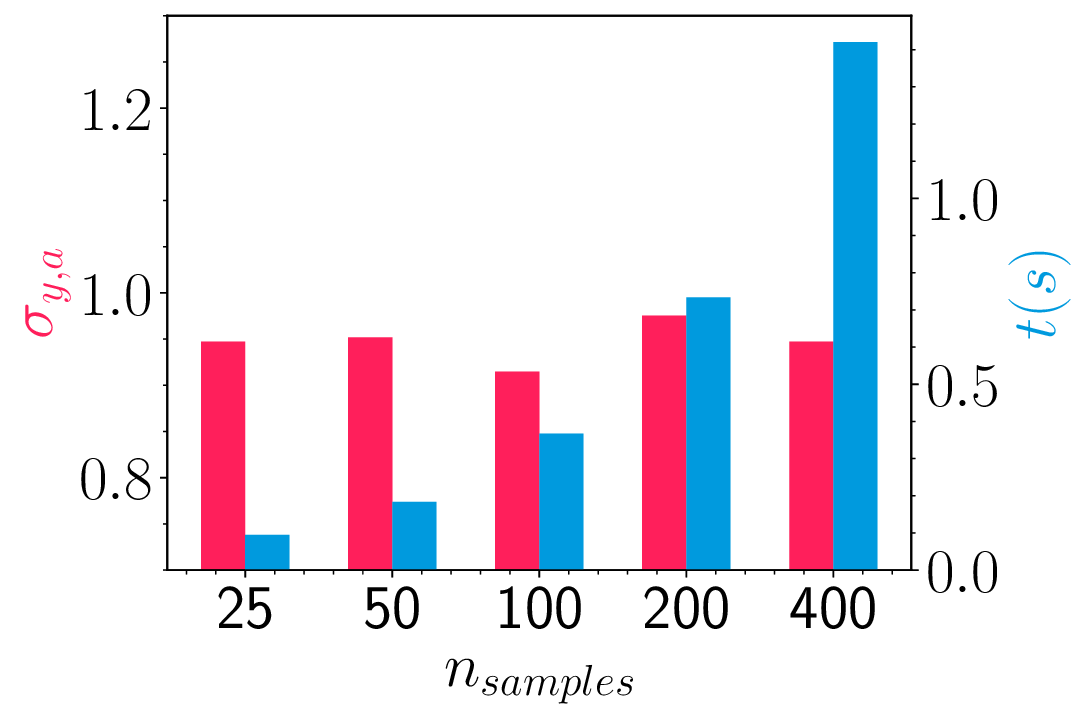}\label{fig:coarse_sim_dif}}
    \caption{The relative simulation/inference time and predicted uncertainty value for a) simulations and b) diffusion models with coarse configurations. (raf30 airfoil, $Re=10.5\times 10^6$, $\alpha=20.00^\circ$). All result were evaluated using CPUs.}
   \label{fig:coarse_sim}
\end{figure}

Increasing the relative tolerance of the simulator improves simulation speed, but only when the tolerance is below 0.3, as shown in Fig. \ref{fig:coarse_sim}. At this threshold, the simulation slightly speeds up. It's duration decreases to 89.8\% compared to the finest configuration (relative tolerance = 0.1) of the reference solve, which results in a relative uncertainty prediction error of 17.90\% . 

In contrast, diffusion models exhibit more significant acceleration with fewer samples, while maintaining relatively stable prediction accuracy. With 5 samples, the relative prediction error is 5.26\%, and the inference time is reduced to 9.4\% of the finest simulation time (both measured on the CPU). 

To conclude, the diffusion model outperforms conventional simulation methods when performing on a similar level of distributional accuracy. Despite being ca. $9.5\times$ faster using the same hardware, the diffusion model still yields a slightly better distribution. This is, clearly outperforms the simulator
both in terms of accuracy and inference speed. Using the GPU version of the diffusion model could potentially yield at least another order of magnitude.

\section{Discussion}

\subsection{The capabilities of BNNs, heteroscedastic models and DDPMs}
The conducted experiments provide an intuitive understanding of the inherent characteristics of the three methods investigated in the present study. BNNs, serving as epistemic uncertainty models, assume a distribution for network parameters during training. The inherent challenge lies in the elusive nature of the true distribution of network parameters, stemming from the difficulty in establishing prior beliefs on parameter space. The assumption of a Gaussian distribution for network parameters is practical but not universally correct, and alternative prior distribution types may be more suitable for specific tasks~\cite{Wenzel2020,fortuin2022bayesian}
In our scenario, despite employing a scaling factor $\lambda$ to adjust reliance on the prior Gaussian distribution of parameters, the predictions of the BNNs still fail to capture the characteristic features of the target distribution. This limitation underscores the complexity of effectively incorporating data uncertainties into BNN predictions. Besides, it's essential to note that the loss function of BNNs, Eq.~\ref{eq:bnn:loss_func}, primarily focuses on maximizing the expectation of the probabilistic distribution in generating the ground truth flow data given the network parameters, i.e., $\mathbb{E}_{q_{\phi}}[\log(p(\mathbf{d}|{\theta}))$. It lacks a guarantee that each sample drawn from $p(\mathbf{d}|{\theta})$ agrees to the ground truth, as indicated in the noise samples in Fig.~\ref{fig:1D_samples}. Another well-known drawback of BNNs is the high training cost, primarily attributed to the requirement of evaluating the distance from the prior distribution across the entire network parameters. In our single-parameter experiment, the BNN achieves a considerably lower training speed using the same training configuration, completing only 1.9 iterations per second. In comparison, both the heteroscedastic model and the DDPM exhibit substantially higher training speeds, completing 7.1 and 6.9 iterations per second, respectively.

The heteroscedastic model relies on the assumption of Gaussian-distributed data to predict the moment of the target distribution. While the Gaussian distribution assumption provides relatively accurate predictions for the mean and standard deviation of the target distribution in the current RANS simulation case, it falls short when applied to the distribution of the drag coefficient. One potential remedy is extending the heteroscedastic model into a mixture density network, utilizing Gaussian mixture distributions to represent intricate distributions, albeit at the cost of increased computational resources. Another drawback of the heteroscedastic model is that it evaluates the moment on each data point independently, resulting in a noisy sampled flow field akin to BNNs' prediction. 

In contrast to the other approaches, DDPMs circumvents a direct modeling of the target distribution by employing a series of transformations from the target distribution to a simple standard Gaussian distribution. The process involves gradually distorting samples from the original distribution with Gaussian noise until a distribution consisting of only standard Gaussians is reached. A neural network is employed to learn the added noise fields. 
Subsequently, starting with samples drawn from the standard Gaussian distribution, DDPMs reconstruct the original target variable step by step with the network-predicted noise.
In contrast to the heteroscedastic model, which relies on an a priori assumptions about the target distribution parameterized with predicted moments, DDPMs deduce the target distribution by iteratively recovering data samples from the noise. This key distinction eliminates the need for assumptions about the original distribution, which is crucial in complex flow scenarios where obtaining the prior distribution is challenging. In our case, the assumption-free DDPMs not only enhance accuracy in predicting distribution moments but also successfully reconstruct the distribution of the drag coefficient. Moreover, the directly generated samples by DDPMs avoid the loss of association among data points within a sample, ensuring noiseless and physically meaningful samples compared to both BNN and heteroscedastic models. The analysis with POD shown above indicates the possibilities that arise from the efficient sampling procedure enabled by DDPM. 

\subsection{Limitations on DDPMs}

The advantages of DDPMs come with associated drawbacks. Firstly, since moments and other static features are derived from a series of samples, the inference procedure of DDPMs must be executed multiple times to generate sufficient samples. While both BNNs and DDPMs require multiple samples to represent the distribution, DDPMs require a more compute-intensive process, typically requiring hundreds of network inference steps to generate a single sample. The estimation of the solution distribution by DDPMs is slower than other models, yielding smaller speed-up factors than reported in previous studies~\cite{Sekar2019,thuerey2020, Xiaosong2021}. Nevertheless, the research community of DDPMs remains highly dynamic, and several recent publications have outlined promising directions to accelerate the sampling process~\cite{meng2022on,song2021denoising}

In this manuscript, the capabilities of DDPM to generalize are assessed by evaluating its performance across datasets of different sizes and testing the accuracy in interpolation/extrapolation regions. Compared with other methods, DDPM's generalization ability exhibits a nuanced profile. On the one hand, it generally excels with various sizes of snapshot samples and simulation cases. On the other hand, it does not demonstrate superiority in extrapolation cases where the range of $Re$ and $\alpha$ differs from the training dataset. Notably, DDPM showcases good generalization ability for new airfoil shapes, as both interpolation and extrapolation regions in the test dataset involve novel airfoil shapes. This may relate to the convolutional UNet's effectiveness in capturing spatial hierarchies rather than magnitude changes in terms of values. Enhancing the generalizability of DDPM poses an interesting and challenging avenue for future work. The challenges stem from the initial application of DDPM in image synthesis, where defining the concept of generalizability is intricate, and relevant research within the DDPM research community is notably scarce~\cite{li2023on}. Moreover, the sampling and training procedures are not solely determined by neural networks; they involve complex mathematical transformations on intermediate distributions of the target variable. These transformations play a crucial role in the generalizability of DDPM. Despite these challenges, the observation in our research can provide valuable insights to improve generalization in the future. While augmenting simulation data, increasing sample sizes, and introducing new airfoil shapes are both beneficial for enhancing the method's generalizability, the range of Reynolds numbers and angles of attack in the training dataset is the primary limiting factor. \replaced{Broadening the range of Reynolds numbers and angles of attack in the dataset is more promising to improve generalization in our context. However, it's also crucial to simultaneously ensure data adequacy in other dimensions. The diversity of airfoil shapes is particularly important, especially in scenarios like aerodynamic shape optimization, where encountering new airfoils beyond the training dataset is common. Therefore, a balanced dataset encompassing diverse parameters, including Reynolds numbers, angles of attack, and airfoil shapes, is essential to comprehensively enhance the model’s capabilities for generalization.}{Rather than focusing on expanding data in other dimensions, broadening the range of Reynolds numbers and angles of attack in the dataset is the most promising direction to improve generalization in our context.}

\begin{table}[]
\caption{The performance comparison between different models}
\label{tab:compare}
\begin{tabular}{cccc}
\hline
                                       & BNNs                             & Heteroscedastic Models      & DDPMs                            \\ \hline
Training cost                          & $\color{myred}\boldsymbol{\uparrow} \boldsymbol{\uparrow}$ & $\color{mygreen}\boldsymbol{\downarrow}$ & $\color{mygreen}\boldsymbol{\downarrow}$      \\
Inference cost                         & $\color{myred} \boldsymbol{\uparrow}$         & $\color{mygreen}\boldsymbol{\downarrow}$ & $\color{myred}\boldsymbol{\uparrow} \boldsymbol{\uparrow}$ \\
Assumption-free                        & $ \color{myred} \usym{2718} $         & $ \color{myred} \usym{2718} $    & $ \color{mygreen}\usym{2714} $         \\
Accurate moment prediction           & $ \color{myred} \usym{2718} $         & $ \color{mygreen}\usym{2714} $    & $ \color{mygreen}\usym{2714} \usym{2714}$    \\
Accurate drag coefficient prediction & $ \color{myred} \usym{2718} $         & $ \color{myred} \usym{2718} $    & $ \color{mygreen}\usym{2714} $         \\
Physical samples                       & $ \color{myred} \usym{2718}  $         & $ \color{myred} \usym{2718} $    & $ \color{mygreen}\usym{2714} $         \\ \hline
\end{tabular}
\end{table}

As a summary, Table.~\ref{tab:compare} shows a comparison of several key aspects between different models. In light of the current status of the respective algorithms, practitioners are faced with a decision: if sampling from the distribution of solutions is essential for an application, DDPM emerges as a fitting choice, offering superior accuracy and complete information on the distribution. On the other hand, if the requirement is limited to moment distributions, alternative models, particularly the heteroscedastic model, provide a more expedient process for estimation.

\section{Conclusions}
Focusing on the inherent uncertainty of RANS simulations, the present study provides a first evaluation of denoising diffusion probabilistic models to train uncertainty-aware surrogate models that provide a complete and accurate distribution of solutions. Our detailed evaluation shows that DDPMs faithfully reconstruct the complex distributions of solutions of the RANS dataset and generate meaningful individual samples from the distribution of solutions. The distribution of drag coefficients in the flow fields predicted by DDPMs also matches the ground truth very well. While the heteroscedastic models can estimate the expectation and the standard deviation of the target distribution, DDPMs nonetheless show a very substantial gain in inference accuracy. The BNN predictions, on the other hand, require a manual adjustment of the training hyperparameters to match the ground truth distributions. Both methods additionally show a significantly lower quality in terms of their samples from the distribution of solutions compared to DDPM.

To ensure reproducibility, the source code and datasets of the present study are published at \url{https://github.com/tum-pbs/Diffusion-based-Flow-Prediction}. The quantification of the accuracy of the learned distribution of solutions is of general importance for DDPMs. As this training dataset is the first non-trivial and high-dimensional case that provides ground truth for the distribution of solutions, we expect that this dataset will have merit beyond applications in aerospace engineering. 

By providing uncertainty-aware and accurate predictions, DDPM-based surrogate models have the potential to serve as a compelling building block for diverse applications, e.g., to accelerate iterative designs~\cite{Sekar2019id, Li2020,chen2021}. Investigating the uncertainty of a scenario and accessing multiple possible solutions is typically a more favorable workflow than obtaining and working with a single prediction. This is especially important in the aerospace research community, where safety and reliability are of paramount importance~\cite{Cook2017, Huyse2002}. 
Meanwhile, exploring the application of DDPM in turbulence modeling is a highly interesting topic for future work. While the large-scale flow captured in RANS/LES simulations is generally deterministic, the unresolved flow details manifest themselves as probabilistic distributions rather than deterministic solutions, owing to their stochastic nature~\cite{Pope_2000} 
Employing the expectation of probabilistic small-scale flows as an estimate for the time-averaged or spatially-filtered large-scale flow was shown to be a very suitable approach for turbulence modeling~\cite{LANGFORD1999,Anudhyan2023}. 
In this context, DDPM could provide a powerful tool to capture the distribution of unresolved small-scale flows in order to capture their effect on larger scales. Its capabilities to capture complex distributions and flexible conditioning position DDPM as a very promising technique to advance turbulence modeling.

\section{\textcolor{darkgreen}{$^\star$Flow Matching: A Competitive Alternative for Diffusion Models}}

Recently, a new generative modeling variant, \textit{flow matching}~\cite{lipman2023flow}, has gained significant attention and emerged as a strong competitor to diffusion models in various fields~\cite{dao2023lfm,liu2024generative,fischer2023boosting,holzschuh2024}. Instead of denoising as central task, flow matching considers a time-dependent differentiable function $\phi:[0,1]\times \mathbb{R}^d \rightarrow \mathbb{R}^d$ which maps samples $x_0 \in \mathbb{R}^d$ from distribution $p_0$ to $x_t$, where $x_t=\phi(x_0)$. This mapping function is defined as the flow of the corresponding transformation. Flow matching learns the time derivative of this flow as a time-dependent vector field $u:[0,1]\times \mathbb{R}^d \rightarrow \mathbb{R}^d$, where  $u_t(\phi_t(x_0))= \frac{d}{dt}\phi_t(x_0)$, or equivalently, $u_t(x_t)=\frac{d}{dt}x_t$. The loss function is then defined as
\begin{equation}
    \mathcal{L}_{\text{FM} }(\theta) = \mathbb{E}_{t\sim {\mathcal{U}}[0,1],x_t \sim p_t}\| v_{\theta,t}(x_t)-u_t(x_t) \|^2
    .
    \label{eq:fm_margin_loss}
\end{equation}
However, this loss function is intractable since we lack information about $p_t$ and $u_t$. In real applications, $x_0$ typically represents samples from a simple distribution, such as a Gaussian distribution, similar to what is used in diffusion models. Meanwhile, $x_1$ corresponds to samples from the target distribution, i.e., the training dataset. This allows for the construction of a conditional vector field $u_t(x_t\vert x_1) $ based on known samples $x_1$. Consequently, the intermediate probability density and vector field can be marginalized as follows~\cite{lipman2023flow}
\begin{equation}
    p_t=\int p(x_t|x_1)q(x_1)dx_1,
\end{equation}
and
\begin{equation}
u_t(x_t) 
    = \int u_t(x_t\vert x_1) \frac{p(x_t\vert x_1)q(x_1)}{p(x_t)}dx_1.
\end{equation}
With this marginalization, it was demonstrated~\cite{lipman2023flow} that learning this conditional flow is mathematically equivalent to learning the original flow:
\begin{equation}
    \mathcal{L}_{\text{CFM}}(\theta) = \mathbb{E}_{t\sim \mathcal{U}[0,1],x_1\sim q(x_1),x_t \sim p(x_t\vert x_1)} \big\| v_{\theta,t}(x_t) - u_t(x_t\vert x_1) \big\|^2,
\end{equation}
\begin{equation}
     \nabla_\theta \mathcal{L}_{\text{FM}}(\theta) = \nabla_\theta \mathcal{L}_{\text{CFM}}(\theta).
\end{equation}

There are many possible designs for the conditional flow, and one of the simplest and most effective approaches is the optimal transport conditional flow, which defines a linear mapping between samples from $p_0$ and $p_1$:
\begin{equation}
    \psi_t(x_0) = \sigma_t(x_1)x_0 + \mu_t(x_1),
\end{equation}
\begin{equation}
    u_t(x_t\vert x_1)=\frac{d}{dt}\psi_t(x_0)=\sigma_t'(x_1)x_0 + \mu_t'(x_1),
\end{equation}
where $\mu_t(x_1)=tx_1$ and $\sigma_t(x_1)=1-(1-\sigma_\text{min})t$. Here $\sigma_\text{min}$ is sufficiently small so that $p(x_1 \vert x_1)$ is a concentrated Gaussian distribution centered at $x_1$. This actually provides a "straight" conditional flow with a constant vector field independent of time $t$. 

With the learned vector field, we can generate samples $x_1$ from $x_0$ via integration in time:
\begin{equation}
x_1=\int_0^1 v_{\theta,t}(x_t) dt.
\label{eq:sample_fm}
\end{equation} 
The above equation is a simple ODE that can be efficiently solved using many ODE solvers. Notably, if the marginal vector field remains time-invariant, Eq. \ref{eq:sample_fm} can be solved with very few steps using the Euler method. While a constant conditional vector field, as provided by the optimal transport flow, does not guarantee a constant marginal vector field, it is reasonable to expect the marginal vector field to remain relatively simple~\cite{lipman2023flow}.

\begin{figure}[tbh]
    \centering
    \includegraphics[scale=0.33]{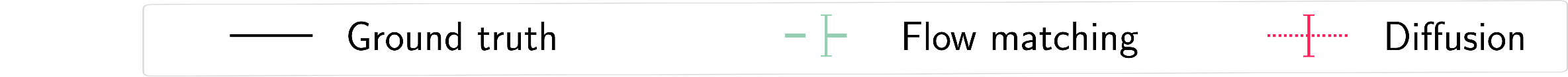}\\
    \sidesubfloat[]{\includegraphics[scale=0.33]{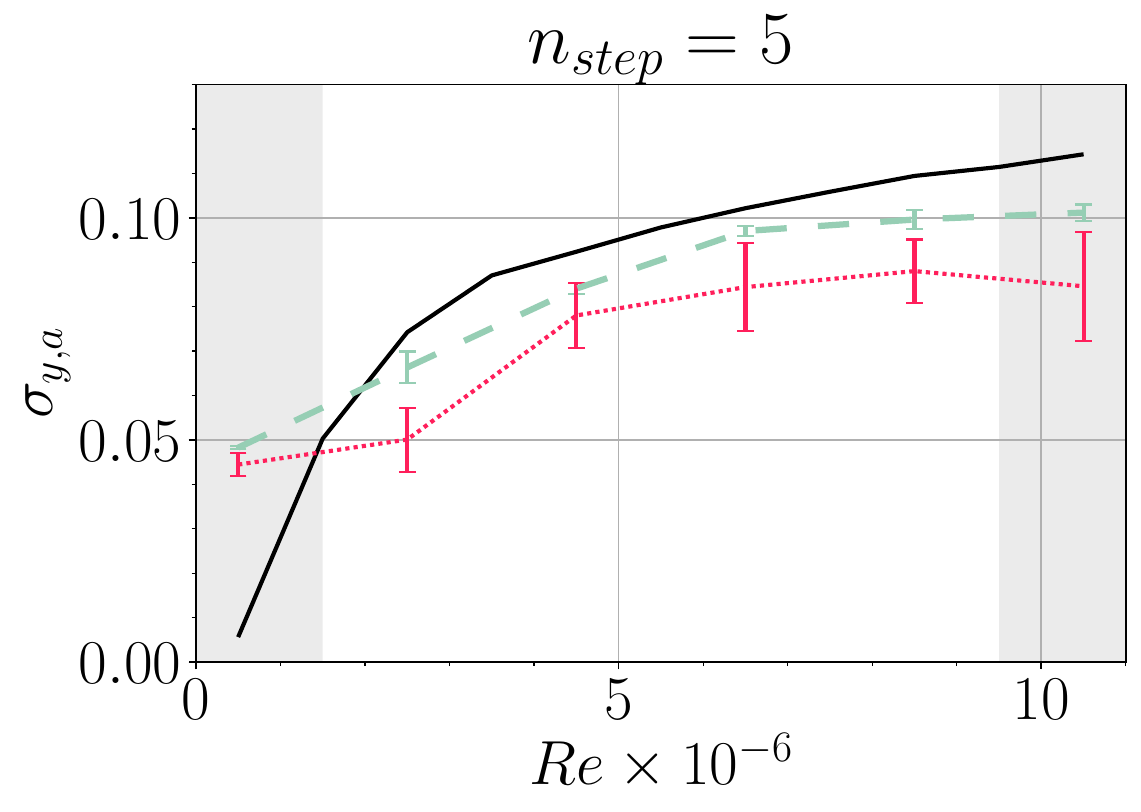}\label{fig:flowmatching_diffusion_5}}
    \sidesubfloat[]{\includegraphics[scale=0.33]{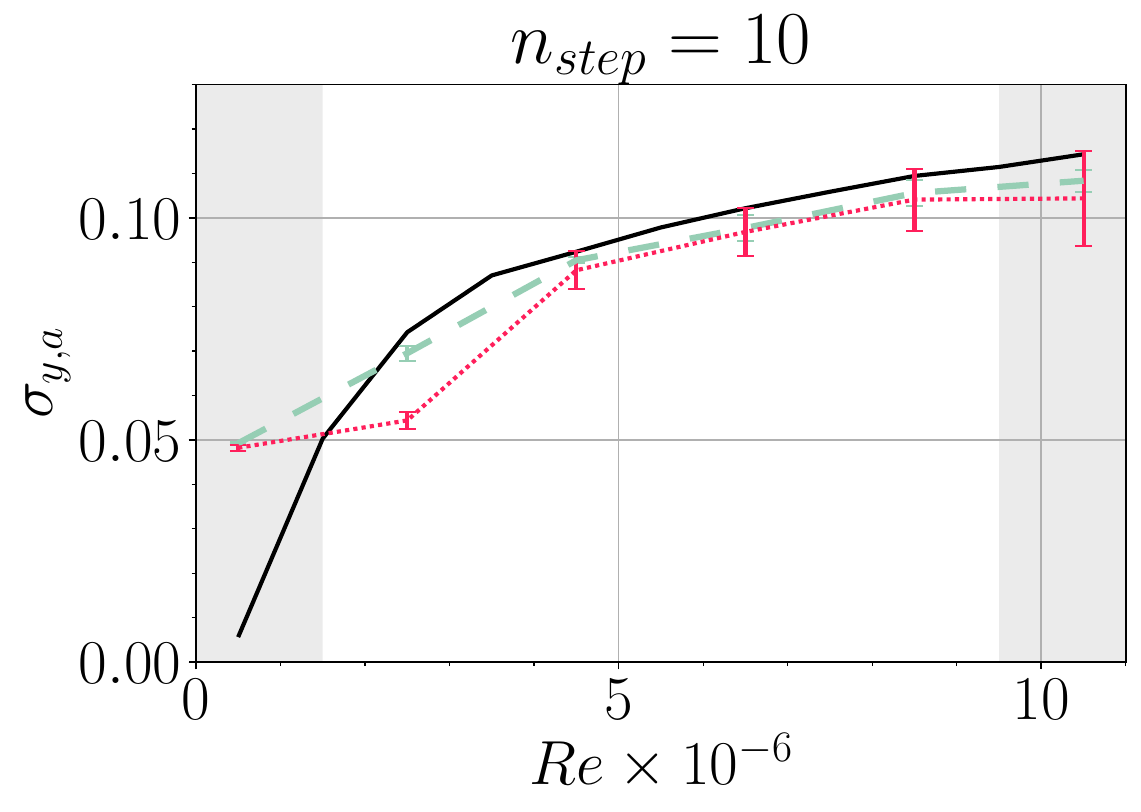}\label{fig:flowmatching_diffusion_10}}\\
    \sidesubfloat[]{\includegraphics[scale=0.33]{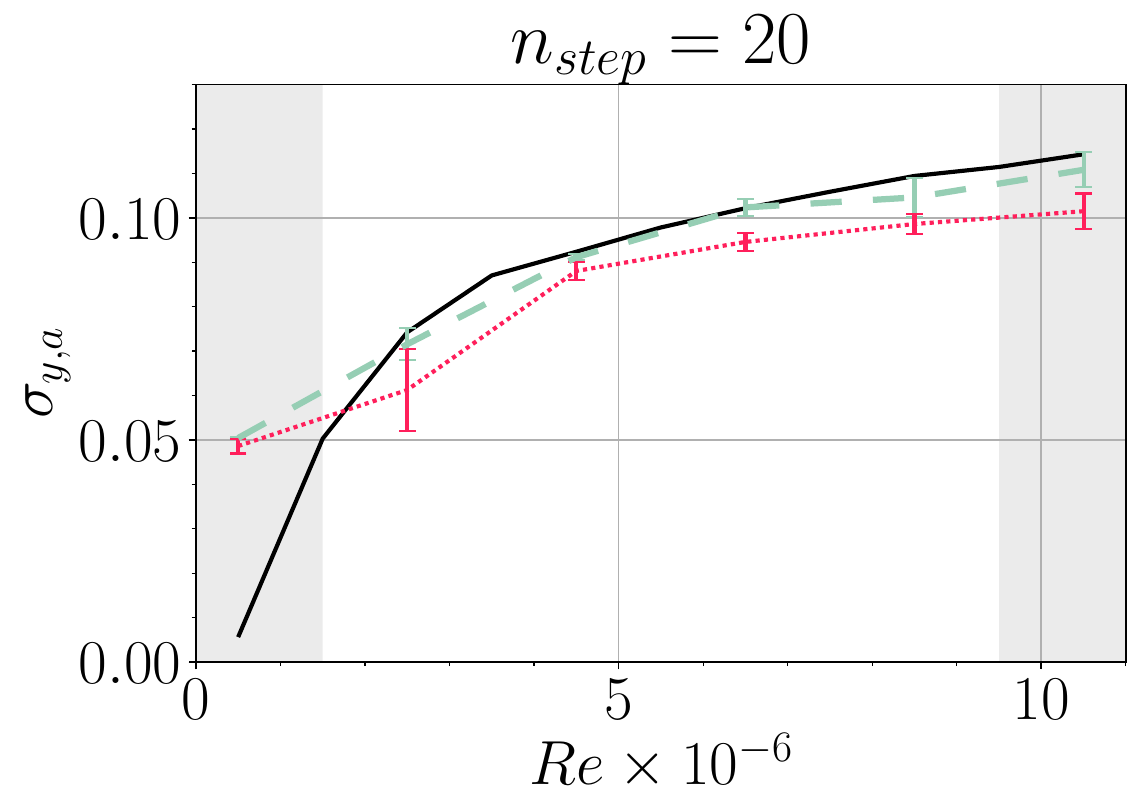}\label{fig:flowmatching_diffusion_20}}
    \sidesubfloat[]{\includegraphics[scale=0.33]{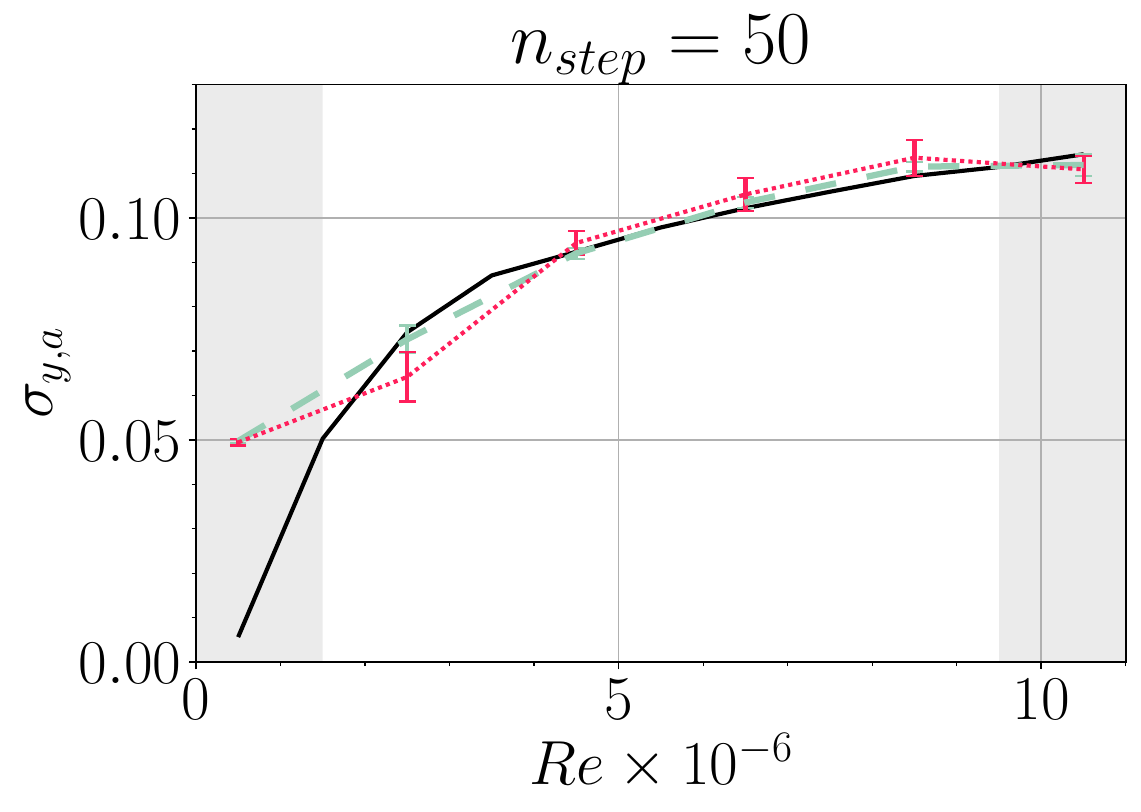}\label{fig:flowmatching_diffusion_50}}\\
    \sidesubfloat[]{\includegraphics[scale=0.33]{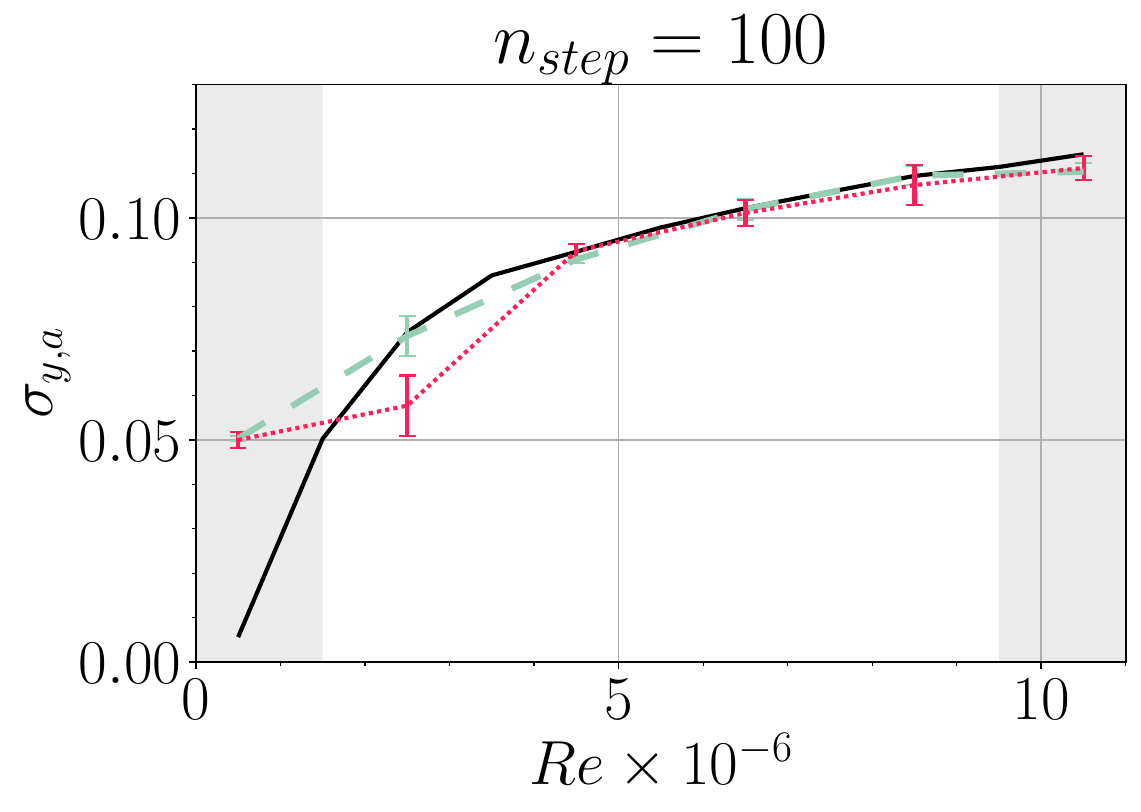}\label{fig:flowmatching_diffusion_100}}
    \sidesubfloat[]{\includegraphics[scale=0.33]{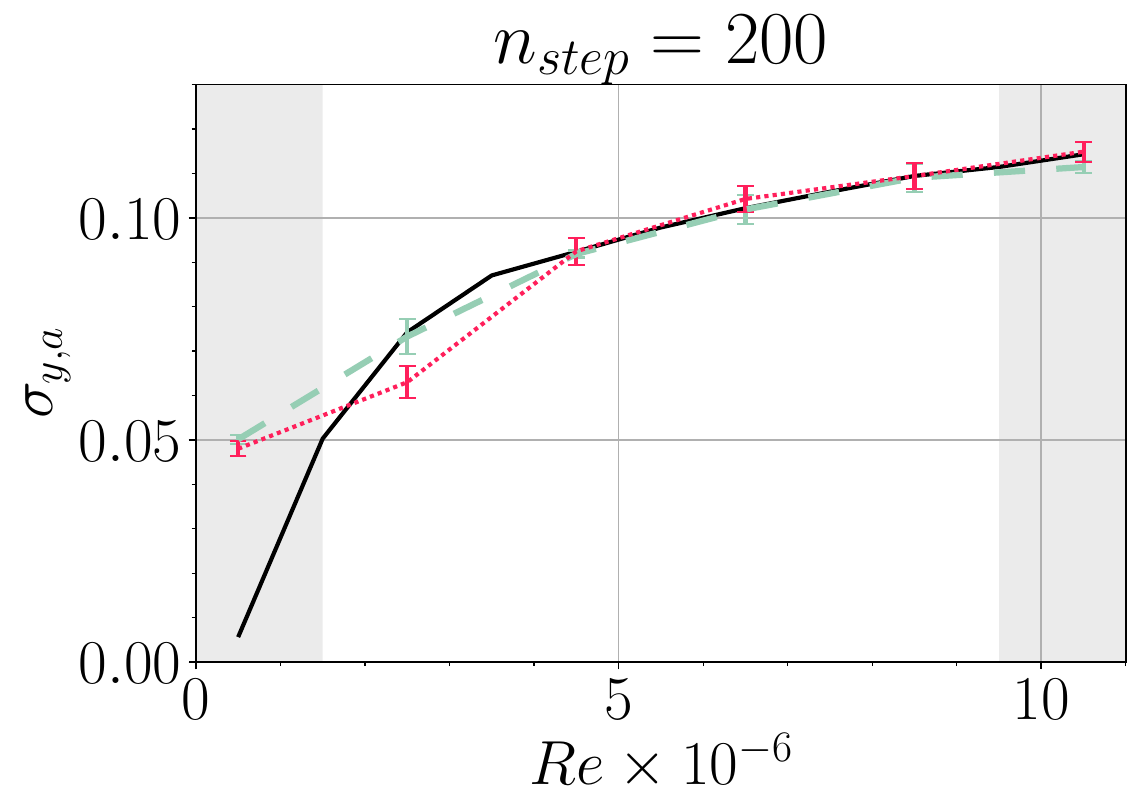}\label{fig:flowmatching_diffusion_200}}\\
    \caption{The performance of flow matching and diffusion models on a single-parameter experiment with different sample steps. }
   \label{fig:1d_flowmatching_dif}
\end{figure}

In this study, we compare flow matching with diffusion models under the same training configurations and neural network sizes. 
Due to the known ground truth, the 1D training scenario of section~\ref{sec:single} provides a very good environment to compare the algorithms.
We employ an Euler scheme to solve the ODE for the sampling in flow matching, ensuring that the number of neural network evaluations corresponds to the number of sampling steps of the diffusion models. As shown in Fig.~\ref{fig:1d_flowmatching_dif}, the flow matching method produces significantly better results than diffusion models with fewer sample steps. With just 20 sample steps, flow matching can accurately predict uncertainty, while diffusion models still exhibit a large prediction error at $Re=2.5\times 10^6$ even with 200 sample steps. Moreover, flow matching yields more stable results, with less variance across different neural network training runs compared to diffusion models, particularly when fewer sample steps are used.

Additionally, flow matching offers greater flexibility in the sampling process after training, as the number of time steps used to solve the sampling ODE can be adjusted. In contrast, diffusion models have fixed sampling steps after training, unless specialized sampling techniques, such as DDIM~\cite{song2021denoising}, are applied. For example, to obtain the results in Fig.~\ref{fig:1d_flowmatching_dif}, we performed 18 training runs of diffusion models  across six parameter groups using three different random seeds, whereas only three flow matching training runs with different random seeds were required. This flexibility allows for a more effective trade-off between inference accuracy and consumption of computational resources.

\begin{figure}[tbh]
    \centering
    \sidesubfloat[]{\includegraphics[scale=0.33]{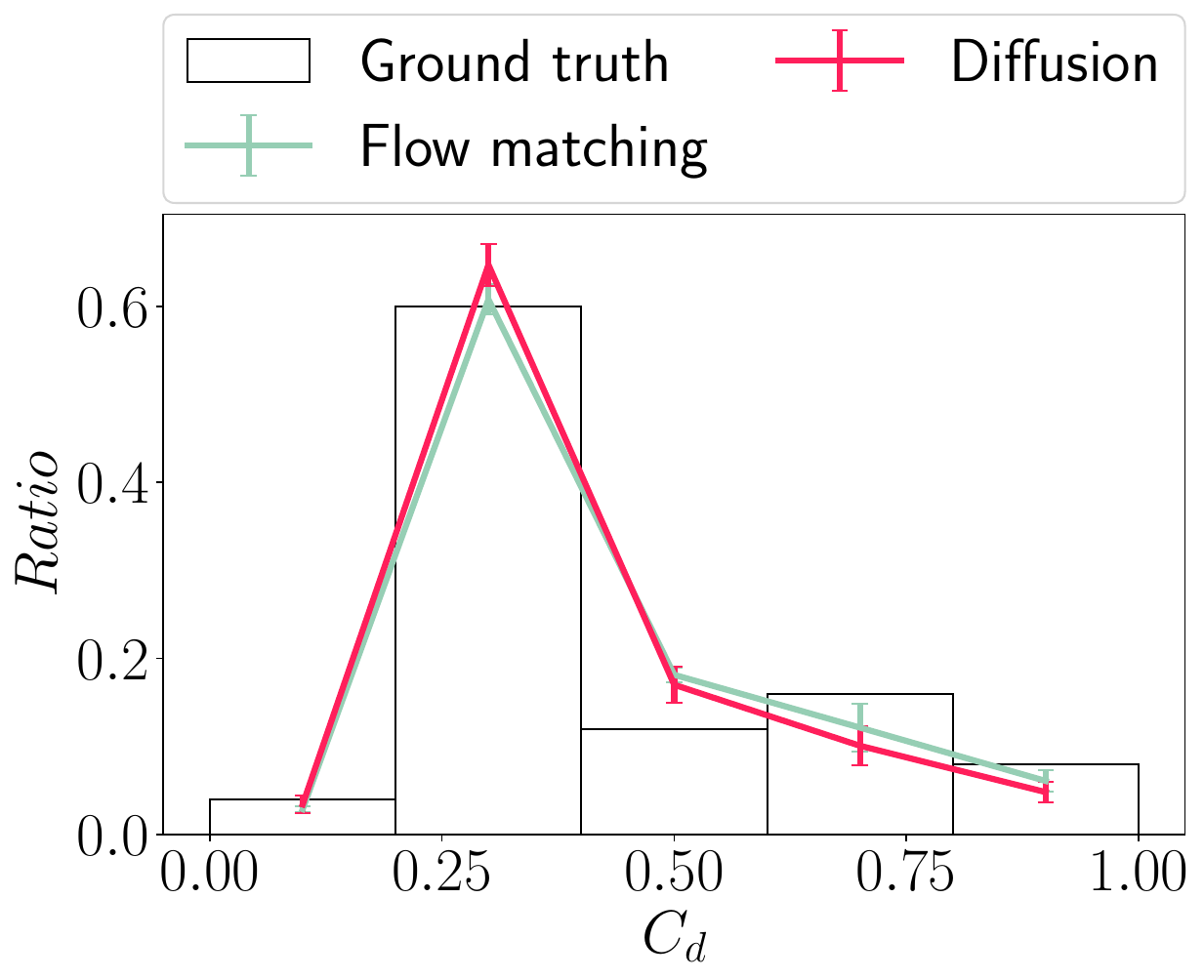}\label{fig:fm_drag_distribution_5}}
    \sidesubfloat[]{\includegraphics[scale=0.33]{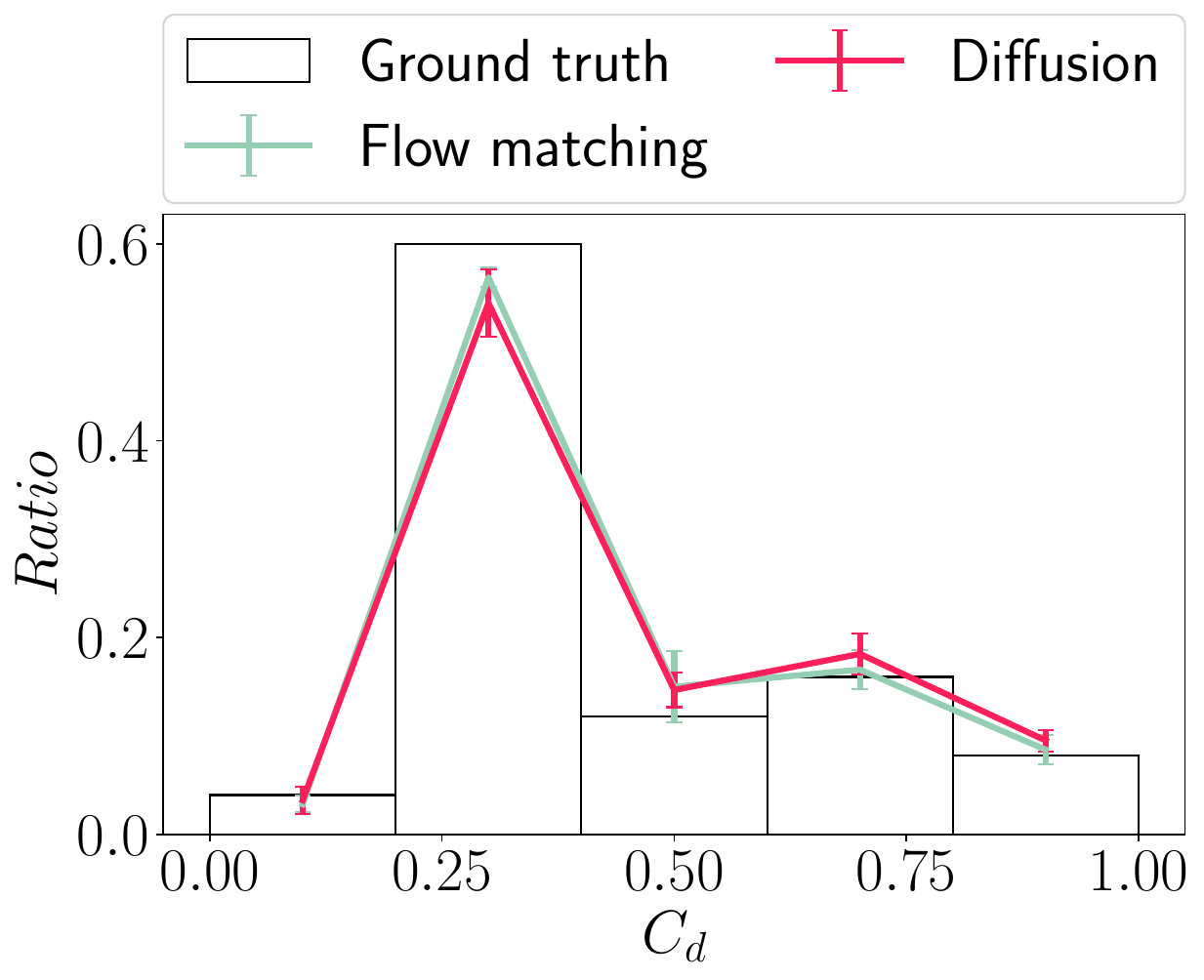}\label{fig:fm_drag_distribution_10}}\\
    \caption{The predicted drag coefficient distribution of flow matching and diffusion models (raf30 airfoil, $\alpha=20.00^\circ$, $Re=6.5\times 10^6$). a) 5 sample steps. b) 200 sample steps.}
   \label{fig:fm_drag_distribution}
\end{figure}

Similar to diffusion models, flow matching evaluates the target distribution by generating samples rather than directly estimating the moments of the distribution. This approach allows for a comprehensive understanding of the distribution, such as obtaining physical samples and accurate drag coefficient distributions. As shown in Fig.~\ref{fig:fm_drag_distribution}, both flow matching and diffusion models provide accurate predictions of the drag coefficient distribution, though flow matching demonstrates slightly better accuracy.

In conclusion, flow matching, as an emerging approach for generative modeling, is highly promising: it offers improved accuracy compared to diffusion models, particularly with fewer sample steps. It retains the advantages of diffusion models in generating complete distributions while addressing the limitation of slow sampling speed. This makes flow matching a promising and efficient tool for uncertainty analysis using generative models.

\appendix
\section*{Appendix\label{sec:app}}
\subsection{Parameter distribution of the dataset\label{sec:app:parameterdistribution}}
The ($Re$,$\alpha$) distribution of the training dataset is chosen to be non-uniform to generate more cases with higher uncertainty: half of the cases in the training dataset are generated with $Re$s and $\alpha$s randomly sampled from the uniform distribution $U(10^6,10^7)$ and $U(-22.5^\circ,22.5^\circ)$, respectively. In contrast, the other half of cases are simulated with $Re$s and $\alpha$s obtained from $f_{sample}(10^6,10^7)$ and $f_{sample}(\pm 22.5^\circ,0)$, respectively. Here, $f_{sample}$ is a sample function:
\begin{equation}
    f_{sample}(a,b)=\left\{\begin{matrix}
 a+(b-a)\frac{e^x-1}{10}  & ,a < b\\
 b+(a-b)\frac{11-e^x}{10}  & ,a >b
\end{matrix}\right. 
    ,
\end{equation}
where $x$ is randomly sampled from $U(0,\mathrm{ln}11)$.

For the test dataset, cases in the interpolation region are generated with the sampling as described above, while cases in the extrapolation region are obtained by  sampling from the enlarged range uniformly.

\subsection{Network Architectures and Training Details\label{sec:app:training}}
\begin{figure}[tbh]
    \centering
    \includegraphics[scale=0.25]{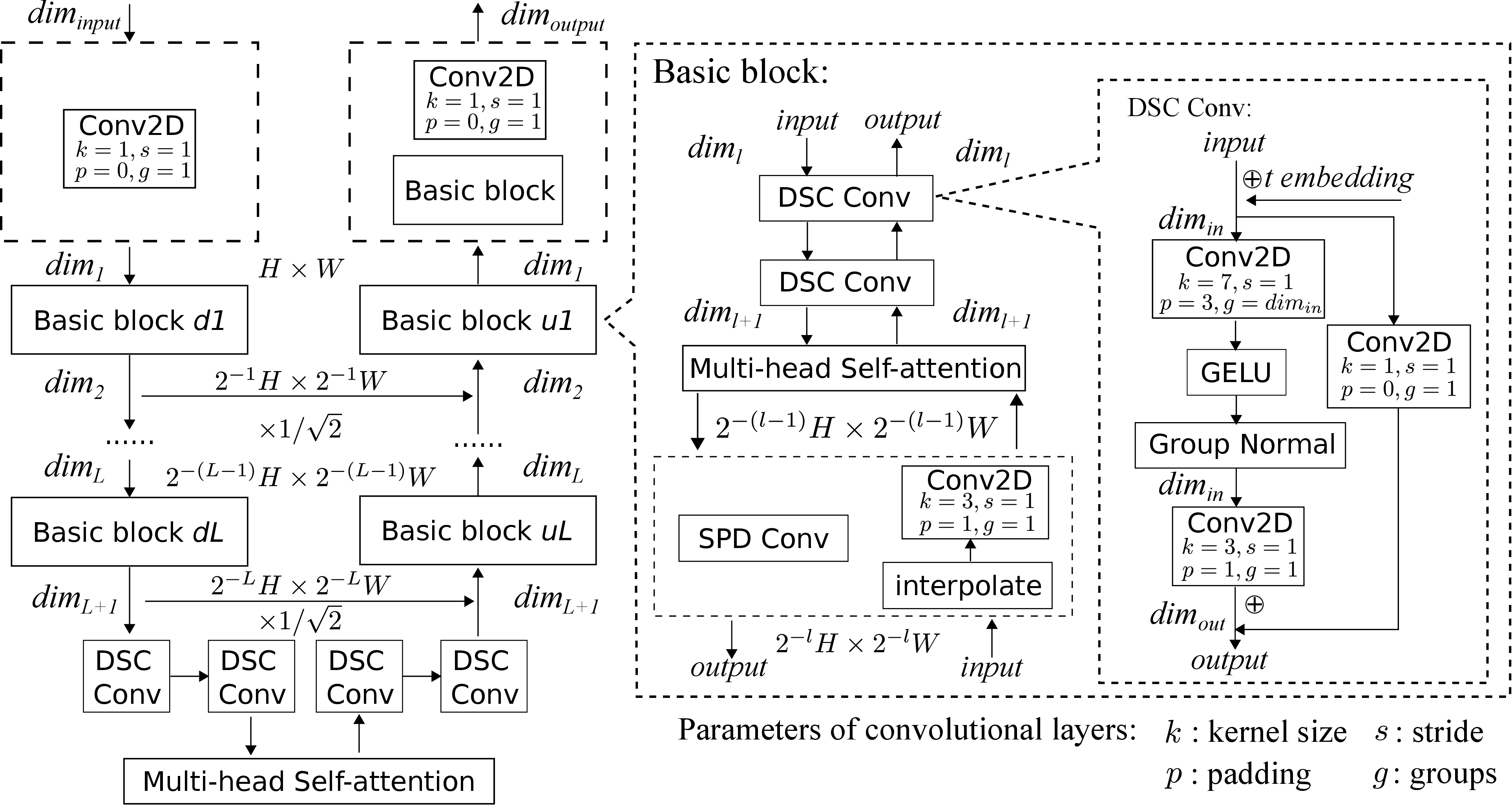}
    \caption{The structure of the U-Net used in the present study.}
    \label{fig:network_structure}
\end{figure}

Following the prevalent DDPM studies, we use a modernized U-Net architecture~\cite{Ho2020,Dhariwal2021}, 
which slightly modifies several components of classic U-Net architectures~\cite{Ronneberger2015,thuerey2020}. 
The network consists of $L$ basic blocks and its structure is shown in Fig.~\ref{fig:network_structure}. Each basic block has two convolutional blocks and one optional multi-head self-attention block~\cite{Vaswani2017} which is activated in the $(L-1)$th and $L$th basic blocks. The convolutional block follows a depthwise separable convolution (DSC) style~\cite{Chollet2017} with a $7\times7$ depthwise convolutional layer and a $3\times3$ pointwise convolutional layer. Besides, an SPD-Conv layer~\cite{Sunkara2023} and an interpolation layer followed by a convolution are used for the downsampling and upsampling, respectively. The initial block of the U-Net is built with a 1$\times$1 convolutional layer to expand the input channels, and the final block is built with a basic block followed by a convolutional layer. In the bottleneck of the U-Net, there are four DSC convolutional layers with a multi-head self-attention block in the middle.

In our experiments, the DDPM model, heteroscedastic model, and BNN model for a certain resolution all use the same network architecture. The major differences between these three models are the number of input and output channels. 
For the network of DDPM, there are 3 channels for the noise field $\mathbf{y}^t_{i}$ and 3 channels for the condition $\mathbf{x}$ in the input. The number of output channels is also 3, representing the predicted noise $ \boldsymbol{\epsilon}_\theta$. 
The input for the heteroscedastic network and BNN are both the 3-channel condition $\mathbf{x}$. While the output of the heteroscedastic model is a $2\times3$-channel tensor representing the predicted $\boldsymbol{\mu}_{\mathbf{y},\theta}$ and $\boldsymbol{\sigma}_{\mathbf{y},\theta}$. For the BNN, the output is only a 3-channel predicted $\mathbf{y}_{i,\theta}$. Besides, the time embedding for the heteroscedastic model and BNN is kept constant $t=200$ as this information is not used in these two variants. In the BNN network, all convolutional layers are replaced with the Flipout Monte Carlo estimator convolutional layers~\cite{Yeming2018}, which we implement with the  BayesianTorch package~\cite{krishnan2022bayesiantorch}. The resulting number of trainable parameters of the networks used in the present study are summarized in Table.~\ref{tab:network_parameter}.

\begin{table}[]
\caption{The network parameters for different models}
\label{tab:network_parameter}
\begin{tabular}{cccccc}
\hline
Data size                       & \multirow{2}{*}{Number of channels in each layer} & \multicolumn{4}{c}{Number of trainable parameters}                                                                                                                              \\
$s\times s$                     &                                                   & DDPM                              & Heteroscedastic model                              & BNN                                                & DFP Net~\cite{thuerey2020}                           \\ \hline
$32\times32$                    & {[}16,32,64,64{]}                                 & 1185218                           & 1185686                                            & 2367332                                            & \textbackslash{} \\
$64\times64$                    & {[}16,32,64,64,128{]}                             & 3208770                           & \textbackslash{}                  & \textbackslash{}                  & \textbackslash{} \\
\multirow{2}{*}{$128\times128$} & {[}32,64,128,128,256,256{]}                       & 19766642                          & \multirow{2}{*}{\textbackslash{}} & \multirow{2}{*}{\textbackslash{}} & \textbackslash{} \\
                                & {[}128,256,256,512,1024,1024,1024{]}              & \textbackslash{} &                                                    &                                                    & 30905859                          \\ \hline
\end{tabular}
\end{table}


All the networks are trained with the AdamW optimizer using $\beta_1=0.5$ and $\beta_2=0.999$. The training uses a batch size of 50 for the data with the resolution of $32\times32$, $64\times64$, and 25 for $128\times128$. The initial learning rate is $1\times10^{-4}$ and the final learning rate is $1\times10^{-5}$ with a learning rate decay every $12.5\times10^4$ iteration for the training of $32\times32$ and $64\times64$ data. We use the same learning rate decay for $128\times128$ data while the initial learning rate is set to $5\times10^{-5}$. 
All networks are trained with $12.5\times10^6$ iterations at which the training loss has largely converged.
However, we found that the heteroscedastic model overfits after $2\times10^6$ iterations in the multi-parameter experiments. Thus all results of the heteroscedastic model are obtained with $2\times10^6$th iterations. The Bayes by Backprop (BBB)~\cite{Blundell2015} method is used to update the parameters distribution of BNN during the backpropagation.

The DFP network used in Sec.~\ref{sec:experiment:large_resolution} is a pre-trained neural network from the RANS airfoil benchmark setup outlined above~\cite{thuerey2020}. It uses a channel exponent factor to control the network size, which was set to 7 to obtain a network with ca. 30m trainable parameters, as shown in Table.~\ref{tab:network_parameter}. 
The details of the network architecture and training procedure of DFP net can be found in Ref~.\cite{thuerey2020}. 

The number of diffusion steps of DDPM used in the present study is $T=200$. We have also tested the performance of DDPM with a varying number of steps, i.e. $T=100$ and $T=400$. However, both models perform similarly to $T=200$ for the predicted expectations  of low uncertainty cases, while the diffusion model with $T=200$ slightly outperformed the other models in the high uncertainty cases. Thus, the experiments in our manuscript focus on DDPM models with $T=200$.

\subsection{Details of the neural network input\label{sec:app:discussion_input}}
The decision to encode parameters $\alpha$, $Re$, and $\Omega$ as constant, three-channel fields is motivated by several factors as outlined below. 
\paragraph{Network architecture compatibility.} The DDPM approach profits from a UNet structure~\cite{Ho2020}, and has been widely employed in published literature for DDPM. The UNet's convolutional nature requires fields as both input and output. 
\paragraph{Fair comparison with BNNs and heteroscedastic models.} In the present study, the input to the UNet in the DDPM consists of a six-channel field, incorporating both noisy fields ($u$, $v$, $p$) and conditioning fields ($\alpha$, $Re$, $\Omega$). An alternative approach could be to utilize only the noisy fields as input, incorporating a separate encoder for the conditioning values ($\alpha$, $Re$, $\Omega$). The encoded scalar information could then be added as an embedding for the UNet, aligning with common practices in text-image generation~\cite{nokey2022,Yang2023}. However, it is pertinent to note that the introduced encoder component is deemed unnecessary for BNNs and heteroscedastic models. These models exclusively require $\alpha$, $Re$, and $\Omega$ as input fields for the UNet. Thus, we have opted to employ field input for $\alpha$, $Re$, and $\Omega$ in DDPM to ensure consistency in input data representation for different methods and avoid unnecessary complexity in the network architecture for BNNs and heteroscedastic models. This maintains a fair and comparable experimental setup across all methods.
\paragraph{Information about airfoil shapes in simulation results.} The decision to directly use the airfoil shape $\Omega$ as a field aligns with the inherent information about the airfoil shape contained in the simulation result. The preprocessing step to obtain the field of airfoil shape from the OpenFOAM simulations is considered natural and straightforward, similar to the extraction of velocity and pressure fields ($u$, $v$, $p$). Another possible alternative solution was to integrate the ($Re$, $\Omega$) field into the airfoil shape, with values $\Omega$ field representing $\alpha$ and $Re$ instead of 0 and 1. However, this approach would introduce challenges in balancing the proportion of $\alpha$ and $Re$ in the single-channel field. Besides, there are no substantial changes in the network size with different numbers of input channels, as shown in Table \ref{tab:size_with_channel}.

\begin{table}[h]
\caption{The size of network with different number of input channels}
\label{tab:size_with_channel}
\begin{tabular}{ccccccc}
\hline
\begin{tabular}[c]{@{}c@{}}Data size\\  ($s \times s$)\end{tabular} & $n_{c,in}=1$ & $n_{c,in}=2$ & $n_{c,in}=3$ & $n_{c,in}=4$ & $n_{c,in}=5$ & $n_{c,in}=6$ \\ \hline
$32 \times 32$                                                      & 1185138      & 1185154      & 1185170      & 1185186      & 1185202      & 1185218      \\
$64 \times 64$                                                      & 3208690      & 3208706      & 3208722      & 3208738      & 3208754      & 3208770      \\
$128 \times 128$                                                    & 19766482     & 19766514     & 19766546     & 19766578     & 19766610     & 19766642     \\ \hline
\end{tabular}
\end{table}

In summary, the choice to encode parameters as three-channel fields serves to harmonize the network architecture requirements, facilitate fair comparisons, and leverage the existing mesh information in the OpenFOAM simulation results.

\subsection{Extended discussion of the single-parameter experiments of BNNs\label{sec:app:discusssion_bnn}}
In the single-parameter experiments, the accuracy of the BNNs' predictions for expectation fields decreases as $\lambda$ increases. Additionally, the magnitude of the predicted standard deviation field amplifies with $\lambda$, while the distribution pattern of the standard deviation always deviates from the ground truth. This observed trend is deeply rooted in the nature of BNNs. The probabilistic nature of BNN predictions is achieved through the probabilistic distribution of network parameters.  Each prediction sample from BNNs results from sampling network parameters from a distribution within the parameter space. 

When the distribution variance of neural network parameters is large, the variance of prediction results using sampled parameters is also substantial. Conversely, decreasing the distribution variance yields predictions with lower variability. As elucidated in the manuscript, the coefficient $\lambda$ adjusts the strength of the loss term which makes the distribution of network parameters conform to the prior distribution, as shown in Eq.~\ref{eq:bnn:loss_func}. When $\lambda$ tends to zero, the network parameters cease to follow a probabilistic distribution. The remaining term in the loss function aims to maximize the log-likelihood of $\mathbb{E}_{q_{\phi}[\log(p(\mathbf{d}|{\theta}))]}$, aligning predictions closely with the ground truth dataset. In this scenario, the standard deviation of BNNs' prediction becomes zero, and the accuracy in expectation predictions is highest.

Conversely, as $\lambda$ increases, the distribution of the network's parameter gradually adheres to the prior distribution. In extreme cases where the KL divergence dominates, the network learns minimally from the data, focusing primarily on matching the prior distribution. This circumstance results in the lowest accuracy for expectation predictions since the network scarcely learns from the data. However, it doesn't imply that standard deviation predictions attain the highest accuracy, as the correctness of the prior distribution is not guaranteed. In our case, the standard practice involves a Gaussian distribution as the prior. Predictions with parameters sampled from a Gaussian distribution may not align well with the ground truth data. 

Fig.~\ref{fig:1D_std_linecompare_veall} demonstrates that the standard deviation magnitude is close to zero for small $\lambda$, but at $\lambda=0.01$, it already surpasses the ground truth magnitude. Further increases in $\lambda$ could lead to even greater deviations from the ground truth. In summary, the intrinsic properties of BNNs make it challenging to definitively assert how standard deviation prediction accuracy changes with the coefficient $\lambda$. Small $\lambda$ renders the neural network deterministic, resulting in zero standard deviation predictions. Conversely, increasing $\lambda$ moves the distribution of network parameters toward the prior distribution. However, ensuring consistency with the real solution using the network's parameter from the prior distribution is challenging without knowledge of the "correct" distribution for the network's parameters. This potentially leads to increased prediction errors.

\subsection{Drag coefficient calculation\label{sec:app:drag_coef}}
The drag coefficient in the present study is calculated as
\begin{equation}
    C_d=\frac{\mathbf{F_d}}{0.5\rho\mathbf{u_f}^2 A}
\approx \frac{\sum_k^{s^2} \left[\mathrm{p} _k \mathbf{n}_k+\mu \mathbf{n}_k\times\left(\nabla\times \mathbf{u} \right)_k \right]h }{0.5\rho\mathbf{u_f}^2 l} \frac{\mathbf{u_f}}{|\mathbf{u_f}| } 
    ,
\end{equation}
where $\mathbf{F_d}$ is the drag force, $\rho$ is the density of air, $A$ is the reference area chosen as the wing area, $h$ is the cell size of the prediction field, subscript $k$ represents the $k$th data in the field, and $\mathbf{n}$ is the unit normal vector field of the airfoil shape calculated as
\begin{equation}
    \mathbf{n}=\frac{\nabla \Omega}{|\nabla \Omega|}
    .
\end{equation}
Here, all gradient calculations are directly performed on the $s \times s$ data using convolutions.
\subsection{Generalization\label{sec:app:generalization}}
\begin{figure}[tbhp]
    \centering
    \sidesubfloat[]{\includegraphics[scale=0.33]{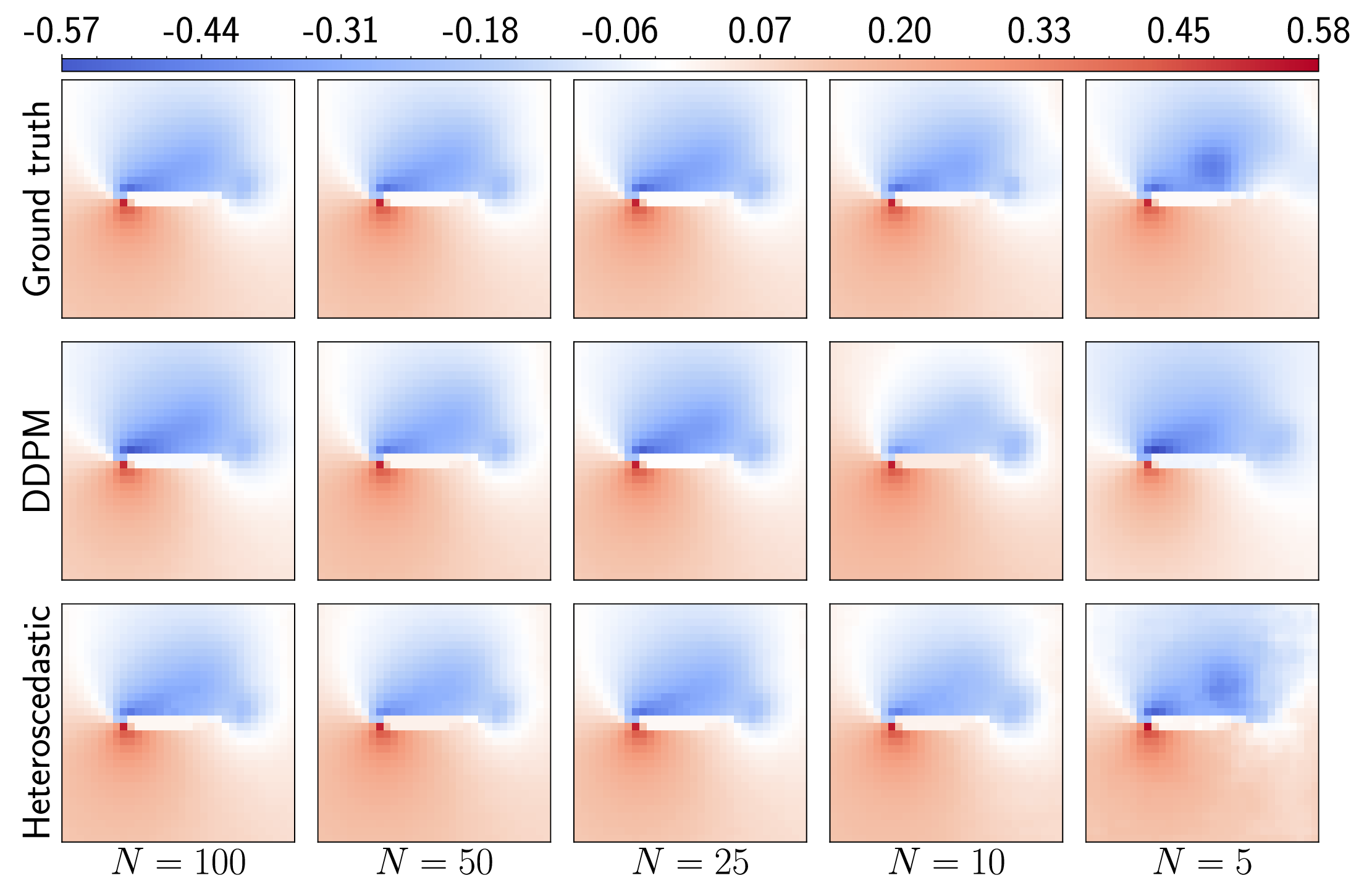}\label{fig:1D_number_of_samples_mean}}
    \\
    \sidesubfloat[]{\includegraphics[scale=0.33]{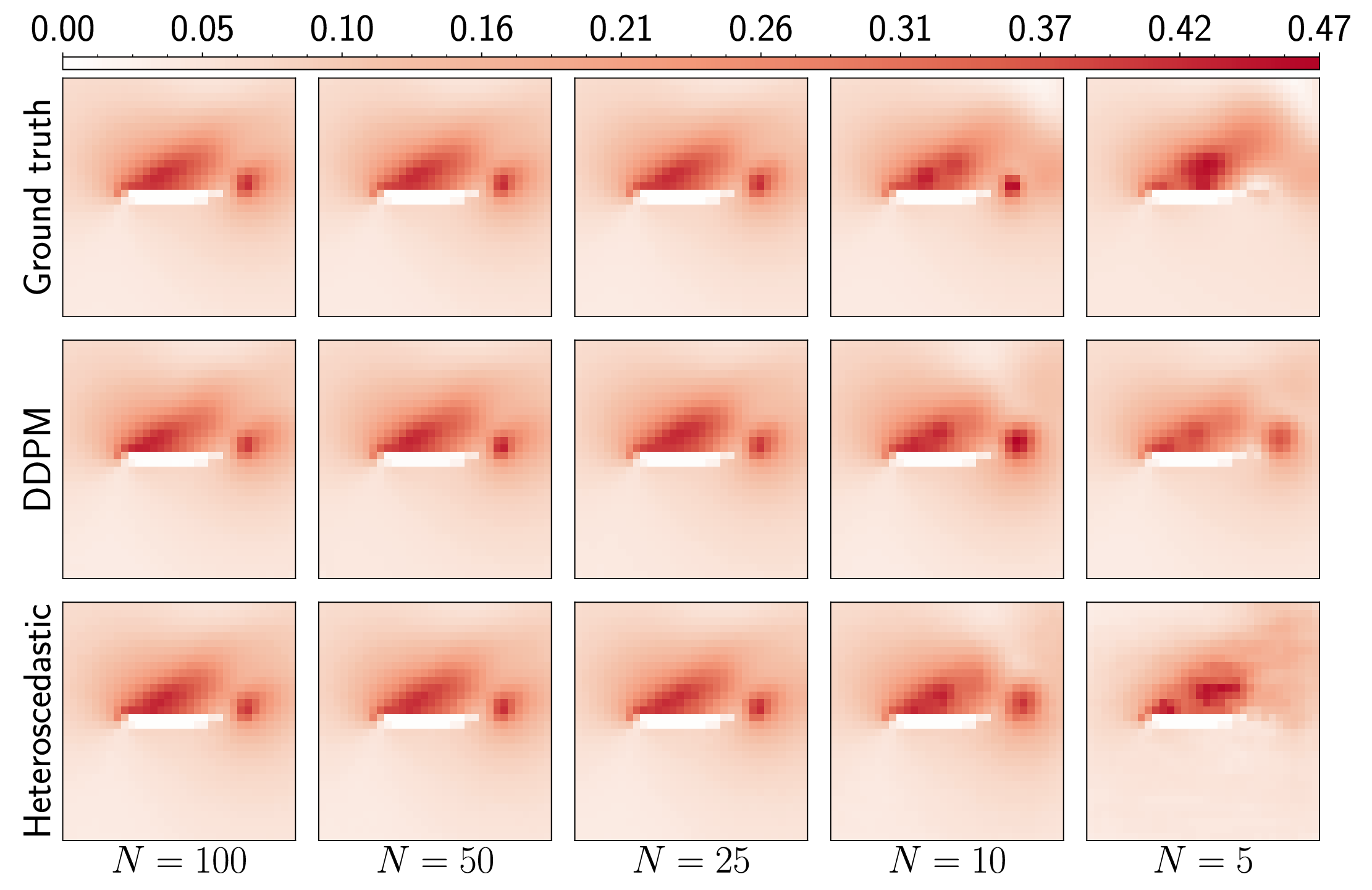}\label{fig:1D_number_of_samples_std}}
    \caption{Distributions of the a) expectation and b) standard deviation for pressure field with different number of snapshots samples (raf30 airfoil, $Re=6.5\times 10^6$, $\alpha=20.00^\circ$).}
    \label{fig:1D_number_of_samples}
\end{figure}

The ground truth and prediction of the expectation and standard deviation distribution for the pressure field with different number of snapshot samples in single-parameter experiments are shown in Fig.~\ref{fig:1D_number_of_samples}. Significant differences in fields only occur when $N<25$ for both ground truth and model predictions.

\subsection{Test set outputs\label{sec:app:testset}}
\begin{figure}[tbhp]
    \centering
    \sidesubfloat[]{\includegraphics[scale=0.26]{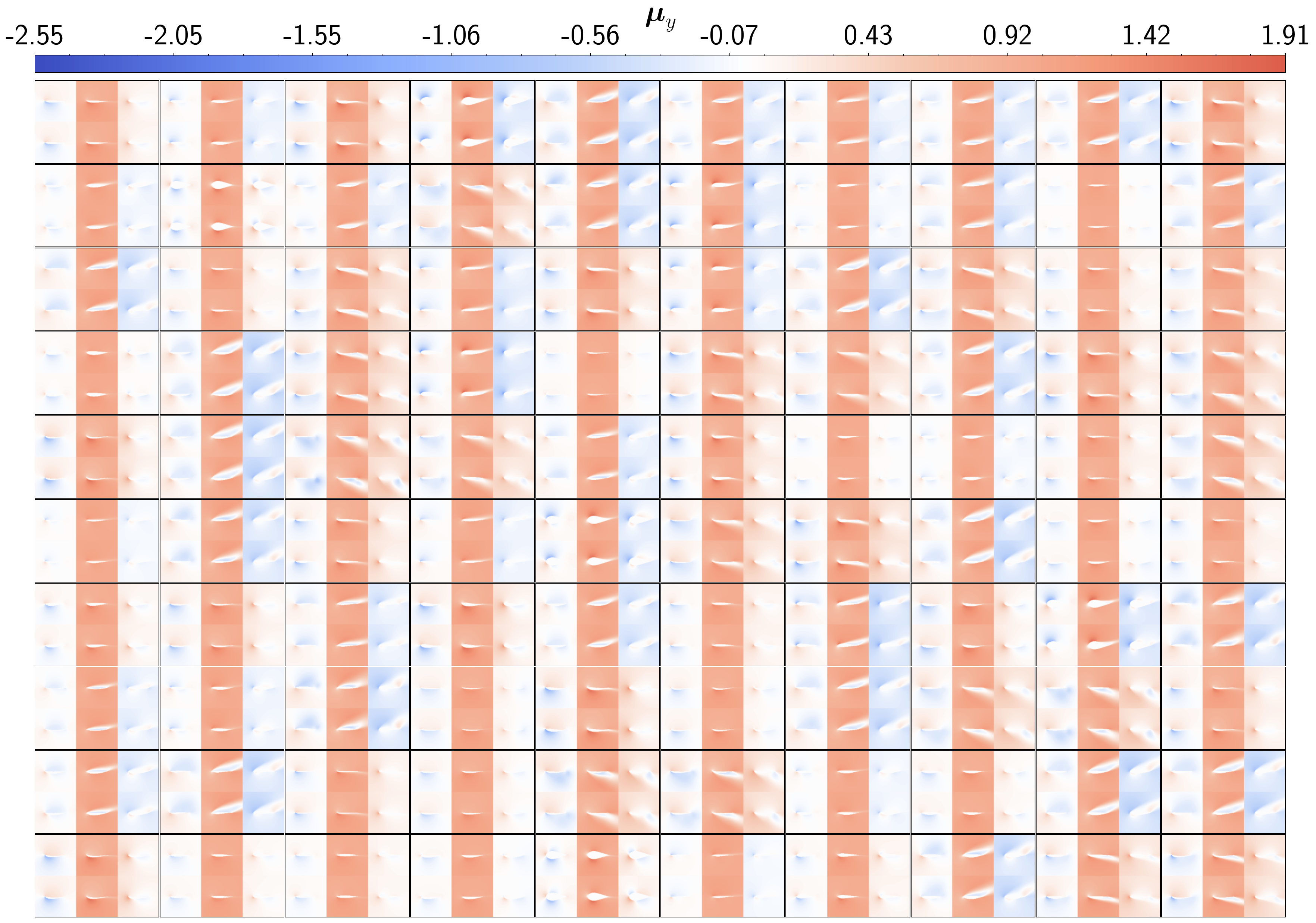}\label{fig:mean_testall_in}}
    \\
    \sidesubfloat[]{\includegraphics[scale=0.26]{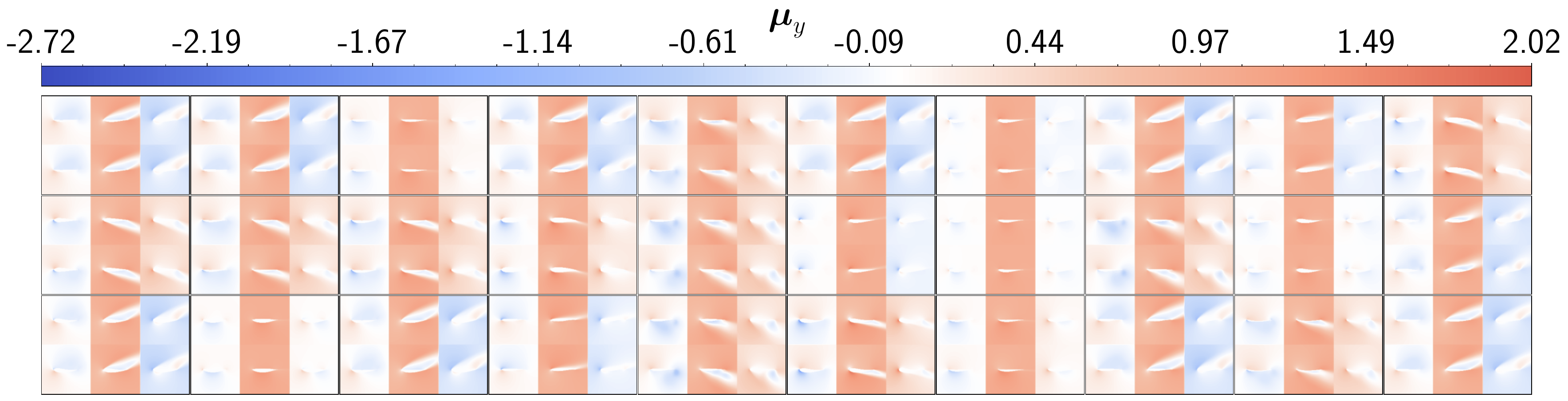}\label{fig:mean_testall_out}}
    \caption{The $(\boldsymbol{\mu}_{\mathrm{p^*}},\boldsymbol{\mu}_{\mathrm{u_x^*}},\boldsymbol{\mu}_{\mathrm{u_y^*}})$ distribution from DDPM (top) and ground truth (bottom) with $128\times 128$ test set. a) Interpolation region. b) Extrapolation region.}
    \label{fig:test_all_mean}
\end{figure}

\begin{figure}[tbhp]
    \centering
    \sidesubfloat[]{\includegraphics[scale=0.26]{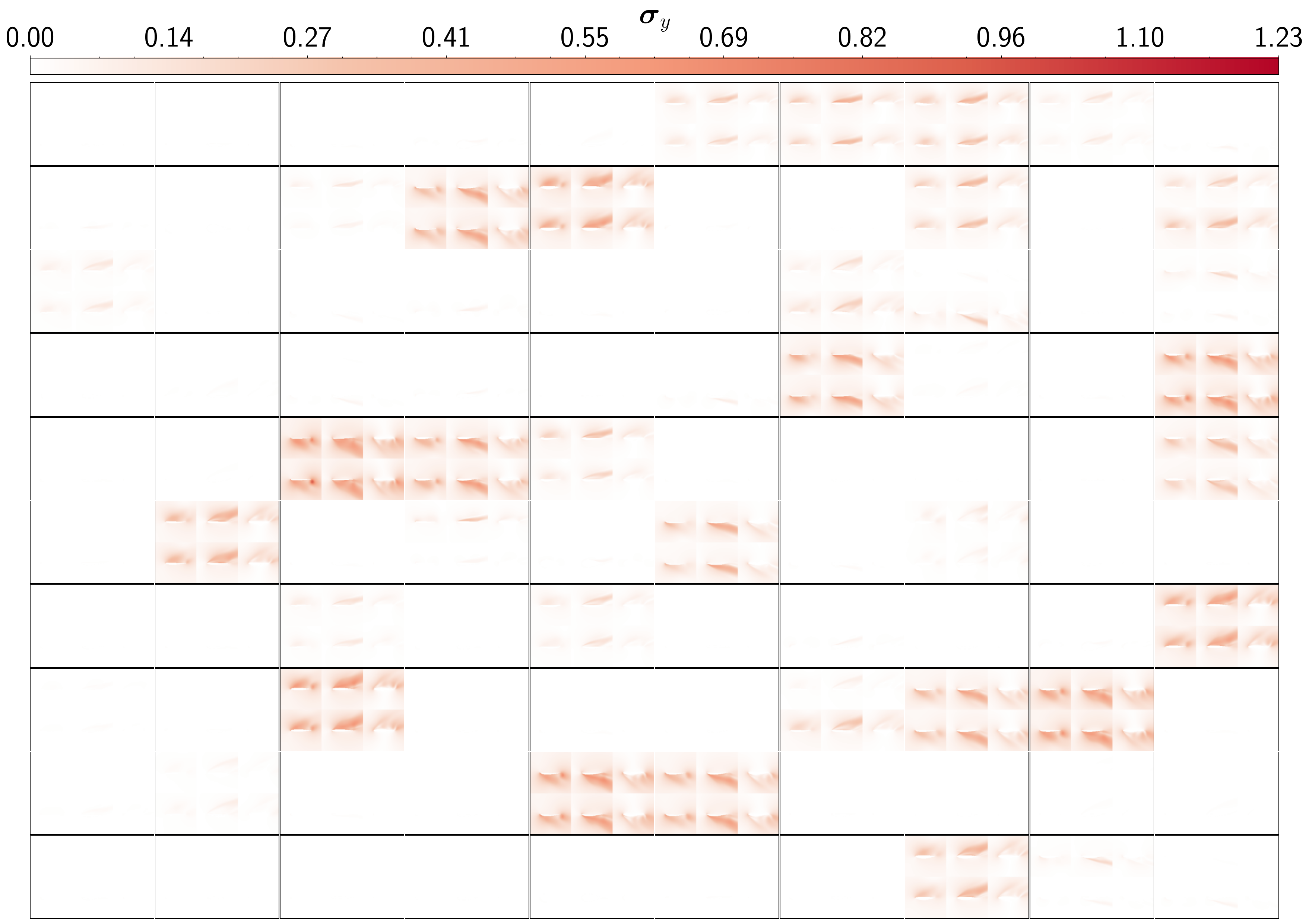}\label{fig:std_testall_in}}
    \\
    \sidesubfloat[]{\includegraphics[scale=0.26]{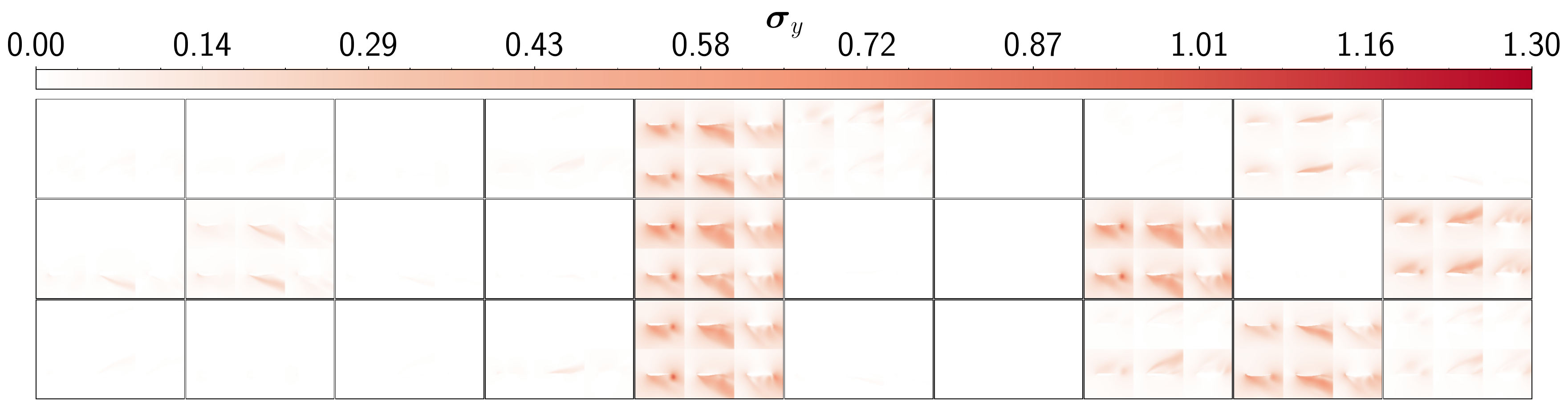}\label{fig:std_testall_out}}
    \caption{The $(\boldsymbol{\sigma}_{\mathrm{p^*}},\boldsymbol{\sigma}_{\mathrm{u_x^*}},\boldsymbol{\sigma}_{\mathrm{u_y^*}})$ distribution from DDPM (top) and ground truth (bottom) with $128\times 128$ test set. a) Interpolation region. b) Extrapolation region.}
    \label{fig:test_all_std}
\end{figure}

The full set of DDPM predictions evaluated on the whole test dataset with the output resolution of $128\times 128$ is shown in Fig.~\ref{fig:test_all_mean} and Fig.~\ref{fig:test_all_std}.

\section*{Funding Sources}
Qiang Liu was supported by the China Scholarship Council (No.202206120036
) for Ph.D research at the Technical University of Munich.

\clearpage
\bibliography{main}

\end{document}